\documentclass[12pt,preprint]{aastex62}
\usepackage{lineno}
\usepackage{natbib}
\usepackage{subfigure}
\usepackage{float}
\usepackage{graphicx}%
\usepackage{threeparttable}
\usepackage{longtable}
\newcommand{\be}{\begin{equation}}
\newcommand{\ee}{\end{equation}}

\begin{document}
\title{The First LHAASO Catalog of Gamma-Ray Sources}
 
\author{Zhen Cao}
\affiliation{Key Laboratory of Particle Astrophyics \& Experimental Physics Division \& Computing Center, Institute of High Energy Physics, Chinese Academy of Sciences, 100049 Beijing, China}
\affiliation{University of Chinese Academy of Sciences, 100049 Beijing, China}
\affiliation{Tianfu Cosmic Ray Research Center, 610000 Chengdu, Sichuan,  China}
 
\author{F. Aharonian}
\affiliation{Dublin Institute for Advanced Studies, 31 Fitzwilliam Place, 2 Dublin, Ireland }
\affiliation{Max-Planck-Institut for Nuclear Physics, P.O. Box 103980, 69029  Heidelberg, Germany}
 
\author{Q. An}
\affiliation{State Key Laboratory of Particle Detection and Electronics, China}
\affiliation{University of Science and Technology of China, 230026 Hefei, Anhui, China}
 
\author{Axikegu}
\affiliation{School of Physical Science and Technology \&  School of Information Science and Technology, Southwest Jiaotong University, 610031 Chengdu, Sichuan, China}
 
\author{Y.X. Bai}
\affiliation{Key Laboratory of Particle Astrophyics \& Experimental Physics Division \& Computing Center, Institute of High Energy Physics, Chinese Academy of Sciences, 100049 Beijing, China}
\affiliation{Tianfu Cosmic Ray Research Center, 610000 Chengdu, Sichuan,  China}
 
\author{Y.W. Bao}
\affiliation{School of Astronomy and Space Science, Nanjing University, 210023 Nanjing, Jiangsu, China}
 
\author{D. Bastieri}
\affiliation{Center for Astrophysics, Guangzhou University, 510006 Guangzhou, Guangdong, China}
 
\author{X.J. Bi}
\affiliation{Key Laboratory of Particle Astrophyics \& Experimental Physics Division \& Computing Center, Institute of High Energy Physics, Chinese Academy of Sciences, 100049 Beijing, China}
\affiliation{University of Chinese Academy of Sciences, 100049 Beijing, China}
\affiliation{Tianfu Cosmic Ray Research Center, 610000 Chengdu, Sichuan,  China}
 
\author{Y.J. Bi}
\affiliation{Key Laboratory of Particle Astrophyics \& Experimental Physics Division \& Computing Center, Institute of High Energy Physics, Chinese Academy of Sciences, 100049 Beijing, China}
\affiliation{Tianfu Cosmic Ray Research Center, 610000 Chengdu, Sichuan,  China}
 
\author{J.T. Cai}
\affiliation{Center for Astrophysics, Guangzhou University, 510006 Guangzhou, Guangdong, China}
 
% ...
\collaboration{The LHAASO Collaboration}

\correspondingauthor{S.Q. Xi, S.C. Hu, S.Z. Chen, M. Zha}
\email{xisq@ihep.ac.cn, hushicong@ihep.ac.cn, chensz@ihep.ac.cn, zham@ihep.ac.cn}
\begin{abstract}
We present the first catalog of very-high energy and ultra-high energy gamma-ray sources detected by the Large High Altitude Air Shower Observatory (LHAASO).
The catalog was compiled using 508 days of  data collected by the Water Cherenkov Detector Array (WCDA) from March 2021 to September 2022 and  933  days  of  data recorded by the Kilometer Squared Array (KM2A) from January 2020 to September 2022. This catalog represents the main result from the most sensitive large coverage gamma-ray survey of the sky above 1 TeV, covering declination from $-$20$^{\circ}$ to 80$^{\circ}$. In total, the catalog contains 90 sources with an extended size smaller than $2^\circ$ and a significance of detection at $> 5\sigma$. Based on our source association criteria, 32 new TeV sources are proposed in this study. Among the 90 sources, 43 sources are detected with ultra-high energy ($E > 100$ TeV) emission at $> 4\sigma$ significance level. 
We provide the position, extension, and spectral characteristics of all the sources in this catalog.

\end{abstract}

\section{Introduction}
Gamma rays, located at the highest energy band of the cosmic electromagnetic radiation, serve as a powerful probe for astrophysics and fundamental physics under extreme conditions. Most gamma rays are produced through the acceleration and propagation of relativistic particles, such as protons or electrons, in astrophysical sources. Relativistic electrons can scatter low-energy photons, including star light, dust scattered light and the Cosmic Microwave Background (CMB), to the gamma-ray band via the inverse Compton process.The whole universe is filled with low energy photons, especially the CMB. Therefore, gamma rays are a unique tool to probe the relativistic electrons when the surrounding magnetic field is weak and the synchrotron radiation is undetectable. Relativistic protons interact with the surrounding medium to create hadronic cascades, which include secondary $\pi^{0}$ mesons that quickly decay into gamma-rays. Hence, gamma rays are also an important tool to study the origin, acceleration and propagation of cosmic rays (CRs).

Thanks to the advancements in space-based and ground-based gamma-ray detectors, our knowledge about the high energy gamma-ray universe has made impressive progress over the past two decades. At high energy (HE, $E > 0.1$ GeV), the Fermi Large Area Telescope~\citep[\textsl{Fermi}-LAT,][]{2009ApJ...697.1071A} has been surveying the whole sky since 2008 and detected 6658 Galactic and extragalactic gamma-ray sources using the first twelve years of observations~\citep{2022ApJS..260...53A}. Compared to its predecessor, EGRET~\citep{1999ApJS..123...79H}, the number of detected sources has increased by a factor of more than 20, and some new categories of gamma-ray emitters have been revealed. Important evidence about the acceleration of GeV--TeV CRs were found in two ancient supernova remnants~\citep{2013Sci...339..807A}. The Dark Matter Particle Explorer (DAMPE) collaboration also reported the detection of more than 200 gamma-ray sources above GeV \citep{DAMPE:2021kao}.
 
At Very High Energy (VHE, $E > 0.1$ TeV), ground-based gamma-ray detectors are necessary due to their large area requirements. The successful operation of the second generation Imaging Atmospheric Cherenkov Telescopes (IACTs), such as H.E.S.S.~\citep{2006A&A...457..899A}, MAGIC~\citep{2016APh....72...76A}, and VERITAS~\citep{2015ICRC...34..792M},
has significantly increased the number of detected VHE gamma-ray sources from about 10 to over 200 since 2004. Several categories of VHE gamma-ray emitters have been firmly established, including Active Galactic Nuclei (AGNs), pulsars and Pulsar Wind Nebulae (PWNe), Supernova Remnants (SNRs), binaries, starburst galaxies, Gamma-Gay Bursts (GRBs) and others. However, due to their narrow Field Of View (FOV, 3$^{\circ}$ $-$ 5$^{\circ}$) and low duty cycle (10\% $-$15\%), IACTs are not suitable for performing long-term comprehensive sky surveys. Most VHE sources are detected while searching for counterparts of sources observed at lower energies, and only certain parts of galactic plane have been surveyed by the IACTs, such as  
H.E.S.S.~\citep{2018A&A...612A...1H} and VERITAS~\citep{2015ICRC...34..868S}. 

To achieve an overall view of the VHE universe, a roughly unbiased sky survey is needed, similar to that carried out by \textsl{Fermi}-LAT in the GeV band. Ground-based Extensive Air Shower (EAS) arrays, with their large field of view and high duty cycle, are ideal detectors for this scientific goal. 
Several VHE gamma-ray surveys with positive results have been conducted to date, including those by Tibet AS$\gamma$~\citep{2005ApJ...633.1005A}, Milagro~\citep{2007ApJ...664L..91A}, and ARGO-YBJ~\citep{2013ApJ...779...27B}. 
However, due to the limitations of detector sensitivity, only a handful of sources have been observed. Nonetheless, impressive progress has been made in observing some typical extended sources~\citep{2014ApJ...790..152B} and variable AGNs~\citep{2016ApJS..222....6B,2012ApJ...758....2B}, highlighting the invaluable role of EAS arrays in VHE gamma-ray observation.
Recently, the sensitivity of EAS arrays has greatly improved thanks to the successful operation of the new generation arrays, such as HAWC and LHAASO. The HAWC Observatory has detected  65 VHE sources including 20 new ones using five years of data in their third catalog~\citep[3HWC;][]{2020ApJ...905...76A}. In particular, HAWC has revealed a new source category,the TeV pulsar halo~\citep{2017Sci...358..911A}, which is a useful tool to probe the CR diffusion in the interstellar medium (ISM) near PWNe.

Another crucial characteristic of the EAS array is that it can extend the observation to the Ultra-High Energy (UHE, $E > 0.1$ PeV) range due to its large detector area and long duty cycle.
The Tibet AS$\gamma$~collaboration first reported a UHE gamma-ray source, the Crab Nebula~\citep{2019PhRvL.123e1101A}, followed by another three sources reported by the HAWC collaboration~\citep{2020PhRvL.124b1102A}. 
Recently, the LHAASO collaboration has made exciting progress by reporting the detection of 12 UHE gamma-ray sources with a statistical significance of over 7$\sigma$ and the maximum energy up to 1.4 PeV~\citep{2021Natur.594...33C}. Additionally, some sources were observed with VHE emission for the first time~\citep{2021ApJ...919L..22C,2021ApJ...917L...4C}. These observations provide crucial candidates to explore the origin of PeV CRs within the Galaxy. LHAASO also detected PeV gamma-ray emission from the Crab Nebula, revealing an extreme electron accelerator with unprecedented high accelerating efficiency~\citep{2021Sci...373..425L}.  These observation involving the highest gamma-ray energy also shed light on exploring the Lorentz Invariance Violation~\citep{2022PhRvL.128e1102C} and dark matter~\citep{2022PhRvL.129z1103C}. It is worth noting that these results were achieved  using only  1 year of data and half of the LHAASO array prior to completion of its construction.


This paper is structured as follows. Section 2 briefly describes the LHAASO detector and the data set. Additionally, it presents the background estimation and significance sky maps. In Section 3, the methods of searching for sources and constructing the catalog are introduced. The characteristics of WCDA and KM2A source components are briefly described and compared. In Section 4, a final source catalog is compiled, including relevant positional, spatial, and spectral information. The results of some typical source populations are discussed in Section 5. Section 6 provides a summary.

\section{Instrument and  Data Analysis}
\subsection{The LHAASO Detector and Data}
LHAASO is a multi-purpose and comprehensive EAS array, designed for the study of CRs and gamma rays across a wide energy range, from sub-TeV to beyond 1 PeV~\citep{2022ChPhC..46c0001M}. It consists of three interconnected detector arrays, a 1.3 km$^2$ array  (KM2A) for gamma-ray detection above 10 TeV, a 78,000 m$^2$  Water Cherenkov Detector Array (WCDA) for TeV gamma-ray detection, and a Wide Field-of-view Cherenkov Telescopes Array (WFCTA) mainly for CR physics. 
 When a high-energy extraterrestrial particle,  gamma ray or CR, enters Earth’s atmosphere, it initiates a cascade consisting of secondary hadrons, muons, leptons, and photons known as an air shower. The WCDA and KM2A detectors record different components of these air showers, which are used to reconstruct the types, energies, and arrival directions of the primary particles.
The WCDA consists of three ponds: WCDA-1 and WCDA-2, both measuring 150~m $\times$ 150~m , and WCDA-3 measuring 300~m $\times$ 110~m. The total area of the array is 78,000 m$^2$. WCDA-1, WCDA-2, WCDA-3 are composed of 900, 900, 1320 detector units, respectively. 
Each detector unit is 5~m $\times$ 5~m separated by non-reflecting black plastic curtains and is  equipped with two upward-facing PMTs (8-inch and 1.5 inch PMT combination) on the bottom at the center of the unit. To further lower the threshold energy,  WCDA-2 and WCDA-3 employ a 20-inch and 3-inch PMT combination. Each pond is filled with purified water up to 4 m above the PMT photo-cathodes. A closed recycling system is implemented to maintain water purity and achieve an attenuation length for near-ultraviolet light longer than 15 meters.

The results presented here for WCDA were obtained using the full-array configuration from 2021 March 5 to 2022 September 30. For each PMT, a hit is formed with the threshold of 1/3 photoelectron (PE) for an 8-inch PMT, and 1~PE for a 20-inch MCP-PMT. A trigger algorithm was implemented to record CR air showers by requiring at least 30 PMTs fired among 12 $\times$ 12 PMT arrays simultaneously within a window of 250 ns. To ensure a reliable data sample, data quality checks were performed based on the DAQ data status and reconstruction quality. The total effective live time used in the data analysis is around 508 days with a trigger rate around 35 kHz. The number of gamma-like events recorded by WCDA is around 1.29$\times 10^{9}$ events after certain data selection and gamma-ray/background discrimination cuts. More details about the array and the reconstruction can be found in~\cite{2021ChPhC..45h5002A}.

KM2A is composed of 5195 electromagnetic particle detectors (EDs) and 1188 muon detectors (MDs) distributed over an area of 1.3 km$^2$. Each ED consists of 1-m$^2$ plastic scintillator covered by a 0.5-cm thick lead plate and a 1.5-inch photomultiplier tube (PMT). The typical detection efficiency is about 98\% and time resolution is about 2 ns. The response time and signal amplitude of each ED is calibrated using offline methods~\citep{2018APh...100...22L,2022PhRvD.106l2004A}. A trigger is generated when 20 EDs are fired within a 400 ns window. The signals recorded by EDs are used to determine the energy and arrival direction of the primary particles. Each MD consists of a cylindrical water tank, with a diameter of 6.8 m and a height of 1.2 m, and a 8-inch PMT,  which is filled with pure water and buried under 2.5 m of soil. The MDs are designed to detect the muon component of showers, which is used to discriminate between gamma-ray and hadron-induced showers. The performance of KM2A for the gamma-rays with energies from 10 TeV to 1.6 PeV has been thoroughly tested using detector simulations and observations of the Crab Nebula~\citep{2021ChPhC..45b5002A}. The resolution is energy- and zenith-dependent. For showers with a zenith angle less than 20°, the angular resolution ranges from 0.5$^{\circ}$ at 20 TeV to 0.2$^{\circ}$ at 100 TeV. The energy resolution is about 24\% at 20 TeV and 13\% at 100 TeV. The rejection power of the hadron-induced showers is about 1000 at 20 TeV and greater than 4000 at energies above 100 TeV.

The  KM2A  detectors  were  constructed  and  merged into  the  data  acquisition  system  (DAQ)  in  stages.  The first half of the KM2A consisting of 2365 EDs and 578 MDs started operating in  December 2019. This partial array was expanded to a 3/4 array, comprising 3978 EDs and 917 MDs, in December 2020. The whole KM2A was completed and operated starting 2021 July 19. The KM2A data collected from 2019 December 27 to 2022 July 31 were used for the analysis in this work. The total live time is $933$ days, corresponding to 730 days of complete KM2A detector. The pipeline of KM2A data analysis presented in~\cite{2021ChPhC..45b5002A} is directly adopted in this analysis with the same event selection conditions. After the pipeline cuts, the number of events  used in this analysis is $\sim 1.35\times10^7$ with reconstructed energies above 25 TeV.

\subsection{Data Binning and Background Estimation}\label{sec::bb}

We use only events with zenith angles less than 50$^\circ$ in this analysis, corresponding to  the survey region in the declination band from $-20^{\circ}$ to 80$^{\circ}$.  For the WCDA data, events are divided into five groups according to the number of hits ($N_{hit}$), i.e., 100-200, 200-300, 300-500, 500-800, $\ge$800. For Crab-like sources, the corresponding energies roughly range from 1 TeV to 25 TeV. It should be noted that events in the same bin for a source with a harder spectrum, or at a larger declination, will tend to have a larger average energy. The PSF width of WCDA data depends on the shower size which is closely related to $N_{hit}$. Therefore, it is insensitive to depend on declination or spectrum. The Crab observation provides a clear measurement of the angular resolution (denoted as $\phi_{68}$), which is $0.73^\circ$,$0.46^\circ$,$0.37^\circ$,$0.29^\circ$ and  $0.25^\circ$ in the five bins, respectively. For the KM2A data, events are divided into five groups per decade according to reconstructed energy ($E_{rec}$), i.e, 25-40 TeV, 40-63 TeV, 63-100 TeV, 100-160 TeV, 160-250 TeV, 250-400 TeV, 400-630 TeV, 630-1000 TeV, 1000-1600 TeV, $>$1600 TeV. The median energy and the angular resolution in each $E_{rec}$ bin slightly vary with the declination and spectrum of the source. The properties of each bin are shown in Table~\ref{tab:irsf}.

\begin{deluxetable}{lcccccccc}[h]
\setlength{\tabcolsep}{0.1in}
\tablecaption{Property of each Analysis Bin \label{tab:irsf}}
\tablehead{
\colhead{Detector}&
\colhead{$N_{hit}/E_{rec}$ (TeV)} &
\colhead{$\phi_{68}\ (^\circ)$} &
\colhead{$E_{\gamma}^{MC}$ (TeV)}
}
\startdata
 WCDA & 100-200 & 0.68 & 1.7  \\ 
 & 200-300 & 0.46 & 3.0  \\ 
 & 300-500 & 0.37 & 4.9  \\ 
 & 500-800 & 0.29 & 8.9  \\ 
 & $\ge$800 & 0.25 & 17.4  \\ 
 \hline
KM2A & 25-40 & 0.43 & 29.5  \\ 
 & 40-63 & 0.36 & 46.8  \\ 
 & 63-100 & 0.30 & 74.1  \\ 
 & 100-160 & 0.26 & 117.5  \\ 
 & 160-250 & 0.23 & 190.5  \\ 
 & 250-400 & 0.20 & 302.0  \\ 
 & 400-630 & 0.18 & 478.6  \\ 
 & 630-1000 & 0.17 & 758.6  \\ 
 & 1000-1600 & 0.16 & 1202.3  \\ 
 & $\ge$1600 & 0.16 & 1949.8  \\ 
 \enddata
\tablecomments{$\phi_{68}$ is the 68\% containment radius of angular resolution. $E_{\gamma}^{MC}$ is median energy in each bin. We consider a reference source with a broken power-law spectrum of an index of 2.5 in 1-25 TeV and an index of 3.5 above 25 TeV,  at a declination of $30^\circ$.}
\end{deluxetable}
The sky maps in celestial coordinates (right ascension and declination) are divided into $0.1^{\circ} \times 0.1^{\circ}$ pixels, and each pixel is filled with the number of the detected events according to their arrival direction. To obtain the excess of gamma-ray induced showers in each pixel, the “direct integration method”~\citep{2004ApJ...603..355F} is adopted to estimate the number of background events. This method uses events with the same direction in local coordinates but different arrival times to estimate the background. In this work, the integrated time is 10 hours and the events within the regions of the Galactic plane ($|b| < 10^{\circ}$) and VHE gamma-ray sources (with space angle less than 5$^{\circ}$) are excluded to estimate the background which is dominated by the residual CRs. The isotropic diffuse gamma rays and electrons may contribute slightly to the background. 

The diffuse  Galactic gamma-ray emission (GDE) resulting from the interaction of CRs with the interstellar medium (ISM) and background photons is an essential component of the gamma-ray sky. In Galactic plane, the VHE and UHE GDE  has already been clearly detected~\citep{2014PhRvD..90l2007A,2021PhRvL.126n1101A,2023arXiv230505372C}. Therefore, the GDE is also an essential background to take into account for detecting and characterizing gamma-ray sources. In VHE and UHE band, the GDE have been interpreted to be mainly pion decay photons generated from hadronic interactions of CR with ISM. Approximately 99\% of the ISM mass is gas.
Modeling the VHE and UHE GDE requires knowledge of CR intensities and spectra, along with the distributions of gas, throughout the Galaxy. However, due to the variation of gas density and CR density across the Galaxy, it is not possible to completely disentangle the GDE. To simplify the analysis, we ignore the variation of CR density.  The flux morphology of GDE is assumed to follow the spatial distribution of gas, which coexist with dust grains. Observations have shown that the gas-to-dust ratio leads to a mass ratio of $M_{gas}/M_{dust}\sim 100$. The dust column density can therefore provide a template for the GDE morphology. In this work,
 we use the GDE template derived by \textsl{Planck} maps of dust optical depth~\citep{2014A&A...571A..11P,2016A&A...596A.109P}. The spectral function is assumed to be identical in all regions. To avoid the influence of the gamma-ray sources, the spectrum of GDE is estimated using the region of Galactic latitude $5^\circ < |b| < 10^\circ$ and then extrapolating it to other Galactic plane regions ($|b| < 5^\circ$). Note that the region of several VHE sources  in the region of $5^\circ < |b| < 10^\circ$ is also masked for the measurement of GDE, effectively excluding the contribution of both Galactic and extragalactic sources.  The GDE spectrum for this region is found to be well reproduced by the diffuse Galactic gamma-ray emission model, which takes into account the local CR spectrum and the gas column density. We add the extrapolated GDE maps from the measurements in the region of $5^\circ < |b| < 10^\circ$ to the background maps for subsequent analysis.

\subsection{Significance Map}\label{sig::map}

\begin{figure}
\centering
\includegraphics[width=0.85\textwidth]{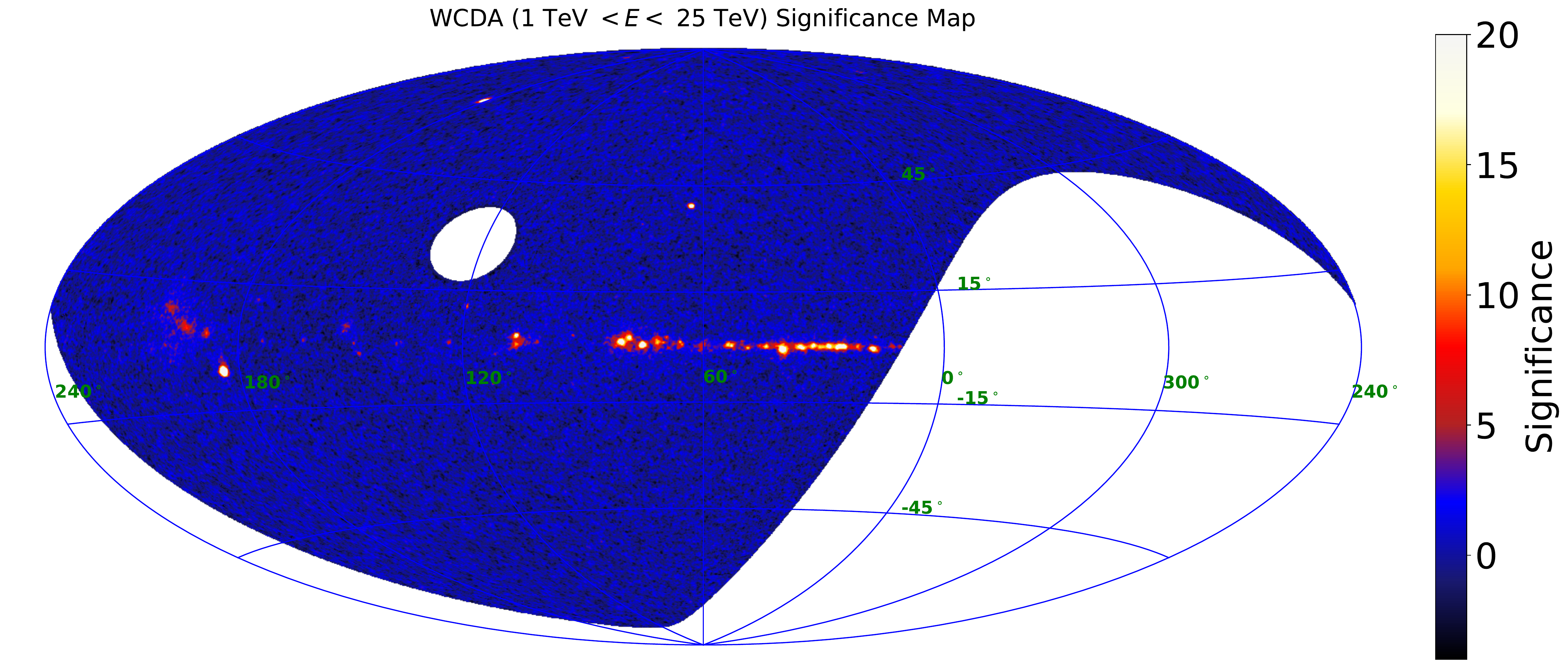}
\includegraphics[width=0.85\textwidth]{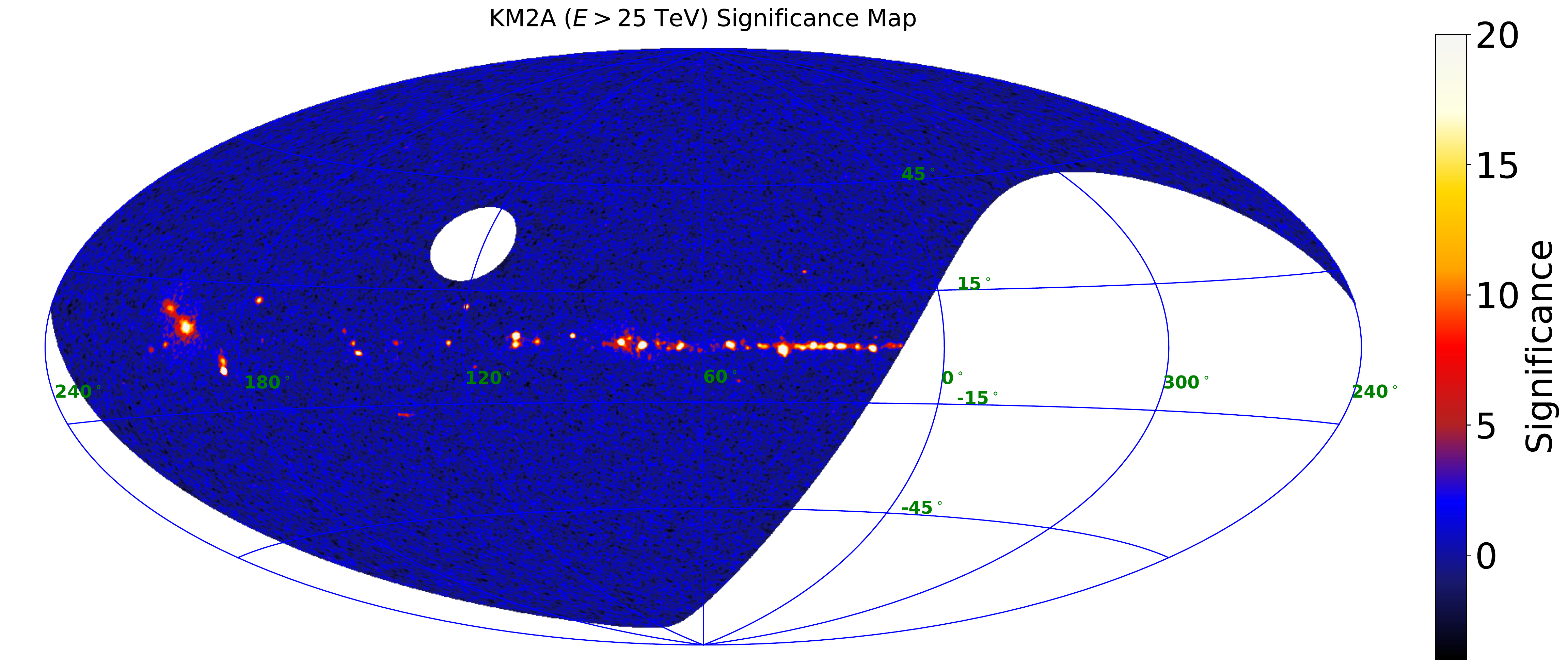}
\caption{Significance maps of the region monitored by LHAASO. A point test source with a spectral index of 2.6 for WCDA data and 3.0 for KM2A data is used.}\label{Fig::sig_allsky}
\end{figure}

We can calculate the significance of excess centered at each pixel using the event and background maps as described in previous section, taking the detector responses into consideration. The WCDA and KM2A data are separately analyzed. To combine the data bins of each instrument, the 3D-likelihood method is used.
A likelihood ratio test is performed between the background-only model and the one point source model. The likelihood calculation assumes that the number of counts in each pixel is distributed according to a Poisson distribution, with the mean given by the number of background counts for the background-only model and the number of background plus expected gamma-ray counts for the source model~\citep{2009A&A...495..989S}. The test statistic (TS), defined as $\rm TS = 2  \ln (\mathcal{L} / \mathcal{L}_0)$, where $\mathcal{L}_0$ is the maximum likelihood value for the null hypothesis and  $\mathcal{L}$ is the maximum value for the source hypothesis, is  used  to  estimate the  significance. In this work, a power-law spectrum is assumed with an index of 2.6 for WCDA data in the energy range 1$-$25 TeV and 3.0 for KM2A data at energies $E >$ 25 TeV as initial conditions. This leaves only one free parameter for the likelihood calculation. According to Wilks’ Theorem, the TS is distributed as $\chi^2$ with one degree of freedom (dof), and the significance can be estimated with S = $\sqrt{\rm TS}$. Figure~\ref{Fig::sig_allsky} shows the significance maps obtained in the energy bands $1 {\rm\ TeV} < E < 25 \rm\ TeV$ and $E > 25\rm\ TeV$ in Galactic coordinates. The signals are clearly visible. However, most sources in the Galactic plane are nearby and overlapping. Hence, further analysis is needed to derive each source separately.

\section{Construction of the Catalog}

The identification of point-like gamma-ray sources and their corresponding significance can be roughly derived from Figure~\ref{Fig::sig_allsky}. However, it is important to note that the significance may be overestimated due to the overlap with nearby sources. Conversely, in the search for point sources, a significant portion of the sources may actually be extended, resulting in an underestimation of their significance. To improve source detection, the significance of a given source is reassessed by coupling the fitting of localization, extension and spectrum, and new potential sources are also explored. In the first step, the WCDA and KM2A data are analyzed separately to achieve two source component catalogs, one for energies ranging from 1$-$25 TeV and another for energies above 25 TeV. These two source component catalogs are merged into the final source catalog following a specific procedure. The sources with clear $E > 100$ TeV emission will also be identified in the final catalog.
 
\subsection{All Sky Fitting and Source Component Detection}\label{itera}

The WCDA data and KM2A data are analyzed following the same strategy to fit the entire sky region and yield the source components with the characteristics of localization, extension and spectrum.

\subsubsection{Determination of  Seeds and ROIs}\label{sec::seeds}

The signals shown in Figure~\ref{Fig::sig_allsky} are potential point sources, denoted as “seeds”. To identify these seeds in the maps,  all local maxima within a 0.5$^{\circ}$ region centered around the candidate, with a significant above 4$\sigma$, are selected. 
Most of the seeds are concentrated in the inner Galactic plane, while in the outer Galactic plane, the seeds are clustering in the region near Geminga. In other areas of sky region, a few individual seeds stand out.

To facilitate the fitting process, all sky data are split into different regions of interest (ROI) based on the seeds and their clustering characteristics. For the inner Galactic plane region, a sliding window  20$^{\circ} \times 20^{\circ}$ ROI is adopted along Galactic longitude with a step length of 10$^{\circ}$. For the Geminga region, a ROI with the radius of 12$^{\circ}$ 
centered at right ascension ($\alpha_{2000}$) $100.2^\circ$ and declination ($\delta_{2000}$) $16.2^\circ$ is used.
For isolated seeds, the ROIs are circular regions with a  3$^{\circ}$ radius, and the overlapping ROIs are merged into one. 

\subsubsection{Source Component Detection Procedure}\label{procedure}

The seeds obtained in Section ~\ref{sec::seeds} may not correspond to a real source component  because the spectrum and extension characteristics are not fully considered. Moreover, the position can be influenced by nearby sources with small angular separations. Thus, an analysis pipeline based on a three-dimensional maximum likelihood algorithm is developed for further analysis. Using this analysis pipeline, the spectral information and spatial morphology and significance for individual source components can be determined simultaneously. Some technical details are described below:

\begin{itemize}
\item The first step of the 3D maximum likelihood algorithm is to build a source model characterized by a 2D-Gaussian morphology based on the initial positions of the seeds. The source spectrum is assumed to follow a power-law shape $dN/dE=N_{0}(E/E_{0})^{-\Gamma}$, where $N_{0}$ is the differential flux at $E_{0}$, $E_{0}$ is the reference energy which is set to 3 TeV for WCDA data and to 50 TeV for KM2A data,  respectively, and $\Gamma$ is the spectral index.

\item Starting with one Gaussian component, we add Gaussian components successively to the source model of each ROI.  A likelihood ratio test is used to compare between two models with $N$ and $N+1$ Gaussian components. We define TS = $2\ln (\mathcal{L}_{N+1} / \mathcal{L}_{N})$, where $\mathcal{L}_{N}$ and $\mathcal{L}_{N+1}$ represents the maximum likelihood of  the model with  $N$ and $N+1$ source components, respectively. During this iterative process, an additional source characterized by 5 free parameters ($\alpha_{2000}$, $\delta_{2000}$, extension, differential flux  and spectral index) is added to the model  if TS $> 25$ (3.8$\sigma$ for 5 dof).

\item  The iteration process is terminated once no significant residual is left in the TS map for each ROI.  Here, no significant residual means that no pixel is with TS $>16$ (4$\sigma$ for 1 dof) in the residual TS map, which is calculated by adding the current sources to the background model and  using the method similar to that in Sec.~\ref{sig::map}.
 
\item During the fitting process, all of the parameters are optimized simultaneously. The Spectral parameter of the GDE is always keep to fixed.

\item After completing the previous steps, the WCDA and KM2A source component lists are formed, respectively, by merging all the list determined in each ROI. Due to the overlap ROIs, the same source component may appear in two or more ROIs. In such cases, the one that was closest to the center of the ROI is selected.

\end{itemize}

This analysis pipeline yields a preliminary source component list, in which most components have good estimates of parameters. However two possible biases may exist for two cases: 1) the target Gaussian component is close to the ROI edge or a bright source; 2) the iteration fitting scheme described above could converge toward a local maximum. Additional checks have been implemented for these two cases using a new ROI centered on the position of each component and re-analyzing the target source component with free parameters for all components in the ROI. If new parameters of the target source component do not significantly differ from the previous iteration fitting, the new parameters of this source component are considered reliable and are adopted. The significant difference refers to the situation where the variance of a single parameter is greater than the corresponding statistical error. If significant variations are observed, a manual fitting process is conducted. This involves adjusting the starting values of the parameters and/or the size of the ROI until the parameters of the target source component converge. This convergence is determined through a visual inspection of likelihood profiles. Once the parameters are obtained, the TS value of each source component is evaluated within each ROI. In this evaluation, the positions and extensions of the other components are fixed, while the spectral parameters are left free for optimization.

\subsubsection{False Positive Expectation}

The trial factor can be roughly estimated as $f_t=\Omega/[2\pi(1-\cos 0.5^\circ)]\sim3.5\times10^4$, where $\Omega$ is the solid angle of the LHAASO survey sky within the declination ranging from $-20^\circ$ to $80^\circ$, and $0.5^\circ$ is the minimum location searching radius for seeds. 
Since the TS value for the above components follows the $\chi^2$ distribution with 5 degrees of freedom, TS = 25 corresponds to the $p$-value of  $6.9\times10^{-5}$. Based on this, we can expect that 2.4 sources in our survey are false detections due to background fluctuations.Thus, we set a higher threshold of TS = 37 (5$\sigma$ for 5 dof) for reported sources, corresponding to an expectation of 0.01 false-positive sources. 
For sources with both WCDA and KM2A components,  the summed TS value of 50 (5.1$\sigma$ for 10 dof) corresponds to  the comparable false-detection to that of the single component of TS = 37. Hence, we also report the sources of which two components are with  $25 < \rm TS < 37$. Furthermore, due to the mismodeling GDE background, the sources in the gas-rich region with a low significance level might still be potential false detection. Therefore, further study is required to determine whether or not they are real sources.

\subsection{ Localization, Extension, and  Energy Spectrum}

\begin{figure}[h]
\centering
\includegraphics[width=0.45\textwidth]{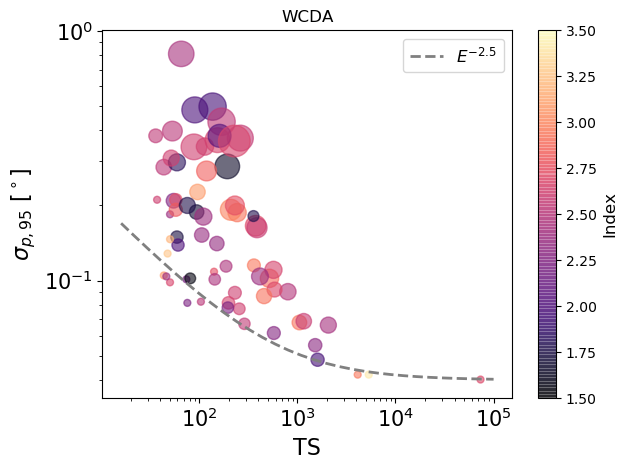}
\includegraphics[width=0.45\textwidth]{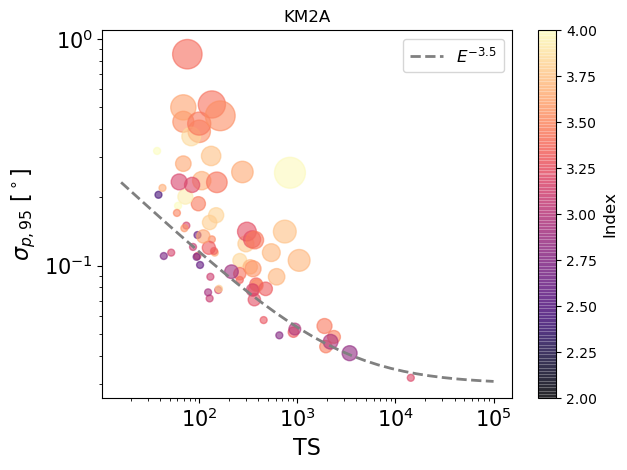}
\caption{Source component location uncertainties ($\sigma_{p,95}$)  as a function of TS values.  $\sigma_{p,95}$ is defined by $\sqrt{\sigma_{p,95,stat}^2+\sigma_{p,sys}^2}$, where $\sigma_{p,sys}$ is  the systematic positional error which  is  $0.03^\circ$ for the KM2A component and $0.04^\circ$ for  the WCDA component, respectively, considering the pointing error only (Detailed in Section \ref{syserror}). The dashed line is a trend for reference, considering a point source at a declination of $20^\circ$ with the index of 2.5 for WCDA component and  of 3.5 for KM2A component. The marker size is scaled by the source radius. The marker color represents the photon spectral index. The source components with the significance of TS $>37$ and the extension size of  $r_{39}<2^\circ$ are plotted. }
\label{Fig::pos}
\end{figure}

The localization of each source component, represented by $\alpha_{2000}$ and $\delta_{2000}$, is given by the procedure in Sec~\ref{procedure}. The statistical errors of the localization  ($\sigma_{\alpha_{2000},stat}$, $\sigma_{\delta_{2000},stat}$) are also estimated by analyzing the shape of the likelihood function with the HESSE tool in the Minuit2 package. 
It is worth noting that the localization of a faint or soft source component is more sensitive to the influence of diffuse emission and nearby sources. As a result, we may find asymmetric position error circles for these components, where  $\sigma_{\alpha_{2000},stat}\cos\delta_{2000}\neq \sigma_{\delta_{2000},stat}$. To account for this, we conservatively consider the bigger error  as the one-dimensional $1\sigma$ position error, i.e., $x_{stat} = max[\sigma_{\delta_{2000},stat},\sigma_{\alpha_{2000},stat}\cos\delta_{2000}]$. In this work, we report the position uncertainties at the 95\% confidence level as $\sigma_{p,95,stat}=2.45\times x_{stat}$, where the 2.45 is the factor  $\sqrt{-2\ln(1-0.95)}$ by which the 95\% confidence level position uncertainty  is related to $x_{stat}$ for  a two-dimensional axisymmetric Gaussian. Figure ~\ref{Fig::pos} illustrates the position uncertainties as a function of the TS values. The relatively large dispersion seen at a given TS is due to variations in local conditions , the source extension and the source spectrum. Considering the systematic error, the error radius (95\% confidence) of the LHAASO sources ranges from $\sim 0.04^\circ$ to $\sim 0.8^\circ$.

\begin{figure}[htbp]
\centering
\includegraphics[width=0.45\textwidth]{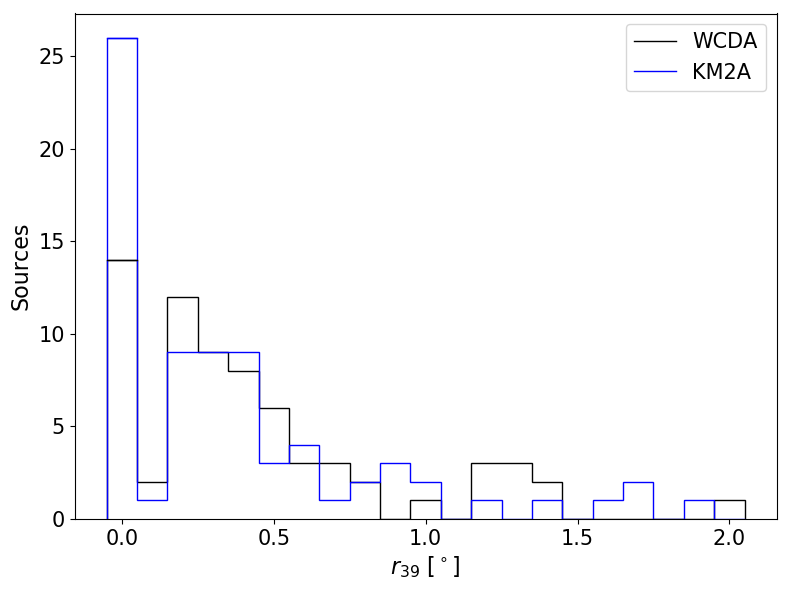}
\caption{The distribution of the extension sizes ($r_{39}$) of the source components detected by LHAASO. The size of the point-like source component is set to 0.
Due to the limitation of statistical level and angular resolution, the measured minimum extension is approximately $0.17^\circ$, leading to a gap of around $0.1^\circ$. }
\label{Fig::ext}
\end{figure}

The extension size of each source component is described by a two-dimensional Gaussian $\sigma$ (namely $r_{39}$ in this work) corresponding to 39\% of the source flux. A statistical error for the measured extension ($\sigma_{r_{39},stat}$) at the $1\sigma$ level can also be determined. The measurement of $r_{39}$ is invalid for the point-like components unresolved by LHAASO. Thus, for all source components, the extension Test Statistic  ($\rm TS_{\rm ext}$) is performed. Here  $\rm TS_{ext}=2 \, \ln (\mathcal{L}_{\rm ext} / \mathcal{L}_{\rm ps})$, where  $\mathcal{L}_{\rm ps}$ and   $\mathcal{L}_{\rm ext}$ are the maximum-likelihood values assuming the source spatial model is a point-like and 2D-Gaussian model, respectively. 
When the $\rm TS_{ext}$ is lower than 9 corresponding to a significance smaller than  $3\sigma$ level, we change the  2D-Gaussian spatial model to a point-like model and re-optimize the position and spectral parameters.
The upper limits on the extension ($r_{39,ul,stat}$) at the 95\% confidence level for these point-like components are derived from the shape of the likelihood function.  Figure~\ref{Fig::ext} shows the extension size of each source component. Roughly 1/3 are unresolved by LHAASO and modeled by a point-like morphology, and roughly 2/3  are modeled by a 2D-Gaussian morphology with an extension size ranging from $\sim 0.2^\circ$ to $> 2^\circ$.

\begin{figure}[htbp]
\centering
\includegraphics[width=0.45\textwidth]{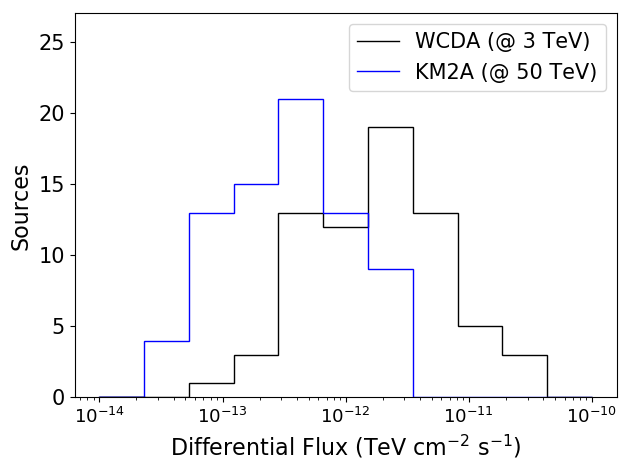}
\includegraphics[width=0.45\textwidth]{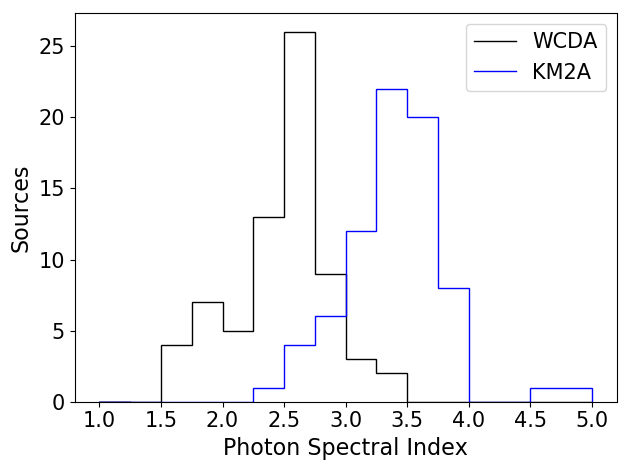}
\caption{The distribution of the SED parameter of the source components. Left: the distribution of the differential flux ($E_0^2N_0$). The reference energy $E_0$ is 3 TeV or 50 TeV for the WCDA or KM2A component, respectively. Right: the distribution of the photon spectral index ($\Gamma$) for WCDA and KM2A components.}
\label{Fig::flux_index}
\end{figure}

The spectral energy distribution (SED) of each source component is modeled by a simple power law (PL). Figure~\ref{Fig::flux_index} shows the distribution of the SED parameters of the source components. The differential flux has a mean of $\sim 3\times10^{-12}\ \rm TeV\ cm^{-2}\ s^{-1}$  for WCDA components at 3 TeV and  $\sim 4\times10^{-13}\ \rm TeV\ cm^{-2}\ s^{-1}$ for KM2A components at 50 TeV. The photon index distributions are obviously different, with an average of $\sim 2.5$ for WCDA components and $\sim 3.5$ for KM2A components, implying that a single power law shape cannot describe the overall SED for the majority of 1LHAASO catalog sources. 

\begin{figure}[htbp]
\centering
\includegraphics[width=0.45\textwidth]{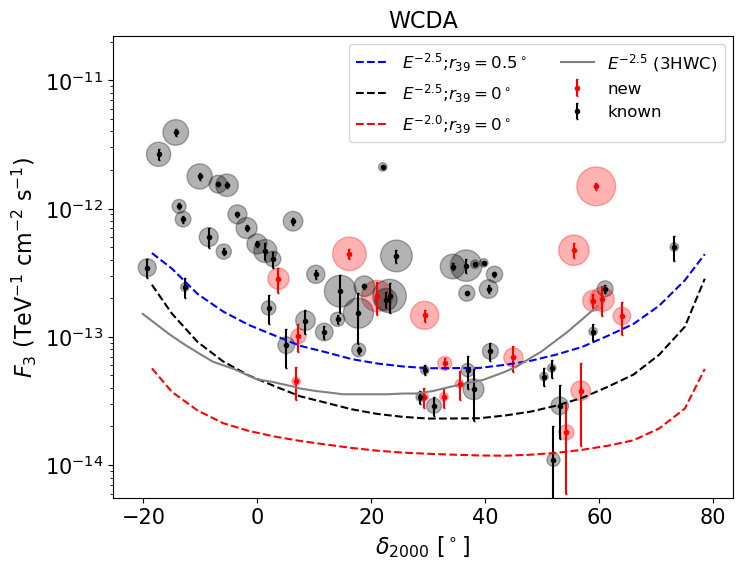}
\includegraphics[width=0.45\textwidth]{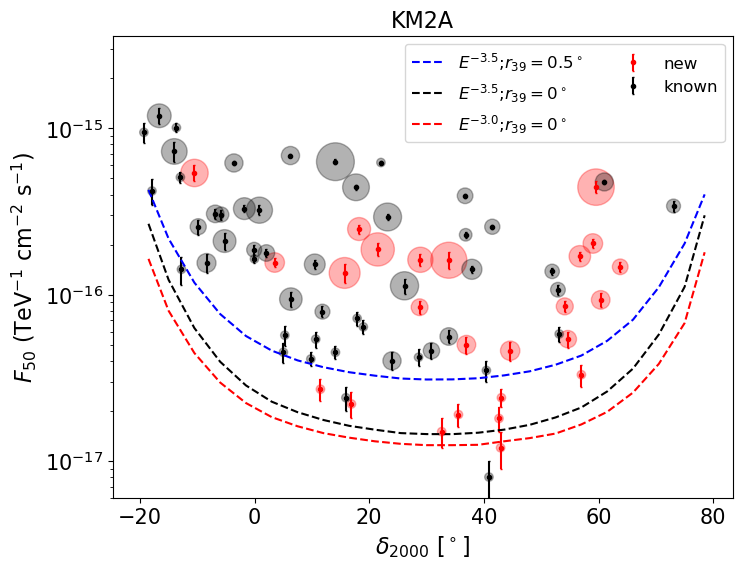}
\caption{1LHAASO source differential flux at 3 TeV ($F_{3}$) or 5 TeV ($F_{50}$) as a function of declination. The marker size represents the source size. The color indicates whether or not these sources are the new sources (seen in Section~\ref{discover}). Left: the integral sensitivity is shown at 3 TeV for  three hypotheses: 1) point source with spectrum of $E^{-2.0}$; 2) point source with spectrum of $E^{-2.5}$; 3)  $0.5^\circ$ gaussian source with spectrum of $E^{-2.5}$. The 3HWC sensitivity  corresponds to the point-source search with the 3HWC data set~\citep{2020ApJ...905...76A}. Right: the integral sensitivity is shown at 50 TeV for three hypotheses: 1) point source with spectrum of $E^{-3.0}$; 2) point source with spectrum of $E^{-3.5}$; 3)  $0.5^\circ$ gaussian source with spectrum of $E^{-3.5}$. }
\label{Fig::sen}
\end{figure}

We compare the flux measurements with the sensitivity of source derived by our used data set. The flux sensitivity is defined as the flux normalization required to have 50\% probability of detecting a source at $5\sigma$ level. As shown in Figure~\ref{Fig::sen}, the sensitivity dependents on the declination, spectral index and source size. Sources passing directly overhead, with a declination of $30^\circ$ for LHAASO ($20^\circ$ for HAWC), exhibit the highest sensitivity.  The sensitivity for a source with a softer spectrum or larger size, or at a larger declination, will tend to decrease. Thus, about 10 WCDA sources, which have fluxes above the 3HWC point source sensitivity, have not been included into the 3HWC catalog due to their large size and/or soft spectrum. In addition, it is possible that the source searching strategy employed in the 3HWC catalog results in the failure to detect several of these sources.

\begin{figure}[htbp]
\centering
\includegraphics[width=0.45\textwidth]{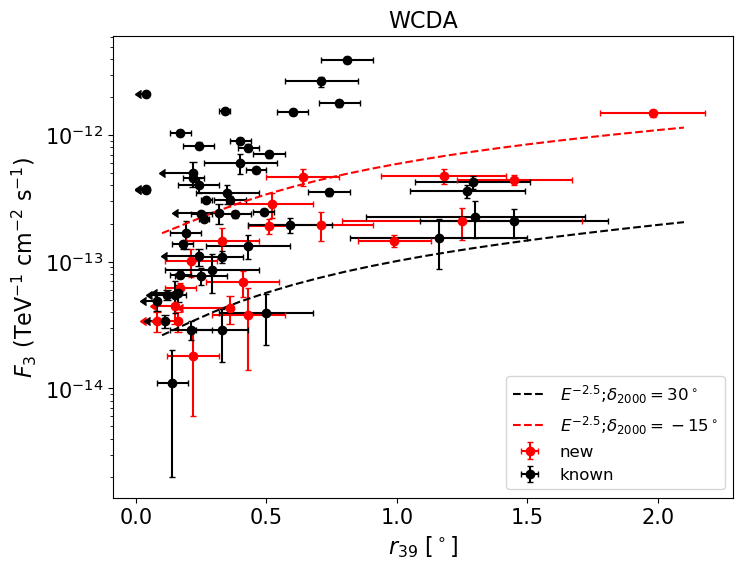}%
\includegraphics[width=0.45\textwidth]{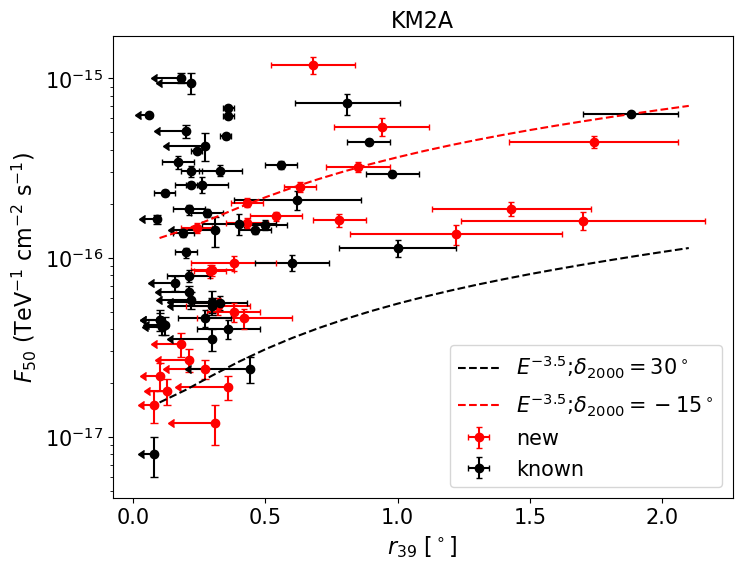}
\includegraphics[width=0.45\textwidth]{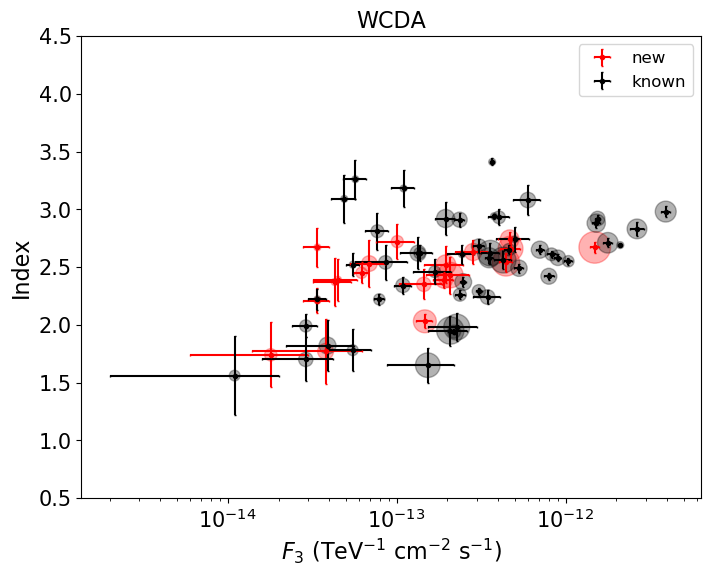}%
\includegraphics[width=0.45\textwidth]{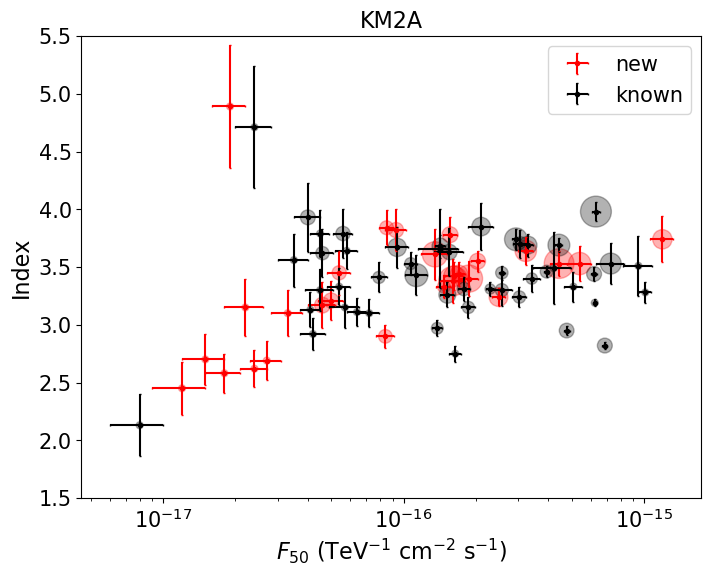}
\caption{ Flux vs. source size and flux vs. photon spectral index.  Top: 1LHAASO source differential flux ($F_{3}$ or $F_{50}$) as a function of source size. The error bar of $r_{39}$ represents the statistical uncertainty at a $95\%$ confidence level. The dashed lines are integral sensitivity shown at 3 TeV or 50 TeV, assuming a point-like source.  Bottom: 1LHAASO source photon spectral index as a function of the differential flux  ($F_{3}$ or $F_{50}$). The definition of the new sources is seen in Section~\ref{discover}.}
\label{Fig::flux_index_ext}
\end{figure}

Figure~\ref{Fig::flux_index_ext} shows the 1LHAASO source fluxes related to source sizes or source photon spectral index. In the flux-size panel, we illustrate the approximate flux sensitivity limit of the 1LHAASO source as a function of source size. It can be observed that the sensitivity worsens as the source size increases, as mentioned earlier. For extended sources with sizes larger than $1^\circ$, we may find a possible lack of sources with flux $F_{3} > 4\times 10^{-13} \rm\ TeV^{-1}\ cm^{-2}\ s^{-1}$ or $F_{50} > 4 \times 10^{-16} \rm\ TeV^{-1}\ cm^{-2}\ s^{-1}$. In the index-flux panel, the photon index in the WCDA band appears to be uncorrelated with the flux $F_3$. However, there might be a broken power-law correlation between the photon index in the KM2A band and the logarithm of $F_{50}$, with a break occurring at a flux of $F_{50} \sim 4 \times 10^{-17} \rm\ TeV^{-1}\ cm^{-2}\ s^{-1}$. It is not yet clear whether this phenomenon is caused by an instrument selection effect or a physical relation. 

\subsection{Source Detection Above 100 TeV}

An exciting characteristic of LHAASO is that it can extend the observation to the UHE regime owing to the large detector area. The UHE sources are crucial to identify the emission mechanism and explore the maximum acceleration energy within the sources, and they are also very important candidates to explore the origin of PeV CRs within the Galaxy. To explore UHE sources, the sources detected at energies above 25 TeV are selected.  The SED parameters of these sources are re-optimized while freezing the spatial model parameters, i.e., the position and extension size, to the values measured at $E > 25\rm\ TeV$. The TS values above 100 TeV (denoted by TS$_{100}$) are calculated. Sources are with TS$_{100} >$ 20 ($> 4\sigma$ significance level based on the $\chi^2$ distribution with two degrees of freedom) are claimed  as UHE sources in this work. 

 \subsection{Merging of the Source Components}

 A WCDA component and a KM2A component are considered to be the same source according to the agreement in position, i.e., $r_d^2 < (\sigma_{p,95,W})^2 + (\sigma_{p,95,K})^2 + \delta_{p}^2$, where $r_d$ is the positional offset between the WCDA component and the KM2A component, and $\sigma_{p,95,W}$ and $\sigma_{p,95,K}$ are the position uncertainties at the $95\%$ confidence level for the WCDA and the KM2A components, respectively. $\delta_{p}$ is the position-offset correction of the components, possibly due to the physical position variation or the changes in morphology at different energy bands. Additionally, incomplete descriptions of the  GDE and/or of nearby sources in the analysis can also contribute to the positional offset. It is difficult to determine the value of  the $\delta_{p}$ for each source. We expect that $\delta_{p}$ is related to the source extension. As a simple criterion, we adopt a loose condition with $\delta_{p} =(r_{39,W}+r_{39,K})$,  where $r_{39,W}$ and $r_{39,K}$ are the extensions of the WCDA and KM2A components, respectively. We merge the closest pairs of the WCDA and KM2A components that satisfy this condition. Ultimately, 54 pairs of components are merged.

\subsection{Known TeV Source Association}\label{asso}

We have conducted a preliminary association of LHAASO sources with known TeV sources based on their position. A searching radius is defined by $r_{sr} = \sqrt{\sigma_{p,95}^2+r_{39}^2+(0.3)^2}$, where $\sigma_{p,95}$ and $r_{39}$ are the  position error and extension size of the source component. For the LHAASO sources with  KM2A and WCDA components,  the position and extension of the component with higher significance is used. For the point-like source components, $r_{39}$ is set to the extension upper limit. The average value of the position errors for the known TeV sources detected by EAS arrays is about 0.3$^\circ$. While for the sources detected by IACTs, the average position error is around 0.05$^\circ$. Roughly, we used the factor of 0.3 degree for the searching radius. If $r_{sr} >1^\circ$, a maximum searching radius of $1^\circ$ is applied. 

The vast majority of the known TeV sources from ground based observatories are included in the TeV online catalog, i.e., TeVCat\footnote{http://TeVCat.uchicago.edu}\citep{2008ICRC....3.1341W}. It is  frequently updated, containing 252 TeV source entries to date. 
When associating LHAASO sources with known TeV sources, we used the canonical name and position reported in TeVCat. We exclude the TeV pulsars for the associations, since the pulsed TeV emission is expected to be null in our source searching procedure. If no TeV source in TeVCat is found for association, we further compare 1LHAASO sources with 3HWC sources, which include some sources not yet included in TeVCat so far.  We list the closest TeV counterparts within the searching radius for each LHAASO source, as shown in table~\ref{tab:1}.

\subsection{Systematic Uncertainties and Caveats}\label{syserror}

Based on the measurement of the standard candle, the Crab Nebula, with a position of $\alpha_{2000}=83.633^{\circ}$ and $\delta_{2000}=22.015^{\circ}$ for the systematic error estimation, a systematic pointing error can be evaluated as $0.03^\circ$ for KM2A data. For the WCDA data, an investigation of the pointing systematic error has been conducted using three point sources: Crab Nebula, Mrk 421, and Mrk 501. The maximum error is from Mrk 501 at about $\sim 0.04^\circ$, which is taken as the pointing error of the WCDA array in this work.  It is important to note that no clear zenith angle dependence of the systematic error is found according to the observation of the Crab Nebula using events at different zenith angles, ranging over 0$^\circ$-15$^\circ$, 15$^\circ$-30$^\circ$, and 30$^\circ$-50$^\circ$. 
Furthermore, no systematic location bias related to declination has been identified based on the observation of the CR Moon shadow at different declination bands. 

The systematic uncertainty of the size of 1LHAASO sources could be contributed by the uncertainties of the Point Spread Function (PSF).  For KM2A data, we obtained a systematic bias for the $\phi_{68}$ at the order of $\sim 0.08^\circ$ by comparing the Crab measurement with our simulated data. For WCDA data, we compared the observed events profile among Crab, Mrk 421, and Mrk 501. The uncertainties can also yield a systematic bias of $\sim 0.05^\circ$. Thus, we can conservatively estimate the systematic error to be at the order of $\sigma_{r_{39},sys}\sim0.05^\circ$ for WCDA and $\sigma_{r_{39},sys}\sim0.08^\circ$ for KM2A component.

The systematic errors affecting the spectrum have been investigated in \cite{2021ChPhC..45h5002A} and in \cite{2021ChPhC..45b5002A}.  The main systematic error is contributed by the atmospheric model in the Monte Carlo simulations.  The total systematic uncertainty is estimated to be  7\% on the flux and 0.02 on the spectral index for KM2A SED measurement. In the case of WCDA data, the overall systematic uncertainty can be as large as $^{+8\%}_{-5\%}$ on the flux, which is estimated by the same method in \cite{2021ChPhC..45h5002A}. The power-law spectral shape can adequately describe the WCDA components, while it is not suitable for about 1/3 KM2A components which shown an evident curved shape at energies above 25 TeV. A detailed spectrum study needs to be carried out in the future.

Mismodeling of GDE could affect the fitted locations, extensions, and SEDs of several source components, especially for those with lower fluxes and larger sizes.
A rough assessment of the impact of the GDE was performed by excluding it from the background maps for the source components in the Galactic plane region. As a result, 11 KM2A source components and 10 WCDA source components exhibited changes (in terms of location, flux, index, or extension) exceeding three times the statistical errors. Source components that were significantly affected by the GDE were tentatively labeled as such. The GDE test is not thorough because we cannot determine all the sources affected by the GDE mismodeling. As a conservative approach, source components with an extended size greater than $2^\circ$ were excluded due to the growing impact of the GDE with increasing component size. Further understanding of very extended sources or obvious GDE-impacted sources requires deeper study of the GDE model, which is beyond the scope of this work.

\begin{figure}[htbp]
\centering
\includegraphics[width=0.45\textwidth]{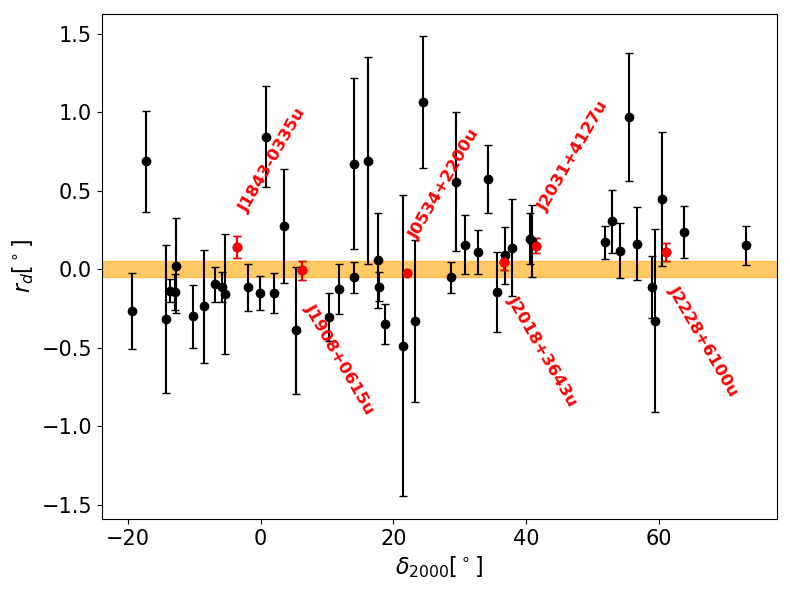}%
\includegraphics[width=0.45\textwidth]{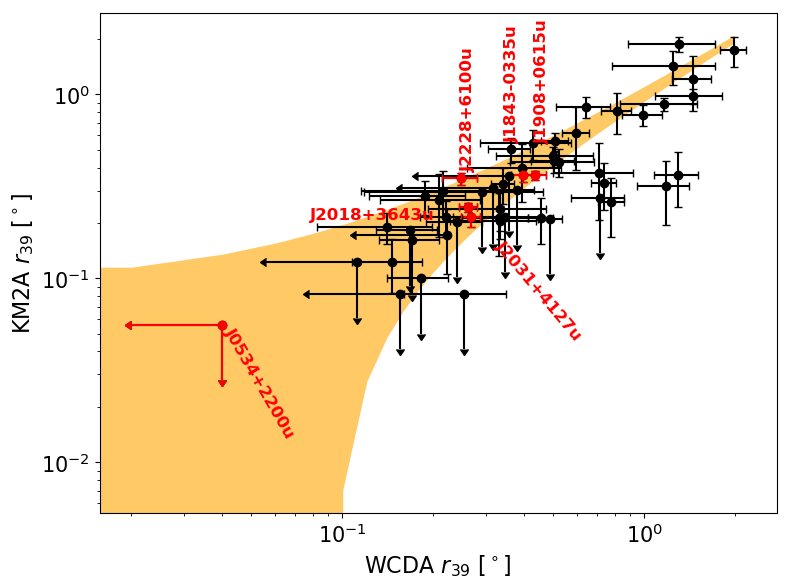}
\caption{The position and extension comparison between the WCDA and KM2A components.The six most significant sources are highlighted in red. Left: the position offsets ($r_d$) relative to the declinations. In the case where the declination of WCDA component is smaller than that of the KM2A component, we take the opposite value.  The error bar is defined as $\sqrt{\sigma_{p,stat,95,W}^2 + \sigma_{p,stat,95,K}^2}$, where $\sigma_{p,stat,95,W}$ and $\sigma_{p,stat,95,K}$ are the statistical positional error  of WCDA and KM2A components, respectively. The orange represents the systematic error just including the pointing error. Right: the extensions of the WCDA components  relative to that of the KM2A components. The error bar represents the statistical uncertainty at a $95\%$ confidence level. The orange band represents the systematic error just considering that of PSF. }
\label{Fig::p_off}
\end{figure}

As shown in Figure~\ref{Fig::p_off}, at 95\% confidence level,  approximately 40\% of the merged sources exhibit a noticeable bias in position or differences in extension. The shift could be an indication of an energy-dependent morphology of these sources. However, it should be noted that due to the poor angular resolution of LHAASO, the offset in position and extension could also be influenced by emission from nearby unresolved sources or mismodeling of GDE. There is a possibility of a false merged source, which could be a combination of WCDA and KM2A components that are physically unrelated. To further examine this, we have roughly conducted individual investigations on the surface brightness distribution and spectrum of each merged source. During this process, conflicts were found in 11 merged sources, which are tentatively labeled as dubious mergers. More comprehensive studies are needed to confirm the physical association between the two components of the  dubious merged sources.  

\section{Results}

Following the above procedure, the source catalog of LHAASO has been constructed. Overall, 90 sources with extension  $< 2^\circ$  are found over the whole LHAASO survey sky. Among them, 65 sources exhibit extended morphology with a confidence level greater than $3\sigma$. A total of 54 sources have been simultaneously detected by both WCDA and KM2A.  Among all the sources, 43 UHE sources have been detected at $> 4\sigma$ confidence level when $E > 100$TeV.

\begin{figure}[htbp]
\centering
\includegraphics[width=0.45\textwidth]{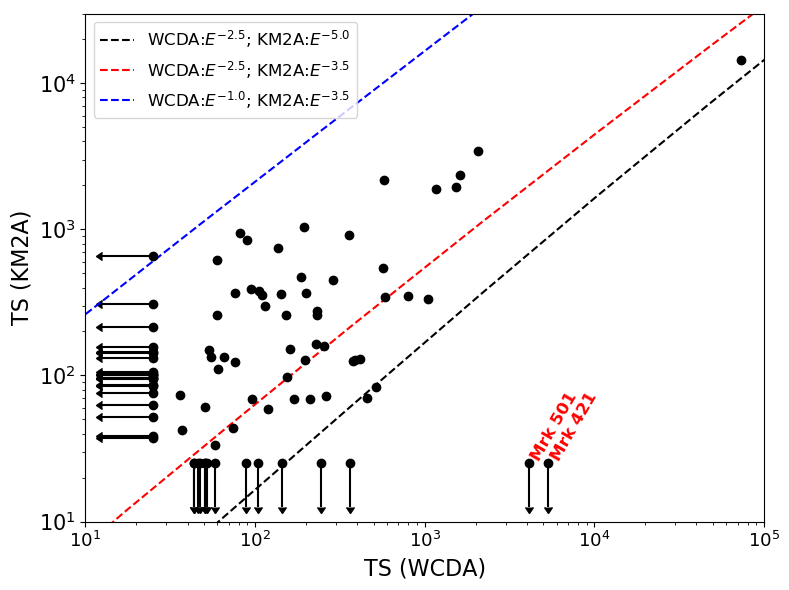}
\caption{TS value of KM2A component versus that of WCDA component. The reference dashed lines indicate the expected TS value for each detector. These values are calculated based on a point source which has a broken power-law spectral shape with a break at an energy of 25 TeV.}
\label{Fig::tswk}
\end{figure}

\begin{figure}
\centering
\includegraphics[width=0.99\textwidth]{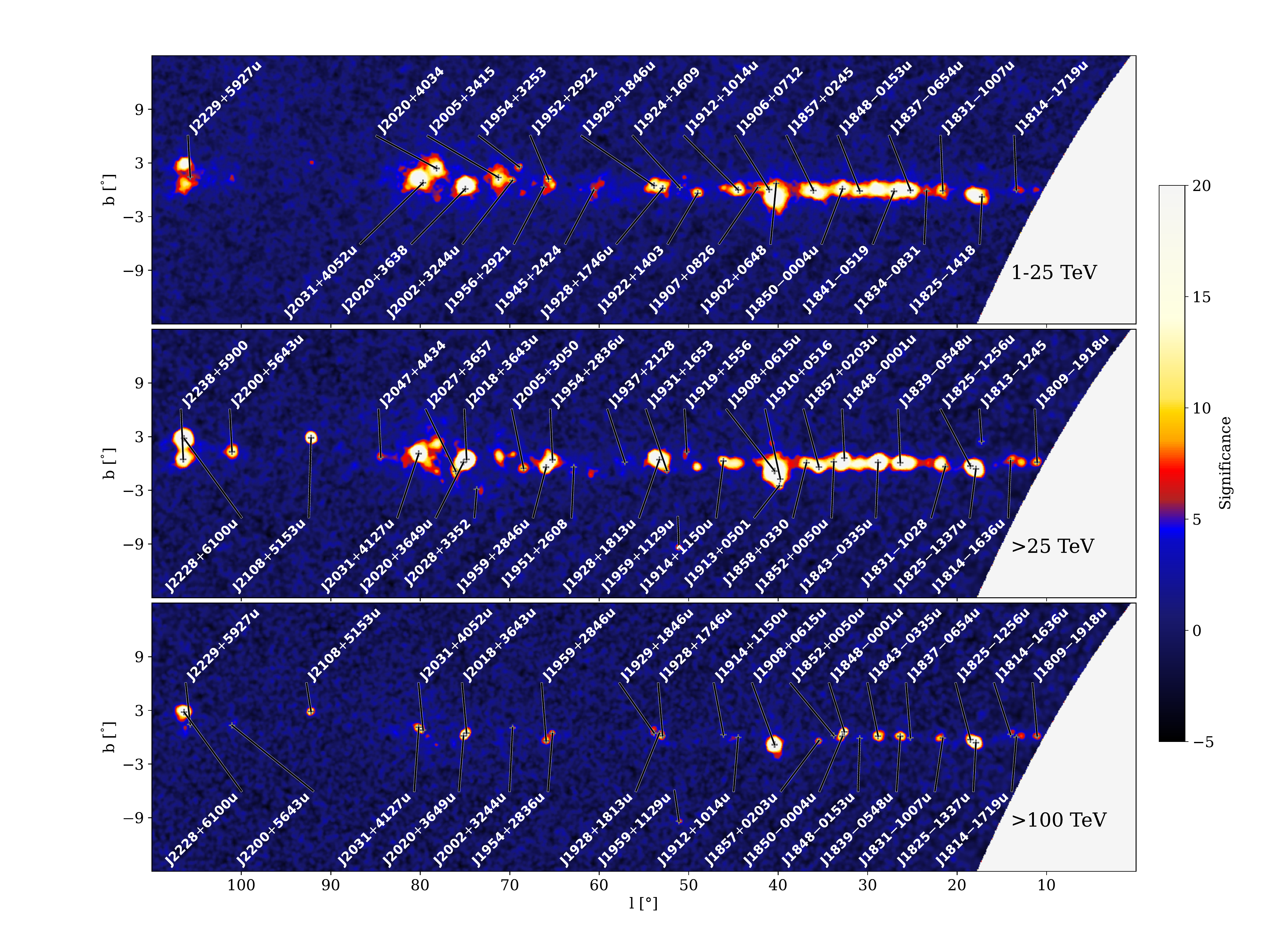}
\caption{LHAASO significance map within the region $ 10^\circ \leq l \leq 115^\circ$ , $\left| b \right| \leq 12^{\circ}$. Top: WCDA ($1 {\rm\ TeV} < E < 25 \rm\ TeV$) significance map. Middle: KM2A ($E > $ 25 TeV) significance map. Bottom: KM2A ($E > $ 100 TeV) significance map.
In this figure and following, the LHAASO source are represented by gray crosses and white labels. The LHAASO sources for which the WCDA component has higher significance are  plotted in the top panel. The LHAASO sources for which the KM2A component has higher significance are plotted in the middle panel. Meanwhile, UHE sources are  shown again in the bottom panel.}
\label{Fig::gal_0_110}
\end{figure}

\begin{figure}
\centering
\includegraphics[width=0.99\textwidth]{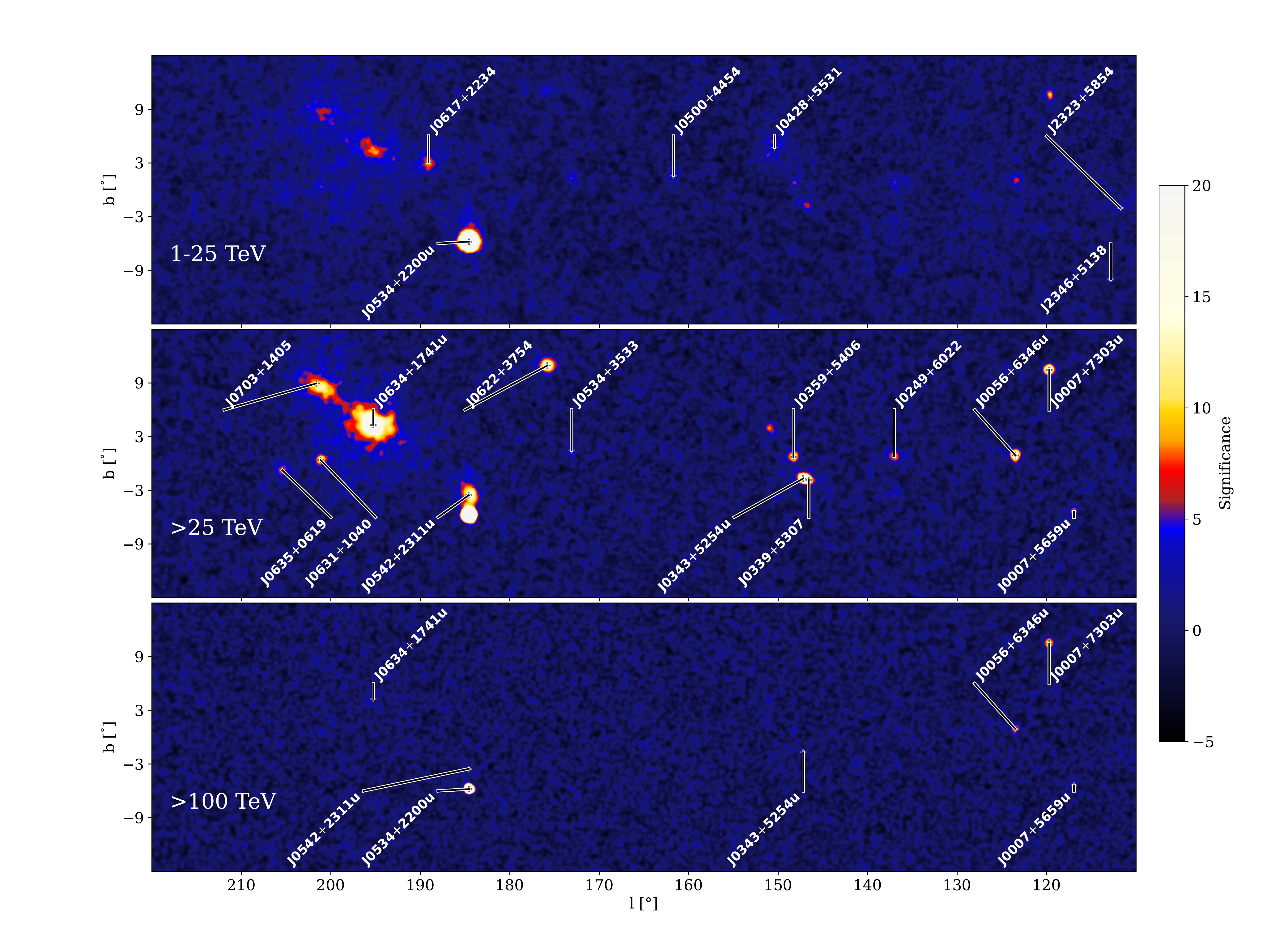}
\caption{LHAASO significance map within region $ 115^\circ \leq l \leq 220^\circ$ , $\left| b \right| \leq 12^{\circ}$.Top: WCDA ($1 {\rm\ TeV} < E < 25 \rm\ TeV$) TeV significance map. Middle: KM2A ($E > $ 25 TeV) significance map. Bottom: KM2A ($E > $ 100 TeV) significance map.}
\label{Fig::gal_110_220}
\end{figure}

Table~\ref{tab:1} presents a comprehensive list of all LHAASO sources obtained through the above procedure, ordered by $\alpha_{2000}$ \footnote{The machine-readable file can be downloaded from the website https://opendata.ihep.ac.cn/\#/Search}. To illustrate these catalog sources, a detailed view of 82 sources with Galactic latitude $|b|<12^\circ$ are shown in Figures~\ref{Fig::gal_0_110}-\ref{Fig::gal_110_220}, in which most sources are concentrated in the inner Galactic plane. Eight individual sources are detected  at Galactic latitude $|b|>12^\circ$, as shown in Figure~\ref{Fig::high_b}. Among them, 1LHAASO J1653+3943 and 1LHAASO J1104+3810, which correspond to Mrk 421 and Mrk 501 respectively, have  been significantly detected by the WCDA detector, however, no gamma-ray emission has been found at E$>$25 TeV by KM2A detector, due to significant absorption from interstellar radiation fields (ISRFs) and CMB.  
In addition to Mrk 421 and Mrk 501, there are another 34 sources detected by only one detector. This finding is reasonable considering the curved spectral shape, as illustrated in Figure~\ref{Fig::tswk}. The different shapes of the spectra can result in various relationships in the TS value between WCDA and KM2A detections.

\section{Discussion}

\subsection{PeVatrons}

\begin{figure}[h]
\centering
\includegraphics[width=0.49\textwidth]{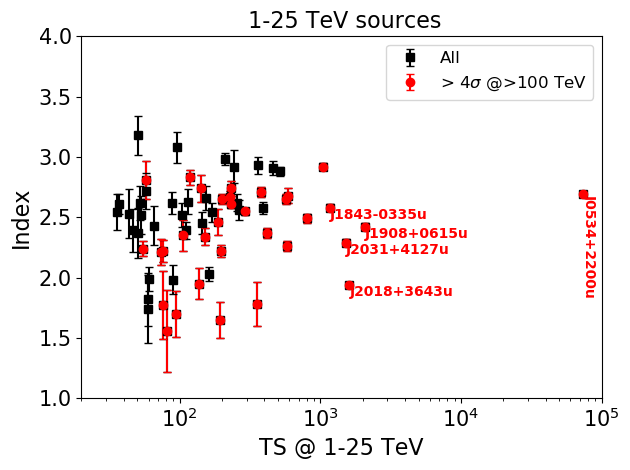}
\includegraphics[width=0.49\textwidth]{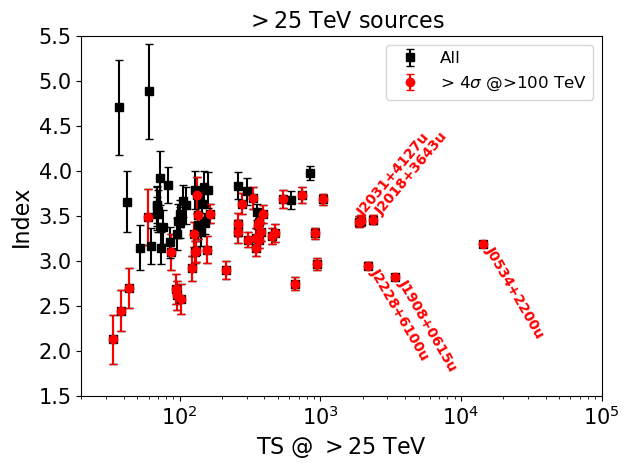}
\caption{Distribution of the spectral index and the significance for sources at  energy range from 1$-$25 TeV (left) and $E > 25 \rm\ TeV$ (right), respectively. The red points indicate the sources also with significance above 4$\sigma$ at $E >$100 TeV. The extragalactic sources are excluded in these figures. The six most significant sources are labeled by their name. }\label{Fig::100}
\end{figure}

The search for and identification of galactic sources capable of accelerating CRs up to PeV energies, known as  PeVatrons, is of great importance.
UHE gamma rays serve as a crucial and highly promising  tool for achieving this target. With its unprecedented sensitivity at UHE, LHAASO represents the best instrument to survey the PeVatrons. In fact, using approximately 300 days of data collected by the half KM2A, LHAASO has already detected 12 UHE sources with significance above 7$\sigma$ and maximum energies reaching up to 1.4 PeV \citep{2021Natur.594...33C}, providing crucial candidates for hadronic PeVatron exploration.  
In this paper, 43 sources with significance above 4$\sigma$ at energy beyond 100 TeV are listed in Table~\ref{tab:1}. Since these sources are well detected at $E > 25\rm\ TeV$ with much higher significance, they are significant enough to be identified as UHE sources. Among the 43 sources, 22 sources are detected with significance above 7$\sigma$ (TS$_{100} \approx$ 54) at $E >$ 100 TeV, improving by a factor of 2 the previous number of sources in \cite{2021Natur.594...33C}.
It is worth noting that 57\% (43 out of 75) $E >$ 25 TeV sources are UHE sources. According to Figure~\ref{Fig::100}, most of these UHE sources have higher significance or harder  spectral index than the other $E >$ 25 TeV sources, which indicates that the remaining $E >$ 25 TeV sources may also be detected  as UHE sources by LHAASO in the future with further accumulation of data. This provides important evidence for the exciting fact that the Milky Way is full of UHE sources and full of PeV particle accelerators.  Further deep analysis focusing on these sources one by one or type by type to identify  the emission mechanism is a straightforward  task to explore and identify the hadronic PeVatrons.

According to Table~\ref{tab:1}, 76\% (33 out of 43) UHE sources are  detected  at energies 1$-$25 TeV by WCDA. Figure~\ref{Fig::100} shows the spectral index versus significance for these sources at energies 1$-$25 TeV.  For comparison, other sources detected at energies 1$-$25 TeV are also presented in the figure, excluding the extragalactic sources listed in Table~\ref{tab:1}. The UHE sources have harder spectral index or higher significance than the others in the energy band 1-25 TeV. Up to now, more than 250 VHE sources have been detected, and more than  100 sources are within  the Galaxy with a significant fraction being located in the southern sky, which is out the FOV of LHAASO. These Galactic sources  would be important UHE candidates and may be revealed with UHE emission in the future observations.

It is worth noting that 10 out of 43 UHE sources were not detected by WCDA in 1-25 TeV band. These sources exhibit dominant gamma-ray emissions at energies around a few tens of TeV or $E>100$ TeV. Among them, sources like 1LHAASO J0216+4237u possibly belong to a new population of gamma-ray sources. These would demonstrate the distinctive importance of the UHE window at higher energy, which could explore new phenomena and new extreme celestial bodies of the Universe. With more accumulation of data, LHAASO will discover more sources with such features in the future. 
These sources should be also important candidates for PeVatron exploration and identification.


\subsection{TeV Sources Discovered by LHAASO}\label{discover}
As shown in Table~\ref{tab:1}, there are 25 sources without any known TeV source association. Additionally, 7 sources have conflicting extensions (with a difference of more than 0.5$^\circ$) compared to the known TeV sources within the searching region. Therefore, tentatively, these 32 sources are announced as new TeV sources detected by LHAASO in this catalog. To determine the physical nature of these sources it is often necessary to analyze their spectral and morphological features, as well as the information about multi-wavelength associations. 
 In this section, we conduct a preliminary analysis of the associations for each new TeV source. The majority of these newly discovered LHAASO sources are expected to be Galactic sources, due to their extended properties or their KM2A component detection. However, several point-like sources detected only by WCDA could potentially have an extra-galactic origin. 
 It is known that there are a number of types of Galactic gamma-ray emitters, such as SNRs, PSRs and their PWNe, massive star clusters (MSCs), star-forming regions(SFRs), superbubbles, binaries, etc. 
 The vast majority of detected Galactic TeV sources are likely to be associated with SNRs and PWNe~\citep{2018A&A...612A...1H}. To search for SNR and PWN associations, we take into account the most complete catalog of SNRs and PWNe to date, SNRcat\footnote{http://www.physics.umanitoba.ca/snr/SNRcat}~\citep{2012AdSpR..49.1313F}.
Although we do not expect  pulsed TeV emission from a pulsar to be detected by LHAASO, the pulsar is a probe to illustrate the characteristic properties of an association SNR and/or PWN,  and is also an indicator of a possible unseen SNR and/or PWN. For the pulsar associations, we used the web-based Australia Telescope National Facility Pulsar catalog (ATNF pulsar catalog \footnote{https://www.atnf.csiro.au/research/pulsar/psrcat/}). There is less association between low spin-down luminosity pulsars and known TeV sources, hence we apply a cut of $\dot{E}>10^{34} \rm\ erg\ s^{-1}$. Additionally, we search for associated high-energy (HE) gamma-ray sources in the 4FGL catalog~\citep{2020ApJS..247...33A}, which covers the sources at energies ranging from  0.1 GeV to $>$500 GeV. Note that in this section we claim an association or a counterpart based on a position less than $0.5^\circ$ away. For the likely extragalactic sources, the AGN catalog is adopted to search for associations within the position error. The association results are listed in Table~\ref{tab:2}.

\begin{figure}
\centering
\includegraphics[width=1.\textwidth]{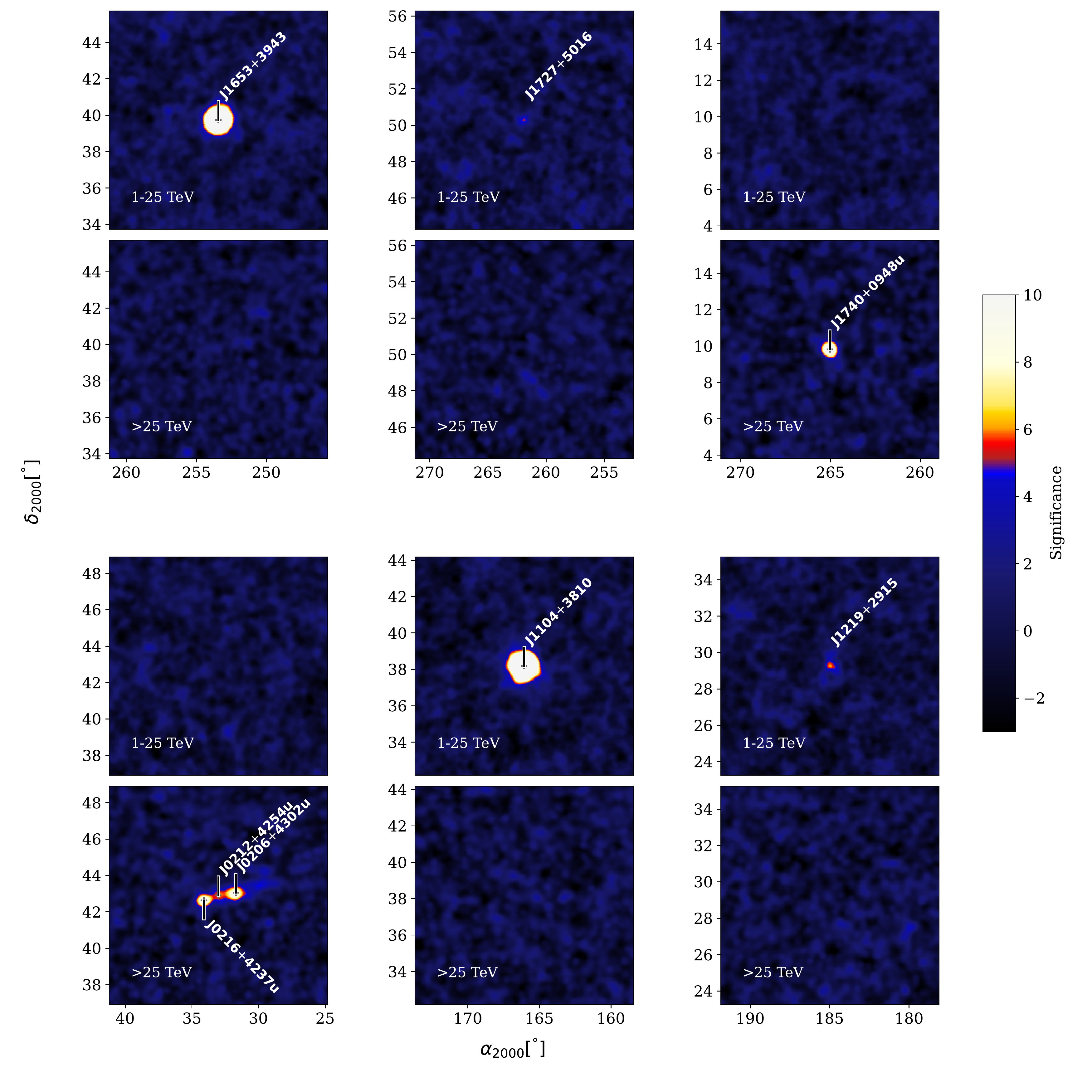}
\caption{LHAASO significance map for eight sources with $|b|>12^\circ$. For each source, WCDA ($1 {\rm\ TeV} < E < 25 \rm\ TeV$) and KM2A  ($E > $ 25 TeV) significance map are shown in top and bottom map, respectively. }
\label{Fig::high_b}
\end{figure}

Seven of the new TeV sources do not have any associations.  They are tentatively classified as ``dark" sources in this work. We list and discuss these ``dark" sources briefly here:

\begin{itemize}
\item 1LHAASO J0007+5659u is a point-like source detected only by KM2A. It is an UHE source with a significance of TS$_{100}$ = 43.5 at energies above 100 TeV, implying a Galactic origin. 

\item 1LHAASO J0206+4302u and 1LHAASO J0212+4254u are the point-like sources only detected by KM2A. Due to both of them being UHE sources, we suggest a Galactic origin, although these two sources are obviously far from the Galactic plane with Galactic latitude  $b \sim -17^\circ$. As shown in Figure~\ref{Fig::high_b}, these two sources and 1LHAASO J0216+4237u are close by and likely to be connected by bridges.  Considering that they have a similar spectral shape (with the index of $\approx 2.5$), we favor a  physical association among these three 1LHAASO sources.

\item 1LHAASO J1937+2128 is an extended source with the size of $\approx 1.3^\circ$ with the WCDA and KM2A components. Although 3HWC J1935+213 ($0.36^\circ$ away) and 3HWC J1936+223 ($0.85^\circ$ away) are found in point searches and are within the $1^\circ$ region around the center position of this 1LHAASO source, we report this source as a new detection due to its larger extension size. 

\item 1LHAASO J1959+1129u is a point-like TeV source detected above 25 TeV with TS = 90.2, and also detected as an UHE source with TS$_{100} = 59.3$. We cannot find any pulsar, SNR or PWN counterparts within a $0.5^\circ$ region around its position. Interestingly, it is  $0.23^\circ$ from the low mass X-ray binary (LXB) 4U 1957+11 which is a black hole candidate with the simplest and cleanest soft, disk-dominated spectra~\citep{2012ApJ...744..107N} in the X-ray regime. The TeV emission from 11LHAASO J1959+1129u is possibly associated with LXB 4U 1957+11. A dedicated study of 1LHAASO J1959+1129u from our LHAASO group is ongoing.

\item 1LHAASO J2200+5643u is an extended TeV source with the size of $\sim 0.5^\circ$  detected by WCDA and KM2A  simultaneously. It is an UHE source with the significance of TS$_{100}$ = 38. The spectral index is  $\approx 1.77$ at energies ranging from $\sim$ 10 TeV to 30 TeV as measured by WCDA and is $\approx 3.44$ at energies above 25 TeV as detected by KM2A, implying a peak or a break energy at a few tens of TeV.

\item 1LHAASO J2229+5927u is a large extended source with the size $\sim 1.8^\circ$ detected by WCDA  and by KM2A.  Within the $1^\circ$ region centered at the position of the WCDA component, we cannot find any counterpart.  The famous SNR G106.3+2.7, which has been extensively studied as a candidate PeVatron, is located at the edge of the extended region of 1LHAASO J2229+5927u. It is possible that this 1LHAASO source is the product of  CRs escaping from SNR G106.3+2.7. More detailed study for this 1LHAASO source is ongoing by our collaboration.
\end{itemize}

Eight of the new TeV sources have GeV counterparts only. The details of these sources are as follows:
\begin{itemize}

\item 1LHAASO 0056+6346u is an extended source  ($r_{39}\approx0.3^\circ$)  detected by WCDA  and KM2A detectors simultaneously. It is also a UHE source with  TS$_{100}$ = 94. The spectral shape is obviously curved with the index $\approx 2.35$ by WCDA and $\approx 3.33$ by KM2A. The unidentified point-like GeV source 4FGL J0057.9+6326 is 0.38$^\circ$ from this 1LHAASO source.
\item  1LHAASO J0500+4454 is an extended source with a size of $\sim 0.4^\circ$, only detected by WCDA in this work. The extension characteristic supports a Galactic origin. An unidentified point-like GeV source 4FGL J0501.7+4459 is 0.32$^\circ$ from this 1LHAASO source.

\item 1LHAASO J1858+0330 is an extended source with a size of $\approx 0.5^\circ$  detected by WCDA with a significance of TS = 114 and by KM2A with a significance of TS = 299. It is located in the Galactic plane ($l=36.8^\circ$ and $b=0.1^\circ$), towards which a complex gas distribution exists. The closest  GeV source is  4FGL J1857.9+0313c which is identified as the a blazar candidate of uncertain type by the \textsl{Fermi}-LAT group. We exclude that the 1LHAASO source  is associated with this extragalactic GeV source due to the extension characteristic. Another unidentified point-like GeV source 4FGL J1858.0+0354 is $0.41^\circ$ from this 1LHAASO source, and an association cannot be ruled out.
\item 1LHAASO J1902+0648 is a point-like source with a significance of TS = 46.2 only detected by WCDA. The unidentified GeV source 4FGL J1902.5+0654 is $0.12^\circ$ from this 1LHAASO source. 

\item 1LHAASO J1924+1609 is an extended source with $r_{39}\approx 1.3^\circ$ detected by WCDA and KM2A simultaneously. Although 3HWC J1923+169, found in a point search by the HAWC group, is $0.86^\circ$ from the center of the WCDA component, we announce 1LHAASO J1924+1609 as a new TeV source due to its large extended size. Within the $0.5^\circ$ region around the centroid of  the WCDA component, there are three unidentified point-like GeV sources, i.e., 4FGL J1924.3+1601c, 4FGL J1925.4+1616 and 4FGL J1924.3+1628. Note that the positional offset between  the WCDA component and the KM2A component is about $0.69^\circ$ with an uncertainty $0.66^\circ$ (at 95\% confidence level). The pulsar PSR J1921+1544 ($\dot{E}=1.31\times 10^{34} \rm\ erg\ s^{-1}$,$\tau_{c}=2320.0\rm\ kyr$,$d=9.04 \rm\ kpc$) is $0.1^\circ$ from the position of the KM2A component.
 \item 1LHAASO J1931+1653 is a point-like source with a significance of TS = 51.8, only detected by KM2A. The unidentified GeV source 4FGL J1931.1+1656 is $0.05^\circ$ from this 1LHAASO source.
\item  1LHAASO J2027+3657 is an extended source with  $r_{39}\approx0.23^\circ$ only detected by KM2A. It is resolved from  the previously published LHAASO source LHASSO J2018+3651, which is one of the brightest sources observed by LHAASO. 
 The point-like GeV source 4FGL J2026.5+3718c is located within a $0.5^\circ$ region around this 1LHAASO source.  
\item 1LHAASO J2047+4434 is an extended TeV source with TS = 62.4, only detected by KM2A. A weak WCDA component with TS $\approx 20$ is found but is not included in this catalog. The unidentified point-like GeV source 4FGL J02049.3+4440c is $0.32^\circ$ from this 1LHAASO source.
\end{itemize}

Sixteen of the new TeV sources have  pulsar or PWN/SNR associations. Each source is briefly described as follows:

\begin{itemize}
\item 1LHAASO J0216+4237u  is  a point-like source located at high Galactic latitude $b \approx -17^\circ$. Interestingly, $0.32^\circ$ from this source,  an energetic millisecond pulsar PSR J0216+4238 ( $\dot{E}=2.44\times 10^{35} \rm\ erg\ s^{-1}$, $\tau_{c}=476\rm\ Myr$, $d=3.15 \rm\ kpc$, $p_0=2.3$ ms) is located. The large position offset reduces the possibility of  an association between 1LHAASO J0216+4237u and PSR J0216+4238. As mentioned above, we favor a physical association between 1LHAASO J0216+4237u, 1LHAASO J0206+4302u and 1LHAASO J0212+4254u. More details will be discussed in a forthcoming report. 

\item 1HAASO J0249+6022 is an extended source with $r_{39}\approx 0.71^\circ$ for the WCDA component and with $r_{39}\approx 0.38^\circ$ for the KM2A component. We can find that the KM2A component overlaps the extended region of the WCDA component, with a positional offset of $\sim 0.45^\circ$.  The WCDA component at energies 1$-$20 TeV could plausibly include the emission of  two TeV sources, resulting in the large position offset compared to the KM2A component. In our searching radius, we just find one pulsar counterpart (PSR J0248+6021, $d=2.0$ kpc $\dot{E}=2.13\times 10^{35} \rm\ erg\ s^{-1}$, $\tau_{c}=62.4\rm\ kyr$, $p_0=217.1$ ms), $0.16^\circ$ from the position of the KM2A component. The 1LHAASO source is likely to correspond to an astrophysical system associated with PSR J0248+6021, such as a composite SNR.\

\item 1LHAASO J0359+5406 is an extended source with a size of $\approx 0.3^\circ$ detected by WCDA and KM2A simultaneously.  Two energetic pulsars, PSR B0355+54 ($d=1$ kpc, $\dot{E}=4.45\times 10^{34} \rm\ erg\ s^{-1}$, $\tau_{c}=564\rm\ kyr$, $p_0=156.4$ ms) and PSR J0359+5414 ($d$ is unknown, $\dot{E}=1.32\times 10^{36} \rm\ erg\ s^{-1}$, $\tau_{c}=75.2\rm\ kyr$, $p_0=79.4$ ms) are  $0.12^\circ$ and $0.16^\circ$ away, respectively, from the position of the KM2A component. The X-ray observations have revealed the presence of a $\sim 30''$ compact nebula  and a fainter tail of emission visible up to $\sim 6'$ southwest of PSR B0355+54~\citep{2016ApJ...833..253K}, and of $\sim 30''$  diffuse X-ray emission extended roughly along the SE-NE direction around PSR J0359+5414~\citep{2018MNRAS.476.2177Z}. Thus, this 1LHAASO source is likely to be a TeV PWN powered by PSR B0355+54 or PSR J0359+5414. On the other hand, it could be a TeV halo associated with the middle-aged pulsar PSR B0355+54. It is worth noting that the HAWC collaboration also reported their detection at the same time~\citep{2023ApJ...944L..29A}.

 \item 1LHAASO J0428+5531 is an extended source  detected by WCDA  and KM2A detectors simultaneously. The WCDA component is  with  $r_{39} \approx 1.18^\circ$.  We can find that  the shell-type SNR G150.3+04.5, with a size of $3.0^\circ \times 2.5^\circ$~\citep{2014A&A...567A..59G}, is $0.26^\circ$ from the center of the WCDA component. In  the GeV band, an extended source 4FGL J0426.5+5522e modeled as a $1.515^\circ$ disk is also associated with SNR G150.3+04.5 \citep{2017ApJS..232...18A}. The WCDA component of 1LHAASO J0428+5531  is  possibly of SNR origin due to the comparable location and extension size in the multi-wavelength observation. The extended KM2A component ($r_{39} \approx 0.23^\circ$) is $\approx 0.97^\circ$ from the central position of the WCDA component and $\approx 0.1^\circ$  away from the radio west shell of  SNR G150.3+04.5~\citep[with a size of  $64.1'\times18.8'$, also named SNR G150.8+03.8 in ][]{2014A&A...566A..76G}. A point-like GeV source 4FGL J0426.5+5434, which is possibly  a pulsar candidate due to the curved SED and the cutoff energy at a few GeV, is $0.06^\circ$ from the position of KM2A component. Whether the KM2A component of 1LHAASO J0428+5531 is associated with the SNR G150.8+03.8 needs more study due to the obvious offset of the position and the extension size.

\item 1LHAASO J0534+3533 is a  point-like source detected by WCDA and KM2A in this work, which is near the centroid position of the shell-type SNR G172.8+01.5,  $0.3^\circ$ away. However, the radio shell size of SNR G172.8+01.5 is estimated as $4.4^\circ\times3.4^\circ$, which is much larger than the TeV emission size of 1LHAASO J0534+3533. If the radio shell is physically associated with 1LHAASO J0534+3533, the TeV emission is likely to be a product of the central pulsar wind nebula.  


\item 1LHAASO J1814-1636u is an extended UHE source with the size of $\approx 0.68^\circ$, only detected by KM2A.  Only two SNRs, i.e., G14.1-00.1 ($0.43^\circ$ away) and SNR G014.3+00.1 ($0.31^\circ$ away)  are found in our searching catalog. However, both SNR candidates are with $<0.2^\circ$ radio shell size, which is inconsistent with that of gamma-ray emission. An unidentified GeV source 4FGL J1816.2-1654c is found at $0.4^\circ$ away from the  position of the TeV emission.

\item Source 1LHAASO J1831-1028 only has an extended KM2A component and with the size of $\approx 0.94^\circ$. In our association procedure, we just find three SNR candidates, i.e., G021.0-00.4 ($0.3^\circ$ away), G021.5-00.1 ($0.36^\circ$  away) and G021.8-00.6 ($0.49^\circ$ away), all of which are with a radio size of $<0.2^\circ$.  Due to the larger conflict between the size of TeV emission and that of radio shell, we disfavor that the same population of electrons produces the TeV emission and the radio emission.

 \item 1LHAASO J1852+0050u is an extended source with a $\sim 0.64^\circ$ WCDA component and $\sim 0.85^\circ$ KM2A component. The nearest known TeV source is 2HWC J1852+013* ($0.55^\circ$ away), which is suffers from GDE impact and is without an extended size measurement. We claim this 1LHAASO source as a new TeV source due to its large extension. The middle-aged pulsar, PSR J1853+0056 ($d$ = 3.84 kpc, $\dot{E}=4.03\times 10^{34} \rm\ erg\ s^{-1}$, $\tau_{c}=204\rm\ kyr$) is the nearest pulsar, $0.31^\circ$ from the position of the KM2A component. The 1LHAASO source is possibly a TeV halo associated with this middle-aged pulsar. On the other hand, an extended GeV source 4FGL J1852.4+0037e (namely Kes 79), modeled by $0.63^\circ$ disk shape and identified as the  SNR or PWN type by the $Fermi$-LAT group, is in agreement with the position and size of the 1LHAASO source. In addition, the SNRcat source G033.6+00.1, which is a small extended radio source with the size of $10'$, is suggested to be associated with the GeV source.

 \item 1LHAASO J1906+0712 is an extended source only detected by WCDA with the size of $\sim 0.21^\circ$. A gamma-ray pulsar PSR J1906+0722 ($\dot{E} = 1.02\times 10^{36} \rm\ erg\ s^{-1}$,$\tau_{c} = 49.2\rm\ kyr$, $d$ is unknown) is  $0.19^\circ$ from this 1LHAASO source. $0.34^\circ$  away, a shell-type SNR G041.1-00.3 (3C 397) is detected in the radio band with the  size of $4.5'\times2.5'$, which is also detected by \textsl{Fermi}-LAT in the GeV band. The pulsar associated with SNR 3C 397 is not identified. An unidentified GeV source 4FGL J1906.9+0712 is also found within the $0.5^\circ$ region of 1LHAASO J1906+0712.

\item 1LHAASO J1928+1813u is an extended source with $r_{39}\approx 0.63^\circ$ only detected by KM2A. It is resolved from the UHE source  LHAASO J1929+1745 seen in previous LHAASO results. Within the $0.5^\circ$ region around the centroid of 1LHAASO J1928+1813u, we can find the source SNR G053.4+00.0 ($0.39^\circ$ away), which is a shell-type SNR with the radio size of $5'$ at a distance $5.6-6.4$ kpc. Two GeV sources listed in 4FGL are found within the association region. At $\sim 0.47^\circ$ away, there is an energetic pulsar PSR J1928+1746 ($d$ = 4.34 kpc, $\dot{E}=1.6\times 10^{36} \rm\ erg\ s^{-1}$, $\tau_{c}=82.6\rm\ kyr$).

\item 1LHAASO J1954+3253  is an  extended TeV source ($r_{39}\approx 0.17^\circ$), only detected by WCDA with the significance of TS = 144. SNR G069.0+02.7 (also named CTB 80) with the size $\sim 80'$ overlaps this 1LHAASO source. SNR CTB 80 is an old SNR with a kinematic distance of $\sim 1.5$ kpc. At the center of SNR CTA 80, the pulsar PSR B1951+32 ($\dot{E} = 3.74\times 10^{36} \rm\ erg\ s^{-1}$,$\tau_{c} = 107.0\rm\ kyr$, $d = 3.00 \rm\ kpc$) is generally regarded as the compact object associated with this old SNR. A X-ray PWN around the pulsar PSR B1951+32 has been identified  by ROSAT, which is shown as a  5 arcmin extended nebula. The pulsar and the SNR/PWN are also detected by $Fermi$-LAT, which are 4FGL J1952.9+3252 and 4FGL J1955.1+3321, respectively. We note that the position of TeV emission is $\sim 0.33^\circ$ from that of the pulsar and its X-ray PWN, for which more study is needed to  confirm the physical association between the 1LHAASO source and CTB 80. 

\item 1LHAASO J1956+2921 is a large extended source with an  $r_{39} \approx 0.99^\circ$ WCDA component and  $r_{39} \approx 0.78^\circ$ KM2A component. It is resolved from the published LHAASO source LHAASO J1958+2845. At $0.36^\circ$ from the position of the WCDA component, a shell type SNR with radio size of $31'\times25'$ is found. 

 \item 1LHAASO J1959+2846u is an UHE TeV source with the extension size of $r_{39}\approx0.3^\circ$, only detected by KM2A, which is also resolved the from previously published source LHAASO J1956+2845.  The  pulsar  PSR J1959+2846 ($0.1^\circ$ away, $\dot{E}=3.42\times 10^{35} \rm\ erg\ s^{-1}$, $d = 1.95\rm\ kpc$, $\tau_{c}=21.7\rm\ kyr$) is the only pulsar counterpart in our searching radius. 
  SNR G065.8-00.5 and SNR G066.0-00.0 are found at $0.16^\circ$ and $0.39^\circ$ from the position of 1LHAASO J1959+2846u, with  radio sizes of  $10'\times6'$ and $30'\times25'$,  respectively.

 \item 1LHAASO J2002+3244u is a point-like source with TS=74.0 as detected by WCDA and with TS = 43.6 as detected by KM2A. It is also an UHE source with TS$_{100}$ = 28.1.  4FGL J2002.3+3246 is spatially coincident with this source, which is identified as a potential association with a SNR or PWN by the  $Fermi$-LAT group. A shell type SNR G069.7+01.0  is possibly associated with  1LHAASO 2002+3244u ($0.04^\circ$ away). We favor that 1LHAASO 2002+3244u has an  SNR origin because the position and size of TeV emission  agree with the radio shell of SNR G069.7+01.0.

\item 1LHAASO J2028+3352 is a large extended source with $r_{39}\approx1.6^\circ$, only detected by KM2A. $0.36^\circ$ from their centroid position, a middle-aged pulsar PSR J2028+3332 ($\dot{E}=3.48\times 10^{34} \rm\ erg\ s^{-1}$,$\tau_{c}=576.0\rm\ kyr$) is found, implying a possible TeV halo identification. Two GeV sources are found within the $0.5^\circ$ region around the 1LHAASO source, of which one is the pulsar PSR J2028+3332 and the other is an unidentified point-like source.

\item 1LHAASO J2238+5900 is an extended  source with the size of  $\approx0.51^\circ$  as detected by  WCDA with TS = 110.2 and with the size of $\approx0.44^\circ$ detected by KM2A with TS$=361$. Just one pulsar PSR J2238+5903 ($\dot{E}=8.99\times 10^{35} \rm\ erg\ s^{-1}$, $d = 2.83\rm\ kpc$, $\tau_{c}=26.6\rm\ kyr$) is found within $0.5^\circ$ region of this source,  at $0.07^\circ$ from the centroid position of the KM2A component.  The young pulsar and  the size of TeV emission shrinking with increasing energy support that  1LHAASO J2238+5900 has a PWN origin.
\end{itemize}

One  possible new extragalactic source is as follows:

\begin{itemize}
\item 1LHAASO J1219+2915 is a  point-like source, only detected by WCDA with a significance of TS = 49.2, and located at high Galactic latitude ($b\sim 82.5^\circ$). We cannot find any pulsar or SNR/PWN counterpart associated with this  source. It is a likely  extragalactic source due to the high Galactic latitude and null detection by KM2A.  The closest AGN counterpart is the LINER-type AGN NGC 4278, $0.05^\circ$ from the 1LHAASO source. It is  the most possible  association even  though we can not firmly identify this 1LHAASO source at present.    
\end{itemize}

\subsection{Pulsar Wind Nebulae or TeV Halos in the 1LHAASO Catalog}\label{p_pulsar}
Pulsars left behind from supernova explosions are rapidly spinning neutron stars with extremely strong magnetic fields. The pulsed gamma-ray emission dominates at GeV energies, and only a few cases can extend to VHE. At VHE, the steady emission is from a PWN produced by the termination of the ultra-relativistic wind. PWNe constitute one of the largest VHE source populations within the Galaxy. 
The Crab nebula PWN has also been identified as an electron PeVatron~\citep{2021Sci...373..425L}.
The diffusion of escaping particles from a PWN would leads to extended VHE gamma-ray emission, which has been extensive discussed as a ``TeV halo” after the discoveries of several cases \citep{2017Sci...358..911A,2021PhRvL.126x1103A}. TeV halos are thought to form around middle-aged pulsars with ages of at least several tens of thousands of years. PWNe or TeV halos should also contribute a significant fraction of the 1LHAASO sources, and an effective method to check for this is to search for spatial coincidence with the known pulsars. 

To search for  pulsars associated with 1LHAASO sources, the ATNF pulsar catalog is adopted. For the hunting of associations  in astronomy, an important work is to estimate the possibility of spatially coincidence by accident. A similar method as that used in \cite{1997ApJ...481...95M} is adopted here. According to a previous empirical result at VHE, the identified PWN or halo type VHE sources are always associated with pulsars with high spindown power. Therefore, for each pulsar within 0.5$^{\circ}$ of a 1LHAASO source, the chance probability $P_{c}$ is estimated according to the space angle $r$ and pulsar spindown power $\dot{E}$ using
\begin{equation}
    P_c=1 -e^{r^2/r_{0}^{2}} ,
\end{equation}
Where $r_0$ is the characteristic angle between confusing sources,
\begin{equation}
    r_{0}=[\pi \rho(\dot{E})]^{-1/2} ,
\end{equation}
and where $\rho(\dot{E})$ is the number density of pulsars with $\dot{E}$  not lower than that of the candidate pulsar. The  $\rho(\dot{E})$ is counted using the pulsars nearby the candidate pulsar with Galactic latitude $|b-b_{c}|<2.5^{\circ}$ and longitude $|l-l_c|<10^{\circ}$, where ($b_c$, $l_c$) denotes the Galactic coordinates of the candidate pulsar.

With this searching, 65 1LHAASO sources are found with at least one pulsar nearby within 0.5$^{\circ}$. To decrease the fake association by accident, the pulsars with chance probability higher than 1\% are excluded. After this filter, 35 1LHAASO sources are found with one associated pulsar each and 2 1LHAASO sources, 1LHAASO J0359+5406 and 1LHAASO J1929+1846, are found with two associated pulsars each. For the source with two associations, the associated pulsar with a lower chance probability is listed. There are also two pairs of 1LHAASO sources, 1LHAASO J1848-0001u vs. 1LHAASO J1850-0004 and 1LHAASO J2020+3638 vs. 1LHAASO J2020+3649u, associated with the same pulsar. Again, the 1LHAASO source with a lower chance probability is listed. Finally, 35 associated pulsars are derived for the 1LHAASO sources.  Detailed information about these associations is listed in Table~\ref{tab:pulsar}.

\begin{figure}[h]
\centering
\includegraphics[width=0.65\textwidth]{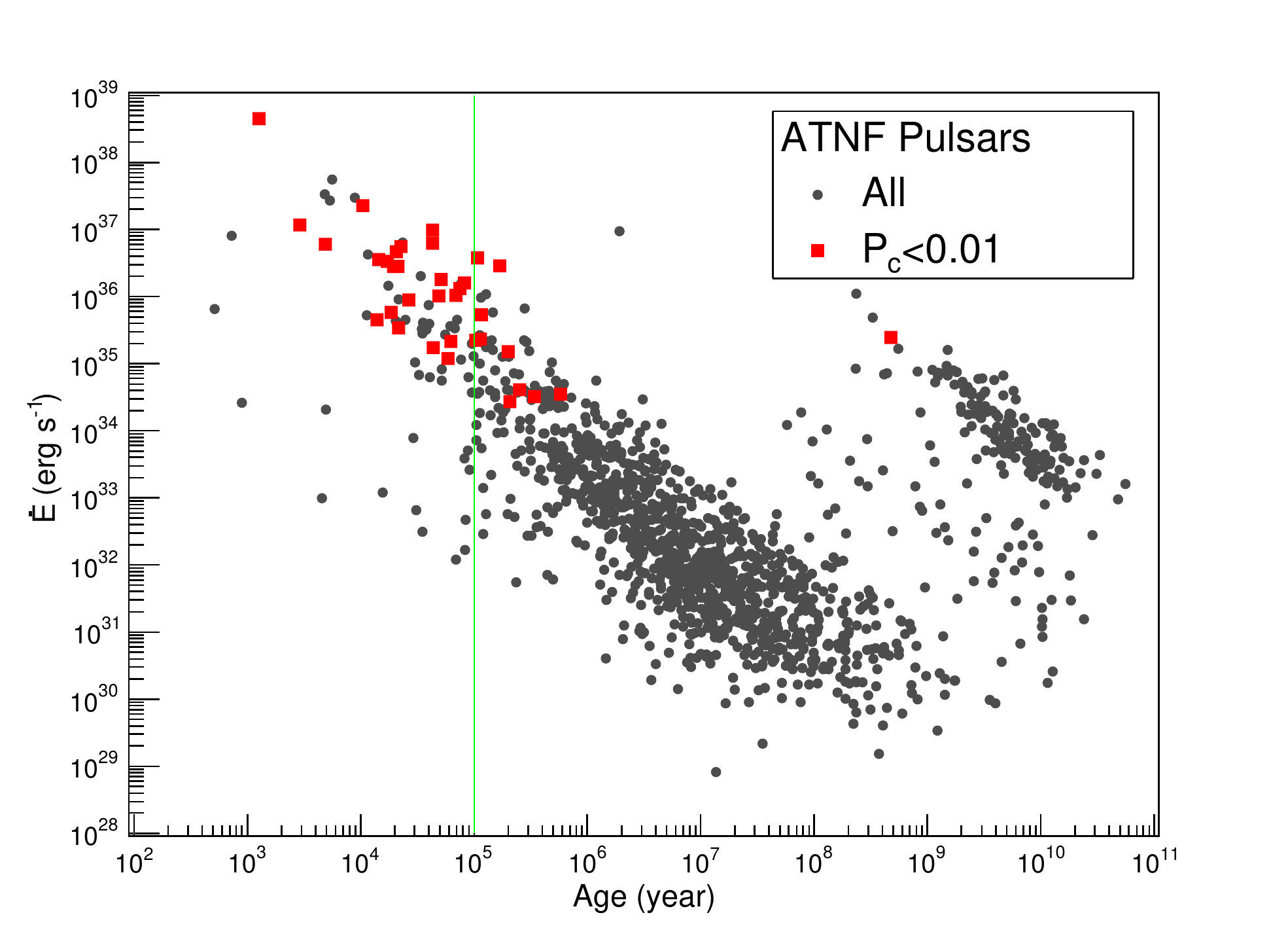}
 \caption{Pulsar spindown power $\dot{E}$ versus age for all pulsars of the ATNF catalog within the FOV of LHAASO and the 1LHAASO-associated pulsars with the chance probability $P_{c}$ less than 0.01.}
\label{Fig::Pulsar}
\end{figure}

Figure~\ref{Fig::Pulsar} shows the spindown power versus age of the associated pulsars. For comparison, all the pulsars of the ATNF catalog within LHAASO FOV are also shown in the figure. As expected, the TeV-gamma associated pulsars are among that with the most energetic spindown power ($\dot{E} > 10^{34}\rm\ erg\ s^{-1}$). Among these associated pulsars, the ages of 24 pulsars are less than 100k years. The corresponding 1LHAASO sources have a high possibility to be PWNe. The ages of 11 pulsars are older than 100k years, and this marks the corresponding 1LHAASO sources to possibly be  PWNe/TeV halos.
It is worth noting that some of them have already been identified as PWNe or PWN/TeV halos in the TeVCat, as marker in the Table~\ref{tab:pulsar}. An exciting result is that one pulsar, i.e., PSR J0218+4232, is a millisecond pulsar, which has been expected to produce VHE emission but there has been a lack of observation evidence. The spatial coincidence between this millisecond pulsar and 1LHAASO J0216+4237u is notable, with a confidence level of  2.9$\sigma$. This finding favor the existence of VHE emission around millisecond pulsars, although a non-physical association cannot be definitively ruled out. 

According to Figure~\ref{Fig::Pulsar}, most of the energetic pulsars with $\dot{E}$ higher than 10$^{36} \rm\ erg\ s^{-1}$. within the FOV of LHAASO are associated with 1LHAASO sources. These shows that the PWNe of energetic pulsars are promising as effective to VHE gamma-ray emitters. It is worth noting that no VHE or UHE emission was found from  a handful of energetic pulsars with $\dot{E}$ higher than 
10$^{37}\rm\ erg\ s^{-1}$. according to Figure~\ref{Fig::Pulsar}. This shows that emission of PWNs may also be affected by other parameters of the pulsars or the surrounding environment, besides the spindown power.
Among the 35 1LHAASO sources with pulsar associations, twenty-two are labeled as UHE sources. This result suggests that PWNe  also contribute a significant portion of the UHE sources and acceleration of electrons close to or up to 1 PeV is common for energetic PWNe. In addition to the Crab Nebula, more electron PeVatrons may be revealed by LHAASO in the future.   

\section{Summary}

The 1LHAASO catalog, with 90 VHE gamma-ray sources, is the first comprehensive search conducted using data from the LHAASO observatory. The catalog utilizes 508 days of data from the full WCDA and 933 days of data from the full KM2A and its partial array. With 32 sources discovered, this survey is the most sensitive and roughly unbiased investigation of large sky regions in the VHE band, covering a declination range from $-20^{\circ}$ to 80$^{\circ}$. Seven new sources do not have counterparts in the GeV gamma-ray, pulsar, and SNR catalogs, and are referred to as ``dark sources". Among the 90 1LHAASO sources, sixty-nine are detected with emissions in the energy range 1-25 TeV, while seventy-five are detected at emission energy of $E > 25$ TeV, both with a significance above 5$\sigma$. The catalog provides information on the extension and spectrum of each source, with 65 sources exhibiting clear extension. To be conservative and avoid confusion with diffuse gamma rays from the Galactic plane, only sources with extensions less than 2$^\circ$ are included in the 1LHAASO catalog. Additionally, 43 sources exhibit clear UHE emission, with a significance above 4$\sigma$ at  $E > 100$ TeV. This significantly expands the current number of known UHE sources, by a factor of 3, and suggests that most Galactic VHE sources are potential UHE sources, indicating a prevalence of PeV particle accelerators in the Milky Way. Ten out of the forty-three UHE sources are not detected at the 1-25 TeV range, possibly representing a new class of gamma-ray sources dominated by emission above tens of TeV. Thirty-five 1LHAASO sources are associated with energetic pulsars, with a chance coincident probability of less than 1\%. Apart from sources already identified as PWNe or TeV halos, the remaining sources associated with energetic ($\dot{E} > 10^{36} \rm\ erg\ s^{-1}$) pulsars are likely PWNs or TeV halos. The majority of energetic pulsars within the LHAASO field of view are associated with 1LHAASO sources, and a significant fraction of these sources are also UHE sources. This suggests that PWNe associated with energetic pulsars are effective emitters of VHE and UHE radiation and have a general capability to accelerate electrons close to or up to PeV energies.

With a large FOV and full duty cycle, the full LHAASO detector configuration has been monitoring the sky since July 2021 and will continue operations for the next 20 years. With the the accumulation of data, the sensitivity in both the VHE and UHE bands will improve. 
Further optimization of data analysis and refining detector response simulation are ongoing, which may also expand its capacity. Deep and multi-wavelength analysis focusing on the 1LHAASO sources one by one is also ongoing. Therefore, LHAASO is expected to unveil more new discoveries and provide a deeper understanding of the VHE and UHE universe in the near future.

\section{Acknowledgments}
We would like to thank all staff members who work at the LHAASO site above 4400 meters above sea level year-round to maintain the detector and keep the water recycling system, electricity power supply and other components of the experiment operating smoothly. We are grateful to Chengdu Management Committee of Tianfu New Area for the constant financial support for research with LHAASO data. 
This research work is  supported by the following grants: The National Key R\&D program of China No.2018YFA0404201, No.2018YFA0404202, No.2018YFA0404203, No.2018YFA0404204,  National Natural Science Foundation of China No.12022502, No.U1831208, No.12205314, No.12105301, No.12261160362, No.12105294, No.U1931201, No.12005246, No.12173039,  Department of Science and Technology of Sichuan Province, China No.2021YFSY0030, Project for Young Scientists in Basic Research of Chinese Academy of Sciences No.YSBR-061,  and in Thailand by the NSRF via the Program Management Unit for Human Resources \& Institutional Development, Research and Innovation (No. B37G660015).

\section{author contributions}
S.Q. Xi and S.Z. Chen led the drafting of text and performed the data analysis of KM2A,  S.C. Hu and M. Zha conducted the data analysis of WCDA. Y.Y. Guo and J.Y. He provided the cross-check. G.M. Xiang illustrated some figures in this article.  Zhen Cao, the spokesperson of the LHAASO Collaboration, coordinated the specific working group for this paper involving all corresponding authors. All other authors participated in data analysis, including detector calibration, data processing, event reconstruction, data quality checks, and various simulations, and provided comments on the manuscript.


\begin{thebibliography}{}
\expandafter\ifx\csname natexlab\endcsname\relax\def\natexlab#1{#1}\fi
\providecommand{\url}[1]{\href{#1}{#1}}
\providecommand{\dodoi}[1]{doi:~\href{http://doi.org/#1}{\nolinkurl{#1}}}
\providecommand{\doeprint}[1]{\href{http://ascl.net/#1}{\nolinkurl{http://ascl.net/#1}}}
\providecommand{\doarXiv}[1]{\href{https://arxiv.org/abs/#1}{\nolinkurl{https://arxiv.org/abs/#1}}}

\bibitem[{{Abdo} {et~al.}(2007){Abdo}, {Allen}, {Berley}, {Casanova}, {Chen},
  {Coyne}, {Dingus}, {Ellsworth}, {Fleysher}, {Fleysher}, {Gonzalez},
  {Goodman}, {Hays}, {Hoffman}, {Hopper}, {H{\"u}ntemeyer}, {Kolterman},
  {Lansdell}, {Linnemann}, {McEnery}, {Mincer}, {Nemethy}, {Noyes}, {Ryan},
  {Saz Parkinson}, {Shoup}, {Sinnis}, {Smith}, {Sullivan}, {Vasileiou},
  {Walker}, {Williams}, {Xu}, \& {Yodh}}]{2007ApJ...664L..91A}
{Abdo}, A.~A., {Allen}, B., {Berley}, D., {et~al.} 2007, \apjl, 664, L91,
  \dodoi{10.1086/520717}

\bibitem[Abdollahi et al.(2020)]{2020ApJS..247...33A} Abdollahi, S., Acero, F., Ackermann, M., et al.\ 2020, \apjs, 247, 33. doi:10.3847/1538-4365/ab6bcb

\bibitem[{{Abdollahi} {et~al.}(2022){Abdollahi}, {Acero}, {Baldini}, {Ballet},
  {Bastieri}, {Bellazzini}, {Berenji}, {Berretta}, {Bissaldi}, {Blandford},
  {Bloom}, {Bonino}, {Brill}, {Britto}, {Bruel}, {Burnett}, {Buson}, {Cameron},
  {Caputo}, {Caraveo}, {Castro}, {Chaty}, {Cheung}, {Chiaro}, {Cibrario},
  {Ciprini}, {Coronado-Bl{\'a}zquez}, {Crnogorcevic}, {Cutini}, {D'Ammando},
  {De Gaetano}, {Digel}, {Di Lalla}, {Dirirsa}, {Di Venere}, {Dom{\'\i}nguez},
  {Fallah Ramazani}, {Fegan}, {Ferrara}, {Fiori}, {Fleischhack}, {Franckowiak},
  {Fukazawa}, {Funk}, {Fusco}, {Galanti}, {Gammaldi}, {Gargano}, {Garrappa},
  {Gasparrini}, {Giacchino}, {Giglietto}, {Giordano}, {Giroletti}, {Glanzman},
  {Green}, {Grenier}, {Grondin}, {Guillemot}, {Guiriec}, {Gustafsson},
  {Harding}, {Hays}, {Hewitt}, {Horan}, {Hou}, {J{\'o}hannesson}, {Karwin},
  {Kayanoki}, {Kerr}, {Kuss}, {Landriu}, {Larsson}, {Latronico},
  {Lemoine-Goumard}, {Li}, {Liodakis}, {Longo}, {Loparco}, {Lott}, {Lubrano},
  {Maldera}, {Malyshev}, {Manfreda}, {Mart{\'\i}-Devesa}, {Mazziotta}, {Mereu},
  {Meyer}, {Michelson}, {Mirabal}, {Mitthumsiri}, {Mizuno}, {Moiseev},
  {Monzani}, {Morselli}, {Moskalenko}, {Negro}, {Nuss}, {Omodei}, {Orienti},
  {Orlando}, {Paneque}, {Pei}, {Perkins}, {Persic}, {Pesce-Rollins},
  {Petrosian}, {Pillera}, {Poon}, {Porter}, {Principe}, {Rain{\`o}}, {Rando},
  {Rani}, {Razzano}, {Razzaque}, {Reimer}, {Reimer}, {Reposeur},
  {S{\'a}nchez-Conde}, {Saz Parkinson}, {Scotton}, {Serini}, {Sgr{\`o}},
  {Siskind}, {Smith}, {Spandre}, {Spinelli}, {Sueoka}, {Suson}, {Tajima},
  {Tak}, {Thayer}, {Thompson}, {Torres}, {Troja}, {Valverde}, {Wood}, \&
  {Zaharijas}}]{2022ApJS..260...53A}
{Abdollahi}, S., {Acero}, F., {Baldini}, L., {et~al.} 2022, \apjs, 260, 53,
  \dodoi{10.3847/1538-4365/ac6751}
\bibitem[Albert et al.(2023)]{2023ApJ...944L..29A} Albert, A., Alfaro, R., Arteaga-Vel{\'a}zquez, J.~C., et al.\ 2023, \apjl, 944, L29. doi:10.3847/2041-8213/acb5ee

\bibitem[{{Abeysekara} {et~al.}(2017){Abeysekara}, {Albert}, {Alfaro},
  {Alvarez}, {{\'A}lvarez}, {Arceo}, {Arteaga-Vel{\'a}zquez}, {Avila Rojas},
  {Ayala Solares}, {Barber}, {Bautista-Elivar}, {Becerril}, {Belmont-Moreno},
  {BenZvi}, {Berley}, {Bernal}, {Braun}, {Brisbois}, {Caballero-Mora},
  {Capistr{\'a}n}, {Carrami{\~n}ana}, {Casanova}, {Castillo}, {Cotti},
  {Cotzomi}, {Couti{\~n}o de Le{\'o}n}, {De Le{\'o}n}, {De la Fuente},
  {Dingus}, {DuVernois}, {D{\'\i}az-V{\'e}lez}, {Ellsworth}, {Engel},
  {Enr{\'\i}quez-Rivera}, {Fiorino}, {Fraija}, {Garc{\'\i}a-Gonz{\'a}lez},
  {Garfias}, {Gerhardt}, {Gonz{\'a}lez Mu{\~n}oz}, {Gonz{\'a}lez}, {Goodman},
  {Hampel-Arias}, {Harding}, {Hern{\'a}ndez}, {Hern{\'a}ndez-Almada}, {Hinton},
  {Hona}, {Hui}, {H{\"u}ntemeyer}, {Iriarte}, {Jardin-Blicq}, {Joshi},
  {Kaufmann}, {Kieda}, {Lara}, {Lauer}, {Lee}, {Lennarz}, {Vargas},
  {Linnemann}, {Longinotti}, {Luis Raya}, {Luna-Garc{\'\i}a}, {L{\'o}pez-Coto},
  {Malone}, {Marinelli}, {Martinez}, {Martinez-Castellanos},
  {Mart{\'\i}nez-Castro}, {Mart{\'\i}nez-Huerta}, {Matthews},
  {Miranda-Romagnoli}, {Moreno}, {Mostaf{\'a}}, {Nellen}, {Newbold}, {Nisa},
  {Noriega-Papaqui}, {Pelayo}, {Pretz}, {P{\'e}rez-P{\'e}rez}, {Ren}, {Rho},
  {Rivi{\`e}re}, {Rosa-Gonz{\'a}lez}, {Rosenberg}, {Ruiz-Velasco}, {Salazar},
  {Salesa Greus}, {Sandoval}, {Schneider}, {Schoorlemmer}, {Sinnis}, {Smith},
  {Springer}, {Surajbali}, {Taboada}, {Tibolla}, {Tollefson}, {Torres},
  {Ukwatta}, {Vianello}, {Weisgarber}, {Westerhoff}, {Wisher}, {Wood},
  {Yapici}, {Yodh}, {Younk}, {Zepeda}, {Zhou}, {Guo}, {Hahn}, {Li}, \&
  {Zhang}}]{2017Sci...358..911A}
{Abeysekara}, A.~U., {Albert}, A., {Alfaro}, R., {et~al.} 2017, Science, 358,
  911, \dodoi{10.1126/science.aan4880}

\bibitem[{{Abeysekara} {et~al.}(2020){Abeysekara}, {Albert}, {Alfaro}, {Angeles
  Camacho}, {Arteaga-Vel{\'a}zquez}, {Arunbabu}, {Avila Rojas}, {Ayala
  Solares}, {Baghmanyan}, {Belmont-Moreno}, {BenZvi}, {Brisbois},
  {Caballero-Mora}, {Capistr{\'a}n}, {Carrami{\~n}ana}, {Casanova}, {Cotti},
  {Cotzomi}, {Couti{\~n}o de Le{\'o}n}, {De la Fuente}, {de Le{\'o}n},
  {Dichiara}, {Dingus}, {DuVernois}, {D{\'\i}az-V{\'e}lez}, {Ellsworth},
  {Engel}, {Espinoza}, {Fleischhack}, {Fraija}, {Galv{\'a}n-G{\'a}mez},
  {Garcia}, {Garc{\'\i}a-Gonz{\'a}lez}, {Garfias}, {Gonz{\'a}lez}, {Goodman},
  {Harding}, {Hernandez}, {Hinton}, {Hona}, {Huang}, {Hueyotl-Zahuantitla},
  {H{\"u}ntemeyer}, {Iriarte}, {Jardin-Blicq}, {Joshi}, {Kaufmann}, {Kieda},
  {Lara}, {Lee}, {Le{\'o}n Vargas}, {Linnemann}, {Longinotti}, {Luis-Raya},
  {Lundeen}, {L{\'o}pez-Coto}, {Malone}, {Marinelli}, {Martinez},
  {Martinez-Castellanos}, {Mart{\'\i}nez-Castro}, {Mart{\'\i}nez-Huerta},
  {Matthews}, {Miranda-Romagnoli}, {Morales-Soto}, {Moreno}, {Mostaf{\'a}},
  {Nayerhoda}, {Nellen}, {Newbold}, {Nisa}, {Noriega-Papaqui}, {Peisker},
  {P{\'e}rez-P{\'e}rez}, {Pretz}, {Ren}, {Rho}, {Rivi{\`e}re},
  {Rosa-Gonz{\'a}lez}, {Rosenberg}, {Ruiz-Velasco}, {Salesa Greus}, {Sandoval},
  {Schneider}, {Schoorlemmer}, {Sinnis}, {Smith}, {Springer}, {Surajbali},
  {Tabachnick}, {Tanner}, {Tibolla}, {Tollefson}, {Torres}, {Torres-Escobedo},
  {Villase{\~n}or}, {Weisgarber}, {Wood}, {Yapici}, {Zhang}, {Zhou}, \& {HAWC
  Collaboration}}]{2020PhRvL.124b1102A}
---. 2020, \prl, 124, 021102, \dodoi{10.1103/PhysRevLett.124.021102}

\bibitem[Abramowski et al.(2014)]{2014PhRvD..90l2007A} Abramowski, A., Aharonian, F., Ait Benkhali, F., et al.\ 2014, \prd, 90, 122007. doi:10.1103/PhysRevD.90.122007

\bibitem[{{Ackermann} {et~al.}(2013){Ackermann}, {Ajello}, {Allafort},
  {Baldini}, {Ballet}, {Barbiellini}, {Baring}, {Bastieri}, {Bechtol},
  {Bellazzini}, {Blandford}, {Bloom}, {Bonamente}, {Borgland}, {Bottacini},
  {Brandt}, {Bregeon}, {Brigida}, {Bruel}, {Buehler}, {Busetto}, {Buson},
  {Caliandro}, {Cameron}, {Caraveo}, {Casandjian}, {Cecchi}, {{\c{C}}elik},
  {Charles}, {Chaty}, {Chaves}, {Chekhtman}, {Cheung}, {Chiang}, {Chiaro},
  {Cillis}, {Ciprini}, {Claus}, {Cohen-Tanugi}, {Cominsky}, {Conrad}, {Corbel},
  {Cutini}, {D'Ammando}, {de Angelis}, {de Palma}, {Dermer}, {do Couto e
  Silva}, {Drell}, {Drlica-Wagner}, {Falletti}, {Favuzzi}, {Ferrara},
  {Franckowiak}, {Fukazawa}, {Funk}, {Fusco}, {Gargano}, {Germani},
  {Giglietto}, {Giommi}, {Giordano}, {Giroletti}, {Glanzman}, {Godfrey},
  {Grenier}, {Grondin}, {Grove}, {Guiriec}, {Hadasch}, {Hanabata}, {Harding},
  {Hayashida}, {Hayashi}, {Hays}, {Hewitt}, {Hill}, {Hughes}, {Jackson},
  {Jogler}, {J{\'o}hannesson}, {Johnson}, {Kamae}, {Kataoka}, {Katsuta},
  {Kn{\"o}dlseder}, {Kuss}, {Lande}, {Larsson}, {Latronico}, {Lemoine-Goumard},
  {Longo}, {Loparco}, {Lovellette}, {Lubrano}, {Madejski}, {Massaro}, {Mayer},
  {Mazziotta}, {McEnery}, {Mehault}, {Michelson}, {Mignani}, {Mitthumsiri},
  {Mizuno}, {Moiseev}, {Monzani}, {Morselli}, {Moskalenko}, {Murgia},
  {Nakamori}, {Nemmen}, {Nuss}, {Ohno}, {Ohsugi}, {Omodei}, {Orienti},
  {Orlando}, {Ormes}, {Paneque}, {Perkins}, {Pesce-Rollins}, {Piron}, {Pivato},
  {Rain{\`o}}, {Rando}, {Razzano}, {Razzaque}, {Reimer}, {Reimer}, {Ritz},
  {Romoli}, {S{\'a}nchez-Conde}, {Schulz}, {Sgr{\`o}}, {Simeon}, {Siskind},
  {Smith}, {Spandre}, {Spinelli}, {Stecker}, {Strong}, {Suson}, {Tajima},
  {Takahashi}, {Takahashi}, {Tanaka}, {Thayer}, {Thayer}, {Thompson},
  {Thorsett}, {Tibaldo}, {Tibolla}, {Tinivella}, {Troja}, {Uchiyama}, {Usher},
  {Vandenbroucke}, {Vasileiou}, {Vianello}, {Vitale}, {Waite}, {Werner},
  {Winer}, {Wood}, {Wood}, {Yamazaki}, {Yang}, \&
  {Zimmer}}]{2013Sci...339..807A}
{Ackermann}, M., {Ajello}, M., {Allafort}, A., {et~al.} 2013, Science, 339,
  807, \dodoi{10.1126/science.1231160}

\bibitem[{{Aharonian} {et~al.}(2006){Aharonian}, {Akhperjanian}, {Bazer-Bachi},
  {Beilicke}, {Benbow}, {Berge}, {Bernl{\"o}hr}, {Boisson}, {Bolz}, {Borrel},
  {Braun}, {Breitling}, {Brown}, {B{\"u}hler}, {B{\"u}sching}, {Carrigan},
  {Chadwick}, {Chounet}, {Cornils}, {Costamante}, {Degrange}, {Dickinson},
  {Djannati-Ata{\"\i}}, {O'C. Drury}, {Dubus}, {Egberts}, {Emmanoulopoulos},
  {Espigat}, {Feinstein}, {Ferrero}, {Fiasson}, {Fontaine}, {Funk}, {Funk},
  {Gallant}, {Giebels}, {Glicenstein}, {Goret}, {Hadjichristidis}, {Hauser},
  {Hauser}, {Heinzelmann}, {Henri}, {Hermann}, {Hinton}, {Hofmann}, {Holleran},
  {Horns}, {Jacholkowska}, {de Jager}, {Kh{\'e}lifi}, {Komin}, {Konopelko},
  {Kosack}, {Latham}, {Le Gallou}, {Lemi{\`e}re}, {Lemoine-Goumard}, {Lohse},
  {Martin}, {Martineau-Huynh}, {Marcowith}, {Masterson}, {McComb}, {de
  Naurois}, {Nedbal}, {Nolan}, {Noutsos}, {Orford}, {Osborne}, {Ouchrif},
  {Panter}, {Pelletier}, {Pita}, {P{\"u}hlhofer}, {Punch}, {Raubenheimer},
  {Raue}, {Rayner}, {Reimer}, {Reimer}, {Ripken}, {Rob}, {Rolland}, {Rowell},
  {Sahakian}, {Saug{\'e}}, {Schlenker}, {Schlickeiser}, {Schwanke}, {Sol},
  {Spangler}, {Spanier}, {Steenkamp}, {Stegmann}, {Superina}, {Tavernet},
  {Terrier}, {Th{\'e}oret}, {Tluczykont}, {van Eldik}, {Vasileiadis}, {Venter},
  {Vincent}, {V{\"o}lk}, {Wagner}, \& {Ward}}]{2006A&A...457..899A}
{Aharonian}, F., {Akhperjanian}, A.~G., {Bazer-Bachi}, A.~R., {et~al.} 2006,
  \aap, 457, 899, \dodoi{10.1051/0004-6361:20065351}

\bibitem[{{Aharonian} {et~al.}(2021{\natexlab{a}}){Aharonian}, {An}, {Axikegu},
  {Bai}, {Bai}, {Bao}, {Bastieri}, {Bi}, {Bi}, {Cai}, {Cai}, {Cao}, {Cao},
  {Chang}, {Chang}, {Chang}, {Chen}, {Chen}, {Chen}, {Chen}, {Chen}, {Chen},
  {Chen}, {Chen}, {Chen}, {Chen}, {Chen}, {Chen}, {Chen}, {Cheng}, {Cheng},
  {Cui}, {Cui}, {Cui}, {Dai}, {Dai}, {Dai}, {Danzengluobu}, {Della Volpe},
  {Piazzoli}, {Dong}, {Fan}, {Fan}, {Fan}, {Fang}, {Fang}, {Feng}, {Feng},
  {Feng}, {Feng}, {Gao}, {Gao}, {Gao}, {Gao}, {Ge}, {Geng}, {Gong}, {Gou},
  {Gu}, {Guo}, {Guo}, {Guo}, {Guo}, {Han}, {He}, {He}, {He}, {He}, {He}, {He},
  {Heller}, {Hor}, {Hou}, {Hou}, {Hu}, {Hu}, {Hu}, {Hu}, {Huang}, {Huang},
  {Huang}, {Huang}, {Huang}, {Ji}, {Ji}, {Jia}, {Jiang}, {Jiang}, {Jin},
  {Kuleshov}, {Levochkin}, {Li}, {Li}, {Li}, {Li}, {Li}, {Li}, {Li}, {Li},
  {Li}, {Li}, {Li}, {Li}, {Li}, {Li}, {Li}, {Li}, {Li}, {Liang}, {Liang},
  {Lin}, {Liu}, {Liu}, {Liu}, {Liu}, {Liu}, {Liu}, {Liu}, {Liu}, {Liu}, {Liu},
  {Liu}, {Liu}, {Liu}, {Liu}, {Liu}, {Long}, {Lu}, {Lv}, {Ma}, {Ma}, {Ma},
  {Mao}, {Masood}, {Mitthumsiri}, {Montaruli}, {Nan}, {Pang},
  {Pattarakijwanich}, {Pei}, {Qi}, {Qiao}, {Ruffolo}, {Rulev}, {S{\'a}iz},
  {Shao}, {Shchegolev}, {Sheng}, {Shi}, {Song}, {Stenkin}, {Stepanov}, {Sun},
  {Sun}, {Sun}, {Tam}, {Tang}, {Tian}, {Wang}, {Wang}, {Wang}, {Wang}, {Wang},
  {Wang}, {Wang}, {Wang}, {Wang}, {Wang}, {Wang}, {Wang}, {Wang}, {Wang},
  {Wang}, {Wang}, {Wang}, {Wang}, {Wang}, {Wang}, {Wang}, {Wei}, {Wei}, {Wei},
  {Wen}, {Wu}, {Wu}, {Wu}, {Wu}, {Wu}, {Xi}, {Xia}, {Xia}, {Xiang}, {Xiao},
  {Xiao}, {Xin}, {Xin}, {Xing}, {Xu}, {Xu}, {Xue}, {Yan}, {Yang}, {Yang},
  {Yang}, {Yang}, {Yang}, {Yang}, {Yang}, {Yao}, {Yao}, {Ye}, {Yin}, {Yin},
  {You}, {You}, {Yu}, {Yuan}, {Zeng}, {Zeng}, {Zeng}, {Zeng}, {Zha}, {Zhai},
  {Zhang}, {Zhang}, {Zhang}, {Zhang}, {Zhang}, {Zhang}, {Zhang}, {Zhang},
  {Zhang}, {Zhang}, {Zhang}, {Zhang}, {Zhang}, {Zhang}, {Zhang}, {Zhang},
  {Zhang}, {Zhang}, {Zhang}, {Zhao}, {Zhao}, {Zhao}, {Zhao}, {Zhao}, {Zheng},
  {Zheng}, {Zhou}, {Zhou}, {Zhou}, {Zhou}, {Zhou}, {Zhou}, {Zhu}, {Zhu}, {Zhu},
  {Zhu}, {Zuo}, \& {Collaboration)}}]{2021ChPhC..45h5002A}
{Aharonian}, F., {An}, Q., {Axikegu}, {et~al.} 2021{\natexlab{a}}, Chinese
  Physics C, 45, 085002, \dodoi{10.1088/1674-1137/ac041b}

\bibitem[{{Aharonian} {et~al.}(2021{\natexlab{b}}){Aharonian}, {An}, {Axikegu},
  {Bai}, {Bai}, {Bao}, {Bastieri}, {Bi}, {Bi}, {Cai}, {Cai}, {Cao}, {Cao},
  {Chang}, {Chang}, {Chang}, {Chen}, {Chen}, {Chen}, {Chen}, {Chen}, {Chen},
  {Chen}, {Chen}, {Chen}, {Chen}, {Chen}, {Chen}, {Chen}, {Cheng}, {Cheng},
  {Cui}, {Cui}, {Cui}, {Dai}, {Dai}, {Dai}, {Danzengluobu}, {Della Volpe},
  {Piazzoli}, {Dong}, {Fan}, {Fan}, {Fan}, {Fang}, {Fang}, {Feng}, {Feng},
  {Feng}, {Feng}, {Gao}, {Gao}, {Gao}, {Gao}, {Ge}, {Geng}, {Gong}, {Gou},
  {Gu}, {Guo}, {Guo}, {Guo}, {Guo}, {Han}, {He}, {He}, {He}, {He}, {He}, {He},
  {Heller}, {Hor}, {Hou}, {Hou}, {Hu}, {Hu}, {Hu}, {Hu}, {Huang}, {Huang},
  {Huang}, {Huang}, {Huang}, {Ji}, {Ji}, {Jia}, {Jiang}, {Jiang}, {Jin},
  {Kuleshov}, {Levochkin}, {Li}, {Li}, {Li}, {Li}, {Li}, {Li}, {Li}, {Li},
  {Li}, {Li}, {Li}, {Li}, {Li}, {Li}, {Li}, {Li}, {Li}, {Liang}, {Liang},
  {Lin}, {Liu}, {Liu}, {Liu}, {Liu}, {Liu}, {Liu}, {Liu}, {Liu}, {Liu}, {Liu},
  {Liu}, {Liu}, {Liu}, {Liu}, {Liu}, {Long}, {Lu}, {Lv}, {Ma}, {Ma}, {Ma},
  {Mao}, {Masood}, {Mitthumsiri}, {Montaruli}, {Nan}, {Pang},
  {Pattarakijwanich}, {Pei}, {Qi}, {Ruffolo}, {Rulev}, {S{\'a}iz}, {Shao},
  {Shchegolev}, {Sheng}, {Shi}, {Song}, {Stenkin}, {Stepanov}, {Sun}, {Sun},
  {Sun}, {Tam}, {Tang}, {Tian}, {Wang}, {Wang}, {Wang}, {Wang}, {Wang}, {Wang},
  {Wang}, {Wang}, {Wang}, {Wang}, {Wang}, {Wang}, {Wang}, {Wang}, {Wang},
  {Wang}, {Wang}, {Wang}, {Wang}, {Wang}, {Wang}, {Wei}, {Wei}, {Wei}, {Wen},
  {Wu}, {Wu}, {Wu}, {Wu}, {Wu}, {Xi}, {Xia}, {Xia}, {Xiang}, {Xiao}, {Xiao},
  {Xin}, {Xin}, {Xing}, {Xu}, {Xu}, {Xue}, {Yan}, {Yang}, {Yang}, {Yang},
  {Yang}, {Yang}, {Yang}, {Yang}, {Yao}, {Yao}, {Ye}, {Yin}, {Yin}, {You},
  {You}, {Yu}, {Yuan}, {Zeng}, {Zeng}, {Zeng}, {Zeng}, {Zha}, {Zhai}, {Zhang},
  {Zhang}, {Zhang}, {Zhang}, {Zhang}, {Zhang}, {Zhang}, {Zhang}, {Zhang},
  {Zhang}, {Zhang}, {Zhang}, {Zhang}, {Zhang}, {Zhang}, {Zhang}, {Zhang},
  {Zhang}, {Zhang}, {Zhao}, {Zhao}, {Zhao}, {Zhao}, {Zhao}, {Zheng}, {Zheng},
  {Zhou}, {Zhou}, {Zhou}, {Zhou}, {Zhou}, {Zhou}, {Zhu}, {Zhu}, {Zhu}, {Zhu},
  {Zuo}, \& {(LHAASO Collaboration)}}]{2021ChPhC..45b5002A}
---. 2021{\natexlab{b}}, Chinese Physics C, 45, 025002,
  \dodoi{10.1088/1674-1137/abd01b}

\bibitem[{{Aharonian} {et~al.}(2021{\natexlab{c}}){Aharonian}, {An}, {Axikegu},
  {Bai}, {Bao}, {Bastieri}, {Bi}, {Bi}, {Cai}, {Cai}, {Cao}, {Cao}, {Chang},
  {Chang}, {Chang}, {Chen}, {Chen}, {Chen}, {Chen}, {Chen}, {Chen}, {Chen},
  {Chen}, {Chen}, {Chen}, {Chen}, {Chen}, {Chen}, {Cheng}, {Cheng}, {Cui},
  {Cui}, {Cui}, {Dai}, {Dai}, {Dai}, {Danzengluobu}, {Della Volpe}, {D'Ettorre
  Piazzoli}, {Dong}, {Fan}, {Fan}, {Fan}, {Fang}, {Fang}, {Feng}, {Feng},
  {Feng}, {Feng}, {Gao}, {Gao}, {Gao}, {Gao}, {Ge}, {Geng}, {Gong}, {Gou},
  {Gu}, {Guo}, {Guo}, {Guo}, {Guo}, {Han}, {He}, {He}, {He}, {He}, {He}, {He},
  {Heller}, {Hor}, {Hou}, {Hou}, {Hu}, {Hu}, {Hu}, {Hu}, {Huang}, {Huang},
  {Huang}, {Huang}, {Huang}, {Ji}, {Ji}, {Jia}, {Jiang}, {Jiang}, {Jin},
  {Kuleshov}, {Levochkin}, {Li}, {Li}, {Li}, {Li}, {Li}, {Li}, {Li}, {Li},
  {Li}, {Li}, {Li}, {Li}, {Li}, {Li}, {Li}, {Li}, {Li}, {Liang}, {Liang},
  {Lin}, {Liu}, {Liu}, {Liu}, {Liu}, {Liu}, {Liu}, {Liu}, {Liu}, {Liu}, {Liu},
  {Liu}, {Liu}, {Liu}, {Liu}, {Liu}, {Long}, {Lu}, {Lv}, {Ma}, {Ma}, {Ma},
  {Mao}, {Masood}, {Mitthumsiri}, {Montaruli}, {Nan}, {Pang},
  {Pattarakijwanich}, {Pei}, {Qi}, {Ruffolo}, {Rulev}, {S{\'a}iz}, {Shao},
  {Shchegolev}, {Sheng}, {Shi}, {Song}, {Stenkin}, {Stepanov}, {Sun}, {Sun},
  {Sun}, {Tam}, {Tang}, {Tian}, {Wang}, {Wang}, {Wang}, {Wang}, {Wang}, {Wang},
  {Wang}, {Wang}, {Wang}, {Wang}, {Wang}, {Wang}, {Wang}, {Wang}, {Wang},
  {Wang}, {Wang}, {Wang}, {Wang}, {Wang}, {Wang}, {Wei}, {Wei}, {Wei}, {Wen},
  {Wu}, {Wu}, {Wu}, {Wu}, {Wu}, {Xi}, {Xia}, {Xia}, {Xiang}, {Xiao}, {Xiao},
  {Xin}, {Xin}, {Xing}, {Xu}, {Xu}, {Xue}, {Yan}, {Yang}, {Yang}, {Yang},
  {Yang}, {Yang}, {Yang}, {Yang}, {Yao}, {Yao}, {Ye}, {Yin}, {Yin}, {You},
  {You}, {Yu}, {Yuan}, {Zeng}, {Zeng}, {Zeng}, {Zeng}, {Zha}, {Zhai}, {Zhang},
  {Zhang}, {Zhang}, {Zhang}, {Zhang}, {Zhang}, {Zhang}, {Zhang}, {Zhang},
  {Zhang}, {Zhang}, {Zhang}, {Zhang}, {Zhang}, {Zhang}, {Zhang}, {Zhang},
  {Zhang}, {Zhang}, {Zhao}, {Zhao}, {Zhao}, {Zhao}, {Zhao}, {Zheng}, {Zheng},
  {Zhou}, {Zhou}, {Zhou}, {Zhou}, {Zhou}, {Zhou}, {Zhu}, {Zhu}, {Zhu}, {Zhu},
  {Zuo}, {LHAASO Collaboration}, \& {Huang}}]{2021PhRvL.126x1103A}
{Aharonian}, F., {An}, Q., {Axikegu}, Bai, L.~X., {et~al.} 2021{\natexlab{c}},
  \prl, 126, 241103, \dodoi{10.1103/PhysRevLett.126.241103}



\bibitem[{{Aharonian} {et~al.}(2022){Aharonian}, {An}, {Axikegu}, {Bai}, {Bao},
  {Bastieri}, {Bi}, {Bi}, {Cai}, {Cao}, {Cao}, {Chang}, {Chang}, {Chen},
  {Chen}, {Chen}, {Chen}, {Chen}, {Chen}, {Chen}, {Chen}, {Chen}, {Chen},
  {Chen}, {Cheng}, {Cheng}, {Cheng}, {Cui}, {Cui}, {Cui}, {D'Ettorre Piazzoli},
  {Dai}, {Dai}, {Dai}, {Danzengluobu}, {Duan}, {Fan}, {Fan}, {Fan}, {Fang},
  {Fang}, {Feng}, {Feng}, {Feng}, {Feng}, {Feng}, {Gao}, {Gao}, {Gao}, {Gao},
  {Gao}, {Gao}, {Ge}, {Geng}, {Gong}, {Gou}, {Gu}, {Guo}, {Guo}, {Guo}, {Guo},
  {Guo}, {Han}, {He}, {He}, {He}, {He}, {He}, {Heller}, {Hor}, {Hou}, {Hou},
  {Hu}, {Hu}, {Hu}, {Hu}, {Hu}, {Huang}, {Huang}, {Huang}, {Huang}, {Huang},
  {Huang}, {Ji}, {Jia}, {Jia}, {Jiang}, {Jiang}, {Jin}, {Kang}, {Ke},
  {Kuleshov}, {Levochkin}, {Li}, {Li}, {Li}, {Li}, {Li}, {Li}, {Li}, {Li},
  {Li}, {Li}, {Li}, {Li}, {Li}, {Li}, {Li}, {Li}, {Li}, {Li}, {Liang}, {Liang},
  {Lin}, {Liu}, {Liu}, {Liu}, {Liu}, {Liu}, {Liu}, {Liu}, {Liu}, {Liu}, {Liu},
  {Liu}, {Liu}, {Liu}, {Liu}, {Liu}, {Long}, {Lu}, {Luo}, {Lv}, {Ma}, {Ma},
  {Ma}, {Mao}, {Masood}, {Min}, {Mitthumsiri}, {Nan}, {Ou}, {Pang},
  {Pattarakijwanich}, {Pei}, {Qi}, {Qi}, {Qiao}, {Qin}, {Ruffolo}, {S{\'a}iz},
  {Shao}, {Shao}, {Shchegolev}, {Sheng}, {Shi}, {Song}, {Stenkin}, {Stepanov},
  {Su}, {Sun}, {Sun}, {Sun}, {Tam}, {Tang}, {Tian}, {Wang}, {Wang}, {Wang},
  {Wang}, {Wang}, {Wang}, {Wang}, {Wang}, {Wang}, {Wang}, {Wang}, {Wang},
  {Wang}, {Wang}, {Wang}, {Wang}, {Wang}, {Wang}, {Wang}, {Wang}, {Wang},
  {Wei}, {Wei}, {Wei}, {Wen}, {Wu}, {Wu}, {Wu}, {Wu}, {Wu}, {Xi}, {Xia}, {Xia},
  {Xiang}, {Xiao}, {Xiao}, {Xin}, {Xin}, {Xing}, {Xiong}, {Xu}, {Xu}, {Xue},
  {Yan}, {Yan}, {Yang}, {Yang}, {Yang}, {Yang}, {Yang}, {Yang}, {Yang}, {Yang},
  {Yao}, {Yao}, {Ye}, {Yin}, {Yin}, {You}, {You}, {Yu}, {Yuan}, {Yue}, {Zeng},
  {Zeng}, {Zeng}, {Zeng}, {Zha}, {Zhai}, {Zhang}, {Zhang}, {Zhang}, {Zhang},
  {Zhang}, {Zhang}, {Zhang}, {Zhang}, {Zhang}, {Zhang}, {Zhang}, {Zhang},
  {Zhang}, {Zhang}, {Zhang}, {Zhang}, {Zhang}, {Zhang}, {Zhang}, {Zhang},
  {Zhao}, {Zhao}, {Zhao}, {Zhao}, {Zhao}, {Zheng}, {Zheng}, {Zhou}, {Zhou},
  {Zhou}, {Zhou}, {Zhou}, {Zhou}, {Zhu}, {Zhu}, {Zhu}, {Zhu}, {Zuo}, \& {LHAASO
  Collaboration}}]{2022PhRvD.106l2004A}
---. 2022, \prd, 106, 122004, \dodoi{10.1103/PhysRevD.106.122004}

\bibitem[{{Ajello} {et~al.}(2017){Ajello}, {Atwood}, {Baldini}, {Ballet},
  {Barbiellini}, {Bastieri}, {Bellazzini}, {Bissaldi}, {Blandford}, {Bloom},
  {Bonino}, {Bregeon}, {Britto}, {Bruel}, {Buehler}, {Buson}, {Cameron},
  {Caputo}, {Caragiulo}, {Caraveo}, {Cavazzuti}, {Cecchi}, {Charles},
  {Chekhtman}, {Cheung}, {Chiaro}, {Ciprini}, {Cohen}, {Costantin}, {Costanza},
  {Cuoco}, {Cutini}, {D'Ammando}, {de Palma}, {Desiante}, {Digel}, {Di Lalla},
  {Di Mauro}, {Di Venere}, {Dom{\'\i}nguez}, {Drell}, {Dumora}, {Favuzzi},
  {Fegan}, {Ferrara}, {Fortin}, {Franckowiak}, {Fukazawa}, {Funk}, {Fusco},
  {Gargano}, {Gasparrini}, {Giglietto}, {Giommi}, {Giordano}, {Giroletti},
  {Glanzman}, {Green}, {Grenier}, {Grondin}, {Grove}, {Guillemot}, {Guiriec},
  {Harding}, {Hays}, {Hewitt}, {Horan}, {J{\'o}hannesson}, {Kensei}, {Kuss},
  {La Mura}, {Larsson}, {Latronico}, {Lemoine-Goumard}, {Li}, {Longo},
  {Loparco}, {Lott}, {Lubrano}, {Magill}, {Maldera}, {Manfreda}, {Mazziotta},
  {McEnery}, {Meyer}, {Michelson}, {Mirabal}, {Mitthumsiri}, {Mizuno},
  {Moiseev}, {Monzani}, {Morselli}, {Moskalenko}, {Negro}, {Nuss}, {Ohsugi},
  {Omodei}, {Orienti}, {Orlando}, {Palatiello}, {Paliya}, {Paneque}, {Perkins},
  {Persic}, {Pesce-Rollins}, {Piron}, {Porter}, {Principe}, {Rain{\`o}},
  {Rando}, {Razzano}, {Razzaque}, {Reimer}, {Reimer}, {Reposeur}, {Saz
  Parkinson}, {Sgr{\`o}}, {Simone}, {Siskind}, {Spada}, {Spandre}, {Spinelli},
  {Stawarz}, {Suson}, {Takahashi}, {Tak}, {Thayer}, {Thayer}, {Thompson},
  {Torres}, {Torresi}, {Troja}, {Vianello}, {Wood}, \&
  {Wood}}]{2017ApJS..232...18A}
{Ajello}, M., {Atwood}, W.~B., {Baldini}, L., {et~al.} 2017, \apjs, 232, 18,
  \dodoi{10.3847/1538-4365/aa8221}

\bibitem[{{Albert} {et~al.}(2020){Albert}, {Alfaro}, {Alvarez}, {Camacho},
  {Arteaga-Vel{\'a}zquez}, {Arunbabu}, {Avila Rojas}, {Ayala Solares},
  {Baghmanyan}, {Belmont-Moreno}, {BenZvi}, {Brisbois}, {Caballero-Mora},
  {Capistr{\'a}n}, {Carrami{\~n}ana}, {Casanova}, {Cotti}, {Couti{\~n}o de
  Le{\'o}n}, {De la Fuente}, {Diaz Hernandez}, {Diaz-Cruz}, {Dingus},
  {DuVernois}, {Durocher}, {D{\'\i}az-V{\'e}lez}, {Ellsworth}, {Engel},
  {Espinoza}, {Fan}, {Fang}, {Alonso}, {Fleischhack}, {Fraija},
  {Galv{\'a}n-G{\'a}mez}, {Garcia}, {Garc{\'\i}a-Gonz{\'a}lez}, {Garfias},
  {Giacinti}, {Gonz{\'a}lez}, {Goodman}, {Harding}, {Hernandez}, {Hinton},
  {Hona}, {Huang}, {Hueyotl-Zahuantitla}, {H{\"u}ntemeyer}, {Iriarte},
  {Jardin-Blicq}, {Joshi}, {Kieda}, {Lara}, {Lee}, {Le{\'o}n Vargas},
  {Linnemann}, {Longinotti}, {Luis-Raya}, {Lundeen}, {L{\'o}pez-Coto},
  {Malone}, {Marandon}, {Martinez}, {Martinez-Castellanos},
  {Mart{\'\i}nez-Castro}, {Matthews}, {Miranda-Romagnoli}, {Morales-Soto},
  {Moreno}, {Mostaf{\'a}}, {Nayerhoda}, {Nellen}, {Newbold}, {Nisa},
  {Noriega-Papaqui}, {Olivera-Nieto}, {Omodei}, {Peisker}, {P{\'e}rez Araujo},
  {P{\'e}rez-P{\'e}rez}, {Ren}, {Rho}, {Rivi{\`e}re}, {Rosa-Gonz{\'a}lez},
  {Ruiz-Velasco}, {Salazar}, {Salesa Greus}, {Sandoval}, {Schneider},
  {Schoorlemmer}, {Serna}, {Sinnis}, {Smith}, {Springer}, {Surajbali},
  {Tollefson}, {Torres}, {Torres-Escobedo}, {Ukwatta}, {Ure{\~n}a-Mena},
  {Weisgarber}, {Werner}, {Willox}, {Zepeda}, {Zhou}, {de Le{\'o}n},
  {{\'A}lvarez}, \& {HAWC Collaboration}}]{2020ApJ...905...76A}
{Albert}, A., {Alfaro}, R., {Alvarez}, C., {et~al.} 2020, \apj, 905, 76,
  \dodoi{10.3847/1538-4357/abc2d8}

\bibitem[{{Aleksi{\'c}} {et~al.}(2016){Aleksi{\'c}}, {Ansoldi}, {Antonelli},
  {Antoranz}, {Babic}, {Bangale}, {Barcel{\'o}}, {Barrio}, {Becerra
  Gonz{\'a}lez}, {Bednarek}, {Bernardini}, {Biasuzzi}, {Biland}, {Bitossi},
  {Blanch}, {Bonnefoy}, {Bonnoli}, {Borracci}, {Bretz}, {Carmona}, {Carosi},
  {Cecchi}, {Colin}, {Colombo}, {Contreras}, {Corti}, {Cortina}, {Covino}, {Da
  Vela}, {Dazzi}, {De Angelis}, {De Caneva}, {De Lotto}, {de O{\~n}a Wilhelmi},
  {Delgado Mendez}, {Dettlaff}, {Dominis Prester}, {Dorner}, {Doro}, {Einecke},
  {Eisenacher}, {Elsaesser}, {Fidalgo}, {Fink}, {Fonseca}, {Font}, {Frantzen},
  {Fruck}, {Galindo}, {Garc{\'\i}a L{\'o}pez}, {Garczarczyk}, {Garrido
  Terrats}, {Gaug}, {Giavitto}, {Godinovi{\'c}}, {Gonz{\'a}lez Mu{\~n}oz},
  {Gozzini}, {Haberer}, {Hadasch}, {Hanabata}, {Hayashida}, {Herrera},
  {Hildebrand}, {Hose}, {Hrupec}, {Idec}, {Illa}, {Kadenius}, {Kellermann},
  {Knoetig}, {Kodani}, {Konno}, {Krause}, {Kubo}, {Kushida}, {La Barbera},
  {Lelas}, {Lemus}, {Lewandowska}, {Lindfors}, {Lombardi}, {Longo},
  {L{\'o}pez}, {L{\'o}pez-Coto}, {L{\'o}pez-Oramas}, {Lorca}, {Lorenz},
  {Lozano}, {Makariev}, {Mallot}, {Maneva}, {Mankuzhiyil}, {Mannheim},
  {Maraschi}, {Marcote}, {Mariotti}, {Mart{\'\i}nez}, {Mazin}, {Menzel},
  {Miranda}, {Mirzoyan}, {Moralejo}, {Munar-Adrover}, {Nakajima}, {Negrello},
  {Neustroev}, {Niedzwiecki}, {Nilsson}, {Nishijima}, {Noda}, {Orito},
  {Overkemping}, {Paiano}, {Palatiello}, {Paneque}, {Paoletti}, {Paredes},
  {Paredes-Fortuny}, {Persic}, {Poutanen}, {Prada Moroni}, {Prandini},
  {Puljak}, {Reinthal}, {Rhode}, {Rib{\'o}}, {Rico}, {Rodriguez Garcia},
  {R{\"u}gamer}, {Saito}, {Saito}, {Satalecka}, {Scalzotto}, {Scapin},
  {Schultz}, {Schlammer}, {Schmidl}, {Schweizer}, {Shore}, {Sillanp{\"a}{\"a}},
  {Sitarek}, {Snidaric}, {Sobczynska}, {Spanier}, {Stamerra}, {Steinbring},
  {Storz}, {Strzys}, {Takalo}, {Takami}, {Tavecchio}, {Tejedor}, {Temnikov},
  {Terzi{\'c}}, {Tescaro}, {Teshima}, {Thaele}, {Tibolla}, {Torres}, {Toyama},
  {Treves}, {Vogler}, {Wetteskind}, {Will}, \& {Zanin}}]{2016APh....72...76A}
{Aleksi{\'c}}, J., {Ansoldi}, S., {Antonelli}, L.~A., {et~al.} 2016,
  Astroparticle Physics, 72, 76, \dodoi{10.1016/j.astropartphys.2015.02.005}

\bibitem[{{Amenomori} {et~al.}(2005){Amenomori}, {Ayabe}, {Chen}, {Cui},
  {Danzengluobu}, {Ding}, {Ding}, {Feng}, {Feng}, {Gao}, {Geng}, {Guo}, {He},
  {He}, {Hibino}, {Hotta}, {Hu}, {Hu}, {Huang}, {Huang}, {Jia}, {Kajino},
  {Kasahara}, {Katayose}, {Kato}, {Kawata}, {Labaciren}, {Le}, {Li}, {Lu},
  {Lu}, {Meng}, {Mizutani}, {Mu}, {Munakata}, {Nagai}, {Nanjo}, {Nishizawa},
  {Ohnishi}, {Ohta}, {Onuma}, {Ouchi}, {Ozawa}, {Ren}, {Saito}, {Sakata},
  {Sasaki}, {Shibata}, {Shiomi}, {Shirai}, {Sugimoto}, {Takashima}, {Takita},
  {Tan}, {Tateyama}, {Torii}, {Tsuchiya}, {Udo}, {Utsugi}, {Wang}, {Wang},
  {Wang}, {Wu}, {Xue}, {Yamamoto}, {Yan}, {Yang}, {Yasue}, {Ye}, {Yu}, {Yuan},
  {Yuda}, {Zhang}, {Zhang}, {Zhang}, {Zhang}, {Zhang}, {Zhang}, {Zhaxisangzhu},
  {Zhou}, \& {Tibet As{\ensuremath{\gamma}}
  Collaboration}}]{2005ApJ...633.1005A}
{Amenomori}, M., {Ayabe}, S., {Chen}, D., {et~al.} 2005, \apj, 633, 1005,
  \dodoi{10.1086/491612}

\bibitem[{{Amenomori} {et~al.}(2019){Amenomori}, {Bao}, {Bi}, {Chen}, {Chen},
  {Chen}, {Chen}, {Chen}, {Cirennima}, {Cui}, {Danzengluobu}, {Ding}, {Fang},
  {Fang}, {Feng}, {Feng}, {Feng}, {Gao}, {Gou}, {Guo}, {He}, {He}, {Hibino},
  {Hotta}, {Hu}, {Hu}, {Huang}, {Jia}, {Jiang}, {Jin}, {Kajino}, {Kasahara},
  {Katayose}, {Kato}, {Kato}, {Kawata}, {Kozai}, {Labaciren}, {Le}, {Li}, {Li},
  {Li}, {Lin}, {Liu}, {Liu}, {Liu}, {Liu}, {Lou}, {Lu}, {Meng}, {Mitsui},
  {Munakata}, {Nakamura}, {Nanjo}, {Nishizawa}, {Ohnishi}, {Ohta}, {Ozawa},
  {Qian}, {Qu}, {Saito}, {Sakata}, {Sako}, {Sengoku}, {Shao}, {Shibata},
  {Shiomi}, {Sugimoto}, {Takita}, {Tan}, {Tateyama}, {Torii}, {Tsuchiya},
  {Udo}, {Wang}, {Wu}, {Xue}, {Yagisawa}, {Yamamoto}, {Yang}, {Yuan}, {Zhai},
  {Zhang}, {Zhang}, {Zhang}, {Zhang}, {Zhang}, {Zhang}, {Zhang},
  {Zhaxisangzhu}, {Zhou}, \& {Tibet AS {\ensuremath{\gamma}}
  Collaboration}}]{2019PhRvL.123e1101A}

\bibitem[Amenomori et al.(2021)]{2021PhRvL.126n1101A} Amenomori, M., Bao, Y.~W., Bi, X.~J., et al.\ 2021, \prl, 126, 141101. doi:10.1103/PhysRevLett.126.141101

{Amenomori}, M., {Bao}, Y.~W., {Bi}, X.~J., {et~al.} 2019, \prl, 123, 051101,
  \dodoi{10.1103/PhysRevLett.123.051101}

\bibitem[{{Atwood} {et~al.}(2009){Atwood}, {Abdo}, {Ackermann}, {Althouse},
  {Anderson}, {Axelsson}, {Baldini}, {Ballet}, {Band}, {Barbiellini},
  {Bartelt}, {Bastieri}, {Baughman}, {Bechtol}, {B{\'e}d{\'e}r{\`e}de},
  {Bellardi}, {Bellazzini}, {Berenji}, {Bignami}, {Bisello}, {Bissaldi},
  {Blandford}, {Bloom}, {Bogart}, {Bonamente}, {Bonnell}, {Borgland},
  {Bouvier}, {Bregeon}, {Brez}, {Brigida}, {Bruel}, {Burnett}, {Busetto},
  {Caliandro}, {Cameron}, {Caraveo}, {Carius}, {Carlson}, {Casandjian},
  {Cavazzuti}, {Ceccanti}, {Cecchi}, {Charles}, {Chekhtman}, {Cheung},
  {Chiang}, {Chipaux}, {Cillis}, {Ciprini}, {Claus}, {Cohen-Tanugi},
  {Condamoor}, {Conrad}, {Corbet}, {Corucci}, {Costamante}, {Cutini}, {Davis},
  {Decotigny}, {DeKlotz}, {Dermer}, {de Angelis}, {Digel}, {do Couto e Silva},
  {Drell}, {Dubois}, {Dumora}, {Edmonds}, {Fabiani}, {Farnier}, {Favuzzi},
  {Flath}, {Fleury}, {Focke}, {Funk}, {Fusco}, {Gargano}, {Gasparrini},
  {Gehrels}, {Gentit}, {Germani}, {Giebels}, {Giglietto}, {Giommi}, {Giordano},
  {Glanzman}, {Godfrey}, {Grenier}, {Grondin}, {Grove}, {Guillemot}, {Guiriec},
  {Haller}, {Harding}, {Hart}, {Hays}, {Healey}, {Hirayama}, {Hjalmarsdotter},
  {Horn}, {Hughes}, {J{\'o}hannesson}, {Johansson}, {Johnson}, {Johnson},
  {Johnson}, {Johnson}, {Kamae}, {Katagiri}, {Kataoka}, {Kavelaars}, {Kawai},
  {Kelly}, {Kerr}, {Klamra}, {Kn{\"o}dlseder}, {Kocian}, {Komin}, {Kuehn},
  {Kuss}, {Landriu}, {Latronico}, {Lee}, {Lee}, {Lemoine-Goumard}, {Lionetto},
  {Longo}, {Loparco}, {Lott}, {Lovellette}, {Lubrano}, {Madejski}, {Makeev},
  {Marangelli}, {Massai}, {Mazziotta}, {McEnery}, {Menon}, {Meurer},
  {Michelson}, {Minuti}, {Mirizzi}, {Mitthumsiri}, {Mizuno}, {Moiseev},
  {Monte}, {Monzani}, {Moretti}, {Morselli}, {Moskalenko}, {Murgia},
  {Nakamori}, {Nishino}, {Nolan}, {Norris}, {Nuss}, {Ohno}, {Ohsugi}, {Omodei},
  {Orlando}, {Ormes}, {Paccagnella}, {Paneque}, {Panetta}, {Parent}, {Pearce},
  {Pepe}, {Perazzo}, {Pesce-Rollins}, {Picozza}, {Pieri}, {Pinchera}, {Piron},
  {Porter}, {Poupard}, {Rain{\`o}}, {Rando}, {Rapposelli}, {Razzano}, {Reimer},
  {Reimer}, {Reposeur}, {Reyes}, {Ritz}, {Rochester}, {Rodriguez}, {Romani},
  {Roth}, {Russell}, {Ryde}, {Sabatini}, {Sadrozinski}, {Sanchez}, {Sander},
  {Sapozhnikov}, {Parkinson}, {Scargle}, {Schalk}, {Scolieri}, {Sgr{\`o}},
  {Share}, {Shaw}, {Shimokawabe}, {Shrader}, {Sierpowska-Bartosik}, {Siskind},
  {Smith}, {Smith}, {Spandre}, {Spinelli}, {Starck}, {Stephens}, {Strickman},
  {Strong}, {Suson}, {Tajima}, {Takahashi}, {Takahashi}, {Tanaka}, {Tenze},
  {Tether}, {Thayer}, {Thayer}, {Thompson}, {Tibaldo}, {Tibolla}, {Torres},
  {Tosti}, {Tramacere}, {Turri}, {Usher}, {Vilchez}, {Vitale}, {Wang},
  {Watters}, {Winer}, {Wood}, {Ylinen}, \& {Ziegler}}]{2009ApJ...697.1071A}
{Atwood}, W.~B., {Abdo}, A.~A., {Ackermann}, M., {et~al.} 2009, \apj, 697,
  1071, \dodoi{10.1088/0004-637X/697/2/1071}

\bibitem[{{Bartoli} {et~al.}(2012){Bartoli}, {Bernardini}, {Bi}, {Bleve},
  {Bolognino}, {Branchini}, {Budano}, {Calabrese Melcarne}, {Camarri}, {Cao},
  {Cardarelli}, {Catalanotti}, {Cattaneo}, {Chen}, {Chen}, {Chen}, {Creti},
  {Cui}, {Dai}, {D'Al{\'\i} Staiti}, {Danzengluobu}, {Dattoli}, {De Mitri},
  {D'Ettorre Piazzoli}, {Di Girolamo}, {Ding}, {Di Sciascio}, {Feng}, {Feng},
  {Feng}, {Galeazzi}, {Giroletti}, {Gou}, {Guo}, {He}, {Hu}, {Hu}, {Huang},
  {Iacovacci}, {Iuppa}, {James}, {Jia}, {Labaciren}, {Li}, {Li}, {Li},
  {Liguori}, {Liu}, {Liu}, {Liu}, {Liu}, {Lu}, {Ma}, {Ma}, {Mancarella},
  {Mari}, {Marsella}, {Martello}, {Mastroianni}, {Montini}, {Ning}, {Pagliaro},
  {Panareo}, {Panico}, {Perrone}, {Pistilli}, {Ruggieri}, {Salvini},
  {Santonico}, {Shen}, {Sheng}, {Shi}, {Stanescu}, {Surdo}, {Tan}, {Vallania},
  {Vernetto}, {Vigorito}, {Wang}, {Wang}, {Wu}, {Wu}, {Xu}, {Xue}, {Yang},
  {Yang}, {Yao}, {Yuan}, {Zha}, {Zhang}, {Zhang}, {Zhang}, {Zhang}, {Zhang},
  {Zhang}, {Zhang}, {Zhao}, {Zhaxiciren}, {Zhaxisangzhu}, {Zhou}, {Zhu}, {Zhu},
  {Zizzi}, \& {ARGO-YBJ Collaboration}}]{2012ApJ...758....2B}
{Bartoli}, B., {Bernardini}, P., {Bi}, X.~J., {et~al.} 2012, \apj, 758, 2,
  \dodoi{10.1088/0004-637X/758/1/2}

\bibitem[{{Bartoli} {et~al.}(2013){Bartoli}, {Bernardini}, {Bi}, {Bolognino},
  {Branchini}, {Budano}, {Calabrese Melcarne}, {Camarri}, {Cao}, {Cardarelli},
  {Catalanotti}, {Chen}, {Chen}, {Chen}, {Creti}, {Cui}, {Dai}, {D'Amone},
  {Danzengluobu}, {De Mitri}, {D'Ettorre Piazzoli}, {Di Girolamo}, {Ding}, {Di
  Sciascio}, {Feng}, {Feng}, {Feng}, {Gou}, {Guo}, {He}, {Hu}, {Hu}, {Huang},
  {Iacovacci}, {Iuppa}, {Jia}, {Labaciren}, {Li}, {Li}, {Li}, {Liguori}, {Liu},
  {Liu}, {Liu}, {Liu}, {Lu}, {Ma}, {Ma}, {Mancarella}, {Mari}, {Marsella},
  {Martello}, {Mastroianni}, {Montini}, {Ning}, {Panareo}, {Panico}, {Perrone},
  {Pistilli}, {Ruggieri}, {Salvini}, {Santonico}, {Sbano}, {Shen}, {Sheng},
  {Shi}, {Surdo}, {Tan}, {Vallania}, {Vernetto}, {Vigorito}, {Wang}, {Wang},
  {Wu}, {Wu}, {Xu}, {Xue}, {Yang}, {Yang}, {Yao}, {Yuan}, {Zha}, {Zhang},
  {Zhang}, {Zhang}, {Zhang}, {Zhang}, {Zhang}, {Zhang}, {Zhao}, {Zhaxiciren},
  {Zhaxisangzhu}, {Zhou}, {Zhu}, {Zhu}, {Zizzi}, \& {ARGO-YBJ
  Collaboration}}]{2013ApJ...779...27B}
---. 2013, \apj, 779, 27, \dodoi{10.1088/0004-637X/779/1/27}

\bibitem[{{Bartoli} {et~al.}(2014){Bartoli}, {Bernardini}, {Bi}, {Branchini},
  {Budano}, {Camarri}, {Cao}, {Cardarelli}, {Catalanotti}, {Chen}, {Chen},
  {Creti}, {Cui}, {Dai}, {D'Amone}, {Danzengluobu}, {De Mitri}, {D'Ettorre
  Piazzoli}, {Di Girolamo}, {Di Sciascio}, {Feng}, {Feng}, {Feng}, {Gou},
  {Guo}, {He}, {Hu}, {Hu}, {Iacovacci}, {Iuppa}, {Jia}, {Labaciren}, {Li},
  {Liguori}, {Liu}, {Liu}, {Liu}, {Lu}, {Ma}, {Ma}, {Mancarella}, {Mari},
  {Marsella}, {Martello}, {Mastroianni}, {Montini}, {Ning}, {Panareo},
  {Perrone}, {Pistilli}, {Ruggieri}, {Salvini}, {Santonico}, {Shen}, {Sheng},
  {Shi}, {Surdo}, {Tan}, {Vallania}, {Vernetto}, {Vigorito}, {Wang}, {Wu},
  {Wu}, {Xue}, {Yang}, {Yang}, {Yao}, {Yuan}, {Zha}, {Zhang}, {Zhang}, {Zhang},
  {Zhang}, {Zhao}, {Zhaxiciren}, {Zhaxisangzhu}, {Zhou}, {Zhu}, {Zhu}, {Zizzi},
  \& {ARGO-YBJ Collaboration}}]{2014ApJ...790..152B}
---. 2014, \apj, 790, 152, \dodoi{10.1088/0004-637X/790/2/152}

\bibitem[{{Bartoli} {et~al.}(2016){Bartoli}, {Bernardini}, {Bi}, {Cao},
  {Catalanotti}, {Chen}, {Chen}, {Cui}, {Dai}, {D'Amone}, {Danzengluobu}, {De
  Mitri}, {D'Ettorre Piazzoli}, {Di Girolamo}, {Di Sciascio}, {Feng}, {Feng},
  {Feng}, {Gou}, {Guo}, {He}, {Hu}, {Hu}, {Iacovacci}, {Iuppa}, {Jia},
  {Labaciren}, {Li}, {Liu}, {Liu}, {Liu}, {Lu}, {Ma}, {Ma}, {Mancarella},
  {Mari}, {Marsella}, {Mastroianni}, {Montini}, {Ning}, {Perrone}, {Pistilli},
  {Salvini}, {Santonico}, {Shen}, {Sheng}, {Shi}, {Surdo}, {Tan}, {Vallania},
  {Vernetto}, {Vigorito}, {Wang}, {Wu}, {Wu}, {Xue}, {Yang}, {Yang}, {Yao},
  {Yuan}, {Zha}, {Zhang}, {Zhang}, {Zhang}, {Zhang}, {Zhao}, {Zhaxiciren},
  {Zhaxisangzhu}, {Zhou}, {Zhu}, {Zhu}, \& {ARGO-YBJ
  Collaboration}}]{2016ApJS..222....6B}
---. 2016, \apjs, 222, 6, \dodoi{10.3847/0067-0049/222/1/6}

\bibitem[{{Cao} {et~al.}(2021{\natexlab{a}}){Cao}, {Aharonian}, {An},
  {Axikegu}, {Bai}, {Bao}, {Bastieri}, {Bi}, {Bi}, {Cai}, {Cai}, {Cao},
  {Chang}, {Chang}, {Chang}, {Chen}, {Chen}, {Chen}, {Chen}, {Chen}, {Chen},
  {Chen}, {Chen}, {Chen}, {Chen}, {Chen}, {Chen}, {Chen}, {Cheng}, {Cheng},
  {Cui}, {Cui}, {Cui}, {Dai}, {Dai}, {Dai}, {Danzengluobu}, {della Volpe},
  {D'Ettorre Piazzoli}, {Dong}, {Fan}, {Fan}, {Fan}, {Fang}, {Fang}, {Feng},
  {Feng}, {Feng}, {Feng}, {Gao}, {Gao}, {Gao}, {Gao}, {Ge}, {Geng}, {Gong},
  {Gou}, {Gu}, {Guo}, {Guo}, {Guo}, {Guo}, {Han}, {He}, {He}, {He}, {He}, {He},
  {He}, {Heller}, {Hor}, {Hou}, {Hou}, {Hu}, {Hu}, {Hu}, {Hu}, {Huang},
  {Huang}, {Huang}, {Huang}, {Huang}, {Ji}, {Ji}, {Jia}, {Jiang}, {Jiang},
  {Jin}, {Kuleshov}, {Levochkin}, {Li}, {Li}, {Li}, {Li}, {Li}, {Li}, {Li},
  {Li}, {Li}, {Li}, {Li}, {Li}, {Li}, {Li}, {Li}, {Li}, {Li}, {Liang}, {Liang},
  {Lin}, {Liu}, {Liu}, {Liu}, {Liu}, {Liu}, {Liu}, {Liu}, {Liu}, {Liu}, {Liu},
  {Liu}, {Liu}, {Liu}, {Liu}, {Liu}, {Long}, {Lu}, {Lv}, {Ma}, {Ma}, {Ma},
  {Mao}, {Masood}, {Mitthumsiri}, {Montaruli}, {Nan}, {Pang},
  {Pattarakijwanich}, {Pei}, {Qi}, {Ruffolo}, {Rulev}, {S{\'a}iz}, {Shao},
  {Shchegolev}, {Sheng}, {Shi}, {Song}, {Stenkin}, {Stepanov}, {Sun}, {Sun},
  {Sun}, {Tam}, {Tang}, {Tian}, {Wang}, {Wang}, {Wang}, {Wang}, {Wang}, {Wang},
  {Wang}, {Wang}, {Wang}, {Wang}, {Wang}, {Wang}, {Wang}, {Wang}, {Wang},
  {Wang}, {Wang}, {Wang}, {Wang}, {Wang}, {Wang}, {Wei}, {Wei}, {Wei}, {Wen},
  {Wu}, {Wu}, {Wu}, {Wu}, {Wu}, {Xi}, {Xia}, {Xia}, {Xiang}, {Xiao}, {Xiao},
  {Xin}, {Xin}, {Xing}, {Xu}, {Xu}, {Xue}, {Yan}, {Yang}, {Yang}, {Yang},
  {Yang}, {Yang}, {Yang}, {Yang}, {Yao}, {Yao}, {Ye}, {Yin}, {Yin}, {You},
  {You}, {Yu}, {Yuan}, {Zeng}, {Zeng}, {Zeng}, {Zeng}, {Zha}, {Zhai}, {Zhang},
  {Zhang}, {Zhang}, {Zhang}, {Zhang}, {Zhang}, {Zhang}, {Zhang}, {Zhang},
  {Zhang}, {Zhang}, {Zhang}, {Zhang}, {Zhang}, {Zhang}, {Zhang}, {Zhang},
  {Zhang}, {Zhang}, {Zhao}, {Zhao}, {Zhao}, {Zhao}, {Zhao}, {Zheng}, {Zheng},
  {Zhou}, {Zhou}, {Zhou}, {Zhou}, {Zhou}, {Zhou}, {Zhu}, {Zhu}, {Zhu}, {Zhu},
  \& {Zuo}}]{2021Natur.594...33C}
{Cao}, Z., {Aharonian}, F.~A., {An}, Q., {et~al.} 2021{\natexlab{a}}, \nat,
  594, 33, \dodoi{10.1038/s41586-021-03498-z}

\bibitem[{{Cao} {et~al.}(2021{\natexlab{b}}){Cao}, {Aharonian}, {An},
  {Axikegu}, {Bai}, {Bai}, {Bao}, {Bastieri}, {Bi}, {Bi}, {Cai}, {Cai}, {Cao},
  {Chang}, {Chang}, {Chen}, {Chen}, {Chen}, {Chen}, {Chen}, {Chen}, {Chen},
  {Chen}, {Chen}, {Chen}, {Chen}, {Chen}, {Chen}, {Cheng}, {Cheng}, {Cui},
  {Cui}, {Cui}, {Piazzoli}, {Dai}, {Dai}, {Dai}, {Dan-Zeng-Luo-Bu}, {Volpe},
  {Dong}, {Duan}, {Fan}, {Fan}, {Fan}, {Fang}, {Fang}, {Feng}, {Feng}, {Feng},
  {Feng}, {Gao}, {Gao}, {Gao}, {Gao}, {Gao}, {Ge}, {Geng}, {Gong}, {Gou}, {Gu},
  {Guo}, {Guo}, {Guo}, {Guo}, {Guo}, {Han}, {He}, {He}, {He}, {He}, {He}, {He},
  {Heller}, {Hor}, {Hou}, {Hu}, {Hu}, {Hu}, {Hu}, {Huang}, {Huang}, {Huang},
  {Huang}, {Huang}, {Huang}, {Ji}, {Ji}, {Jia}, {Jiang}, {Jiang}, {Jin}, {Ke},
  {Kuleshov}, {Levochkin}, {Li}, {Li}, {Li}, {Li}, {Li}, {Li}, {Li}, {Li},
  {Li}, {Li}, {Li}, {Li}, {Li}, {Li}, {Li}, {Li}, {Li}, {Liang}, {Liang},
  {Lin}, {Liu}, {Liu}, {Liu}, {Liu}, {Liu}, {Liu}, {Liu}, {Liu}, {Liu}, {Liu},
  {Liu}, {Liu}, {Liu}, {Liu}, {Liu}, {Liu}, {Long}, {Lu}, {Lv}, {Ma}, {Ma},
  {Ma}, {Mao}, {Masood}, {Min}, {Mitthumsiri}, {Montaruli}, {Nan}, {Pang},
  {Pattarakijwanich}, {Pei}, {Qi}, {Qi}, {Qiao}, {Qin}, {Ruffolo}, {Rulev},
  {S{\'a}iz}, {Shao}, {Shchegolev}, {Sheng}, {Shi}, {Song}, {Stenkin},
  {Stepanov}, {Su}, {Sun}, {Sun}, {Sun}, {Tam}, {Tang}, {Tian}, {Wang}, {Wang},
  {Wang}, {Wang}, {Wang}, {Wang}, {Wang}, {Wang}, {Wang}, {Wang}, {Wang},
  {Wang}, {Wang}, {Wang}, {Wang}, {Wang}, {Wang}, {Wang}, {Wang}, {Wang},
  {Wang}, {Wang}, {Wei}, {Wei}, {Wei}, {Wen}, {Wu}, {Wu}, {Wu}, {Wu}, {Wu},
  {Xi}, {Xia}, {Xia}, {Xiang}, {Xiao}, {Xiao}, {Xiao}, {Xin}, {Xin}, {Xing},
  {Xu}, {Xu}, {Xue}, {Yan}, {Yan}, {Yang}, {Yang}, {Yang}, {Yang}, {Yang},
  {Yang}, {Yang}, {Yao}, {Yao}, {Ye}, {Yin}, {Yin}, {You}, {You}, {Yu}, {Yuan},
  {Zeng}, {Zeng}, {Zeng}, {Zeng}, {Zha}, {Zhai}, {Zhang}, {Zhang}, {Zhang},
  {Zhang}, {Zhang}, {Zhang}, {Zhang}, {Zhang}, {Zhang}, {Zhang}, {Zhang},
  {Zhang}, {Zhang}, {Zhang}, {Zhang}, {Zhang}, {Zhang}, {Zhang}, {Zhang},
  {Zhao}, {Zhao}, {Zhao}, {Zhao}, {Zhao}, {Zheng}, {Zheng}, {Zhou}, {Zhou},
  {Zhou}, {Zhou}, {Zhou}, {Zhou}, {Zhu}, {Zhu}, {Zhu}, {Zhu}, \&
  {Zuo}}]{2021ApJ...919L..22C}
{Cao}, Z., {Aharonian}, F., {An}, Q., {et~al.} 2021{\natexlab{b}}, \apjl, 919,
  L22, \dodoi{10.3847/2041-8213/ac2579}

\bibitem[{{Cao} {et~al.}(2021{\natexlab{c}}){Cao}, {Aharonian}, {An},
  {Axikegu}, {Bai}, {Bai}, {Bao}, {Bastieri}, {Bi}, {Bi}, {Cai}, {Cai}, {Cao},
  {Chang}, {Chang}, {Chen}, {Chen}, {Chen}, {Chen}, {Chen}, {Chen}, {Chen},
  {Chen}, {Chen}, {Chen}, {Chen}, {Chen}, {Chen}, {Chen}, {Cheng}, {Cheng},
  {Cui}, {Cui}, {Cui}, {D'Ettorre Piazzoli}, {Dai}, {Dai}, {Dai},
  {Danzengluobu}, {Volpe}, {Dong}, {Duan}, {Fan}, {Fan}, {Fan}, {Fang}, {Fang},
  {Feng}, {Feng}, {Feng}, {Feng}, {Gao}, {Gao}, {Gao}, {Gao}, {Gao}, {Ge},
  {Geng}, {Gong}, {Gou}, {Gu}, {Guo}, {Guo}, {Guo}, {Guo}, {Guo}, {Han}, {He},
  {He}, {He}, {He}, {He}, {He}, {Heller}, {Hor}, {Hou}, {Hu}, {Hu}, {Hu}, {Hu},
  {Huang}, {Huang}, {Huang}, {Huang}, {Huang}, {Huang}, {Ji}, {Ji}, {Jia},
  {Jiang}, {Jiang}, {Jin}, {Ke}, {Kuleshov}, {Levochkin}, {Li}, {Li}, {Li},
  {Li}, {Li}, {Li}, {Li}, {Li}, {Li}, {Li}, {Li}, {Li}, {Li}, {Li}, {Li}, {Li},
  {Li}, {Liang}, {Liang}, {Lin}, {Liu}, {Liu}, {Liu}, {Liu}, {Liu}, {Liu},
  {Liu}, {Liu}, {Liu}, {Liu}, {Liu}, {Liu}, {Liu}, {Liu}, {Liu}, {Liu}, {Long},
  {Lu}, {Lv}, {Ma}, {Ma}, {Ma}, {Mao}, {Masood}, {Min}, {Mitthumsiri},
  {Montaruli}, {Nan}, {Pang}, {Pattarakijwanich}, {Pei}, {Qi}, {Qi}, {Qiao},
  {Qin}, {Ruffolo}, {Rulev}, {S{\'a}iz}, {Shao}, {Shchegolev}, {Sheng}, {Shi},
  {Song}, {Stenkin}, {Stepanov}, {Su}, {Sun}, {Sun}, {Sun}, {Tam}, {Tang},
  {Tian}, {Wang}, {Wang}, {Wang}, {Wang}, {Wang}, {Wang}, {Wang}, {Wang},
  {Wang}, {Wang}, {Wang}, {Wang}, {Wang}, {Wang}, {Wang}, {Wang}, {Wang},
  {Wang}, {Wang}, {Wang}, {Wang}, {Wang}, {Wei}, {Wei}, {Wei}, {Wen}, {Wu},
  {Wu}, {Wu}, {Wu}, {Wu}, {Xi}, {Xia}, {Xia}, {Xiang}, {Xiao}, {Xiao}, {Xiao},
  {Xin}, {Xin}, {Xing}, {Xu}, {Xu}, {Xue}, {Yan}, {Yan}, {Yang}, {Yang},
  {Yang}, {Yang}, {Yang}, {Yang}, {Yang}, {Yao}, {Yao}, {Ye}, {Yin}, {Yin},
  {You}, {You}, {Yu}, {Yuan}, {Zeng}, {Zeng}, {Zeng}, {Zeng}, {Zha}, {Zhai},
  {Zhang}, {Zhang}, {Zhang}, {Zhang}, {Zhang}, {Zhang}, {Zhang}, {Zhang},
  {Zhang}, {Zhang}, {Zhang}, {Zhang}, {Zhang}, {Zhang}, {Zhang}, {Zhang},
  {Zhang}, {Zhang}, {Zhang}, {Zhao}, {Zhao}, {Zhao}, {Zhao}, {Zhao}, {Zheng},
  {Zheng}, {Zhou}, {Zhou}, {Zhou}, {Zhou}, {Zhou}, {Zhou}, {Zhu}, {Zhu}, {Zhu},
  {Zhu}, \& {Zuo}}]{2021ApJ...917L...4C}
---. 2021{\natexlab{c}}, \apjl, 917, L4, \dodoi{10.3847/2041-8213/ac0fd5}

\bibitem[{{Cao} {et~al.}(2022{\natexlab{a}}){Cao}, {Aharonian}, {An},
  {Axikegu}, {Bai}, {Bao}, {Bastieri}, {Bi}, {Bi}, {Cai}, {Cai}, {Cao},
  {Chang}, {Chang}, {Chen}, {Chen}, {Chen}, {Chen}, {Chen}, {Chen}, {Chen},
  {Chen}, {Chen}, {Chen}, {Chen}, {Chen}, {Chen}, {Chen}, {Cheng}, {Cheng},
  {Cui}, {Cui}, {Cui}, {Piazzoli}, {Dai}, {Dai}, {Dai}, {Danzengluobu}, {Dong},
  {Duan}, {Fan}, {Fan}, {Fan}, {Fang}, {Fang}, {Feng}, {Feng}, {Feng}, {Feng},
  {Gao}, {Gao}, {Gao}, {Gao}, {Gao}, {Ge}, {Geng}, {Gong}, {Gou}, {Gu}, {Guo},
  {Guo}, {Guo}, {Guo}, {Guo}, {Han}, {He}, {He}, {He}, {He}, {He}, {He},
  {Heller}, {Hor}, {Hou}, {Hou}, {Hu}, {Hu}, {Hu}, {Hu}, {Huang}, {Huang},
  {Huang}, {Huang}, {Huang}, {Huang}, {Ji}, {Ji}, {Jia}, {Jiang}, {Jiang},
  {Jin}, {Ke}, {Kuleshov}, {Levochkin}, {Li}, {Li}, {Li}, {Li}, {Li}, {Li},
  {Li}, {Li}, {Li}, {Li}, {Li}, {Li}, {Li}, {Li}, {Li}, {Li}, {Li}, {Li},
  {Liang}, {Liang}, {Lin}, {Liu}, {Liu}, {Liu}, {Liu}, {Liu}, {Liu}, {Liu},
  {Liu}, {Liu}, {Liu}, {Liu}, {Liu}, {Liu}, {Liu}, {Liu}, {Liu}, {Long}, {Lu},
  {Lv}, {Ma}, {Ma}, {Ma}, {Mao}, {Masood}, {Min}, {Mitthumsiri}, {Montaruli},
  {Nan}, {Pang}, {Pattarakijwanich}, {Pei}, {Qi}, {Qi}, {Qiao}, {Qin},
  {Ruffolo}, {Rulev}, {S{\'a}iz}, {Shao}, {Shchegolev}, {Sheng}, {Shi}, {Song},
  {Stenkin}, {Stepanov}, {Su}, {Sun}, {Sun}, {Sun}, {Tam}, {Tang}, {Tian},
  {Wang}, {Wang}, {Wang}, {Wang}, {Wang}, {Wang}, {Wang}, {Wang}, {Wang},
  {Wang}, {Wang}, {Wang}, {Wang}, {Wang}, {Wang}, {Wang}, {Wang}, {Wang},
  {Wang}, {Wang}, {Wang}, {Wang}, {Wei}, {Wei}, {Wei}, {Wen}, {Wu}, {Wu}, {Wu},
  {Wu}, {Wu}, {Xi}, {Xia}, {Xia}, {Xiang}, {Xiao}, {Xiao}, {Xiao}, {Xin},
  {Xin}, {Xing}, {Xu}, {Xu}, {Xue}, {Yan}, {Yan}, {Yang}, {Yang}, {Yang},
  {Yang}, {Yang}, {Yang}, {Yang}, {Yao}, {Yao}, {Ye}, {Yin}, {Yin}, {You},
  {You}, {Yu}, {Yuan}, {Zeng}, {Zeng}, {Zeng}, {Zeng}, {Zha}, {Zhai}, {Zhang},
  {Zhang}, {Zhang}, {Zhang}, {Zhang}, {Zhang}, {Zhang}, {Zhang}, {Zhang},
  {Zhang}, {Zhang}, {Zhang}, {Zhang}, {Zhang}, {Zhang}, {Zhang}, {Zhang},
  {Zhang}, {Zhang}, {Zhao}, {Zhao}, {Zhao}, {Zhao}, {Zhao}, {Zheng}, {Zheng},
  {Zhou}, {Zhou}, {Zhou}, {Zhou}, {Zhou}, {Zhou}, {Zhu}, {Zhu}, {Zhu}, {Zhu},
  {Zuo}, \& {LHAASO Collaboration}}]{2022PhRvL.128e1102C}
---. 2022{\natexlab{a}}, \prl, 128, 051102,
  \dodoi{10.1103/PhysRevLett.128.051102}

\bibitem[{{Cao} {et~al.}(2022{\natexlab{b}}){Cao}, {Aharonian}, {An},
  {Axikegu}, {Bai}, {Bao}, {Bastieri}, {Bi}, {Bi}, {Cai}, {Cao}, {Chang},
  {Chang}, {Chen}, {Chen}, {Chen}, {Chen}, {Chen}, {Chen}, {Chen}, {Chen},
  {Chen}, {Chen}, {Chen}, {Cheng}, {Cheng}, {Cheng}, {Cui}, {Cui}, {Cui},
  {D'Ettorre Piazzoli}, {Dai}, {Dai}, {Dai}, {Danzengluobu}, {Duan}, {Fan},
  {Fan}, {Fan}, {Fang}, {Fang}, {Feng}, {Feng}, {Feng}, {Feng}, {Feng}, {Gao},
  {Gao}, {Gao}, {Gao}, {Gao}, {Gao}, {Ge}, {Geng}, {Gong}, {Gou}, {Gu}, {Guo},
  {Guo}, {Guo}, {Guo}, {Guo}, {Han}, {He}, {He}, {He}, {He}, {He}, {Heller},
  {Hor}, {Hou}, {Hou}, {Hu}, {Hu}, {Hu}, {Hu}, {Hu}, {Huang}, {Huang}, {Huang},
  {Huang}, {Huang}, {Huang}, {Ji}, {Jia}, {Jia}, {Jiang}, {Jiang}, {Jin},
  {Kang}, {Ke}, {Kuleshov}, {Levochkin}, {Li}, {Li}, {Li}, {Li}, {Li}, {Li},
  {Li}, {Li}, {Li}, {Li}, {Li}, {Li}, {Li}, {Li}, {Li}, {Li}, {Li}, {Li},
  {Liang}, {Liang}, {Lin}, {Liu}, {Liu}, {Liu}, {Liu}, {Liu}, {Liu}, {Liu},
  {Liu}, {Liu}, {Liu}, {Liu}, {Liu}, {Liu}, {Liu}, {Liu}, {Long}, {Lu}, {Luo},
  {Lv}, {Ma}, {Ma}, {Ma}, {Mao}, {Masood}, {Min}, {Mitthumsiri}, {Nan}, {Ou},
  {Pang}, {Pattarakijwanich}, {Pei}, {Qi}, {Qi}, {Qiao}, {Qin}, {Ruffolo},
  {S{\'a}iz}, {Shao}, {Shao}, {Shchegolev}, {Sheng}, {Shi}, {Song}, {Stenkin},
  {Stepanov}, {Su}, {Sun}, {Sun}, {Sun}, {Tam}, {Tang}, {Tian}, {Wang}, {Wang},
  {Wang}, {Wang}, {Wang}, {Wang}, {Wang}, {Wang}, {Wang}, {Wang}, {Wang},
  {Wang}, {Wang}, {Wang}, {Wang}, {Wang}, {Wang}, {Wang}, {Wang}, {Wang},
  {Wang}, {Wei}, {Wei}, {Wei}, {Wen}, {Wu}, {Wu}, {Wu}, {Wu}, {Wu}, {Xi},
  {Xia}, {Xia}, {Xiang}, {Xiao}, {Xiao}, {Xin}, {Xin}, {Xing}, {Xiong}, {Xu},
  {Xu}, {Xue}, {Yan}, {Yan}, {Yang}, {Yang}, {Yang}, {Yang}, {Yang}, {Yang},
  {Yang}, {Yang}, {Yao}, {Yao}, {Ye}, {Yin}, {Yin}, {You}, {You}, {Yu}, {Yuan},
  {Yue}, {Zeng}, {Zeng}, {Zeng}, {Zeng}, {Zha}, {Zhai}, {Zhang}, {Zhang},
  {Zhang}, {Zhang}, {Zhang}, {Zhang}, {Zhang}, {Zhang}, {Zhang}, {Zhang},
  {Zhang}, {Zhang}, {Zhang}, {Zhang}, {Zhang}, {Zhang}, {Zhang}, {Zhang},
  {Zhang}, {Zhang}, {Zhao}, {Zhao}, {Zhao}, {Zhao}, {Zhao}, {Zheng}, {Zheng},
  {Zhou}, {Zhou}, {Zhou}, {Zhou}, {Zhou}, {Zhou}, {Zhu}, {Zhu}, {Zhu}, {Zhu},
  {Zuo}, {Ando}, {Chianese}, {Fiorillo}, {Miele}, {Ng}, \& {LHAASO
  Collaboration}}]{2022PhRvL.129z1103C}
---. 2022{\natexlab{b}}, \prl, 129, 261103,
  \dodoi{10.1103/PhysRevLett.129.261103}

\bibitem[{{Cao} {et~al.}(2023){Cao}, {Aharonian}, {An}, {Axikegu}, {Bai},
  {Bao}, {Bastieri}, {Bi}, {Bi}, {Cai}, {Cao}, {Cao}, {Cao}, {Chang}, {Chang},
  {Chen}, {Chen}, {Chen}, {Chen}, {Chen}, {Chen}, {Chen}, {Chen}, {Chen},
  {Chen}, {Chen}, {Chen}, {Cheng}, {Cheng}, {Cui}, {Cui}, {Cui}, {Cui}, {Dai},
  {Dai}, {Dai}, {Danzengluobu}, {della Volpe}, {Dong}, {Duan}, {Fan}, {Fan},
  {Fang}, {Fang}, {Feng}, {Feng}, {Feng}, {Feng}, {Feng}, {Gabici}, {Gao},
  {Gao}, {Gao}, {Gao}, {Gao}, {Gao}, {Ge}, {Geng}, {Giacinti}, {Gong}, {Gou},
  {Gu}, {Guo}, {Guo}, {Guo}, {Guo}, {Han}, {He}, {He}, {He}, {He}, {He},
  {Heller}, {Hor}, {Hou}, {Hou}, {Hou}, {Hu}, {Hu}, {Hu}, {Huang}, {Huang},
  {Huang}, {Huang}, {Huang}, {Huang}, {Huang}, {Ji}, {Jia}, {Jia}, {Jiang},
  {Jiang}, {Jiang}, {Jin}, {Kang}, {Ke}, {Kuleshov}, {Kurinov}, {Li}, {Li},
  {Li}, {Li}, {Li}, {Li}, {Li}, {Li}, {Li}, {Li}, {Li}, {Li}, {Li}, {Li}, {Li},
  {Li}, {Li}, {Li}, {Li}, {Liang}, {Liang}, {Lin}, {Liu}, {Liu}, {Liu}, {Liu},
  {Liu}, {Liu}, {Liu}, {Liu}, {Liu}, {Liu}, {Liu}, {Liu}, {Liu}, {Liu}, {Lu},
  {Luo}, {Lv}, {Ma}, {Ma}, {Ma}, {Mao}, {Min}, {Mitthumsiri}, {Mu}, {Nan},
  {Neronov}, {Ou}, {Pang}, {Pattarakijwanich}, {Pei}, {Qi}, {Qi}, {Qiao},
  {Qin}, {Ruffolo}, {Saiz}, {Semikoz}, {Shao}, {Shao}, {Shchegolev}, {Sheng},
  {Shu}, {Song}, {Stenkin}, {Stepanov}, {Su}, {Sun}, {Sun}, {Sun}, {Tam},
  {Tang}, {Tang}, {Tian}, {Wang}, {Wang}, {Wang}, {Wang}, {Wang}, {Wang},
  {Wang}, {Wang}, {Wang}, {Wang}, {Wang}, {Wang}, {Wang}, {Wang}, {Wang},
  {Wang}, {Wang}, {Wang}, {Wang}, {Wang}, {Wang}, {Wei}, {Wei}, {Wei}, {Wen},
  {Wu}, {Wu}, {Wu}, {Wu}, {Wu}, {Xi}, {Xia}, {Xia}, {Xiang}, {Xiao}, {Xiao},
  {Xin}, {Xin}, {Xing}, {Xiong}, {Xu}, {Xu}, {Xu}, {Xu}, {Xue}, {Yan}, {Yan},
  {Yan}, {Yang}, {Yang}, {Yang}, {Yang}, {Yang}, {Yang}, {Yang}, {Yang},
  {Yang}, {Yao}, {Yao}, {Ye}, {Yin}, {Yin}, {You}, {You}, {Yu}, {Yuan}, {Yue},
  {Zeng}, {Zeng}, {Zeng}, {Zha}, {Zhang}, {Zhang}, {Zhang}, {Zhang}, {Zhang},
  {Zhang}, {Zhang}, {Zhang}, {Zhang}, {Zhang}, {Zhang}, {Zhang}, {Zhang},
  {Zhang}, {Zhang}, {Zhang}, {Zhang}, {Zhang}, {Zhao}, {Zhao}, {Zhao}, {Zhao},
  {Zhao}, {Zheng}, {Zhou}, {Zhou}, {Zhou}, {Zhou}, {Zhou}, {Zhou}, {Zhou},
  {Zhu}, {Zhu}, {Zhu}, {Zhu}, \& {Zuo}}]{2023arXiv230505372C}
---. 2023, arXiv e-prints, arXiv:2305.05372, \dodoi{10.48550/arXiv.2305.05372}

\bibitem[{Duan {et~al.}(2021)}]{DAMPE:2021kao}
Duan, K.-K., {et~al.} 2021, PoS, ICRC2021, 631, \dodoi{10.22323/1.395.0631}

\bibitem[{{Ferrand} \& {Safi-Harb}(2012)}]{2012AdSpR..49.1313F}
{Ferrand}, G., \& {Safi-Harb}, S. 2012, Advances in Space Research, 49, 1313,
  \dodoi{10.1016/j.asr.2012.02.004}

\bibitem[{{Fleysher} {et~al.}(2004){Fleysher}, {Fleysher}, {Nemethy}, {Mincer},
  \& {Haines}}]{2004ApJ...603..355F}
{Fleysher}, R., {Fleysher}, L., {Nemethy}, P., {Mincer}, A.~I., \& {Haines},
  T.~J. 2004, \apj, 603, 355, \dodoi{10.1086/381384}

\bibitem[{{Gao} \& {Han}(2014)}]{2014A&A...567A..59G}
{Gao}, X.~Y., \& {Han}, J.~L. 2014, \aap, 567, A59,
  \dodoi{10.1051/0004-6361/201424128}

\bibitem[{{Gerbrandt} {et~al.}(2014){Gerbrandt}, {Foster}, {Kothes},
  {Geisb{\"u}sch}, \& {Tung}}]{2014A&A...566A..76G}
{Gerbrandt}, S., {Foster}, T.~J., {Kothes}, R., {Geisb{\"u}sch}, J., \& {Tung},
  A. 2014, \aap, 566, A76, \dodoi{10.1051/0004-6361/201423679}

\bibitem[{{H.~E.~S.~S. Collaboration}(2018){H.~E.~S.~S.
  Collaboration}, {Abdalla}, {Abramowski}, {Aharonian}, {Ait Benkhali},
  {Ang{\"u}ner}, {Arakawa}, {Arrieta}, {Aubert}, {Backes}, {Balzer}, {Barnard},
  {Becherini}, {Becker Tjus}, {Berge}, {Bernhard}, {Bernl{\"o}hr}, {Blackwell},
  {B{\"o}ttcher}, {Boisson}, {Bolmont}, {Bonnefoy}, {Bordas}, {Bregeon},
  {Brun}, {Brun}, {Bryan}, {B{\"u}chele}, {Bulik}, {Capasso}, {Carrigan},
  {Caroff}, {Carosi}, {Casanova}, {Cerruti}, {Chakraborty}, {Chaves}, {Chen},
  {Chevalier}, {Colafrancesco}, {Condon}, {Conrad}, {Davids}, {Decock}, {Deil},
  {Devin}, {deWilt}, {Dirson}, {Djannati-Ata{\"\i}}, {Domainko}, {Donath},
  {Drury}, {Dutson}, {Dyks}, {Edwards}, {Egberts}, {Eger}, {Emery},
  {Ernenwein}, {Eschbach}, {Farnier}, {Fegan}, {Fernandes}, {Fiasson},
  {Fontaine}, {F{\"o}rster}, {Funk}, {F{\"u}{\ss}ling}, {Gabici}, {Gallant},
  {Garrigoux}, {Gast}, {Gat{\'e}}, {Giavitto}, {Giebels}, {Glawion},
  {Glicenstein}, {Gottschall}, {Grondin}, {Hahn}, {Haupt}, {Hawkes},
  {Heinzelmann}, {Henri}, {Hermann}, {Hinton}, {Hofmann}, {Hoischen}, {Holch},
  {Holler}, {Horns}, {Ivascenko}, {Iwasaki}, {Jacholkowska}, {Jamrozy},
  {Jankowsky}, {Jankowsky}, {Jingo}, {Jouvin}, {Jung-Richardt}, {Kastendieck},
  {Katarzy{\'n}ski}, {Katsuragawa}, {Katz}, {Kerszberg}, {Khangulyan},
  {Kh{\'e}lifi}, {King}, {Klepser}, {Klochkov}, {Klu{\'z}niak}, {Komin},
  {Kosack}, {Krakau}, {Kraus}, {Kr{\"u}ger}, {Laffon}, {Lamanna}, {Lau},
  {Lees}, {Lefaucheur}, {Lemi{\`e}re}, {Lemoine-Goumard}, {Lenain}, {Leser},
  {Lohse}, {Lorentz}, {Liu}, {L{\'o}pez-Coto}, {Lypova}, {Marandon},
  {Malyshev}, {Marcowith}, {Mariaud}, {Marx}, {Maurin}, {Maxted}, {Mayer},
  {Meintjes}, {Meyer}, {Mitchell}, {Moderski}, {Mohamed}, {Mohrmann},
  {Mor{\r{a}}}, {Moulin}, {Murach}, {Nakashima}, {de Naurois}, {Ndiyavala},
  {Niederwanger}, {Niemiec}, {Oakes}, {O'Brien}, {Odaka}, {Ohm}, {Ostrowski},
  {Oya}, {Padovani}, {Panter}, {Parsons}, {Paz Arribas}, {Pekeur}, {Pelletier},
  {Perennes}, {Petrucci}, {Peyaud}, {Piel}, {Pita}, {Poireau}, {Poon},
  {Prokhorov}, {Prokoph}, {P{\"u}hlhofer}, {Punch}, {Quirrenbach}, {Raab},
  {Rauth}, {Reimer}, {Reimer}, {Renaud}, {de los Reyes}, {Rieger}, {Rinchiuso},
  {Romoli}, {Rowell}, {Rudak}, {Rulten}, {Safi-Harb}, {Sahakian}, {Saito},
  {Sanchez}, {Santangelo}, {Sasaki}, {Schandri}, {Schlickeiser},
  {Sch{\"u}ssler}, {Schulz}, {Schwanke}, {Schwemmer}, {Seglar-Arroyo},
  {Settimo}, {Seyffert}, {Shafi}, {Shilon}, {Shiningayamwe}, {Simoni}, {Sol},
  {Spanier}, {Spir-Jacob}, {Stawarz}, {Steenkamp}, {Stegmann}, {Steppa},
  {Sushch}, {Takahashi}, {Tavernet}, {Tavernier}, {Taylor}, {Terrier},
  {Tibaldo}, {Tiziani}, {Tluczykont}, {Trichard}, {Tsirou}, {Tsuji}, {Tuffs},
  {Uchiyama}, {van der Walt}, {van Eldik}, {van Rensburg}, {van Soelen},
  {Vasileiadis}, {Veh}, {Venter}, {Viana}, {Vincent}, {Vink}, {Voisin},
  {V{\"o}lk}, {Vuillaume}, {Wadiasingh}, {Wagner}, {Wagner}, {Wagner}, {White},
  {Wierzcholska}, {Willmann}, {W{\"o}rnlein}, {Wouters}, {Yang}, {Zaborov},
  {Zacharias}, {Zanin}, {Zdziarski}, {Zech}, {Zefi}, {Ziegler}, {Zorn}, \&
  {{\.Z}ywucka}}]{2018A&A...612A...1H}
{H.~E.~S.~S. Collaboration}, {Abdalla}, H., {Abramowski}, A., {et~al.} 2018,
  \aap, 612, A1, \dodoi{10.1051/0004-6361/201732098}

\bibitem[{{Hartman} {et~al.}(1999){Hartman}, {Bertsch}, {Bloom}, {Chen},
  {Deines-Jones}, {Esposito}, {Fichtel}, {Friedlander}, {Hunter}, {McDonald},
  {Sreekumar}, {Thompson}, {Jones}, {Lin}, {Michelson}, {Nolan}, {Tompkins},
  {Kanbach}, {Mayer-Hasselwander}, {M{\"u}cke}, {Pohl}, {Reimer}, {Kniffen},
  {Schneid}, {von Montigny}, {Mukherjee}, \& {Dingus}}]{1999ApJS..123...79H}
{Hartman}, R.~C., {Bertsch}, D.~L., {Bloom}, S.~D., {et~al.} 1999, \apjs, 123,
  79, \dodoi{10.1086/313231}

\bibitem[{{Klingler} {et~al.}(2016){Klingler}, {Rangelov}, {Kargaltsev},
  {Pavlov}, {Romani}, {Posselt}, {Slane}, {Temim}, {Ng}, {Bucciantini},
  {Bykov}, {Swartz}, \& {Buehler}}]{2016ApJ...833..253K}
{Klingler}, N., {Rangelov}, B., {Kargaltsev}, O., {et~al.} 2016, \apj, 833,
  253, \dodoi{10.3847/1538-4357/833/2/253}

\bibitem[{{LHAASO Collaboration}(2021){LHAASO Collaboration}, {Cao},
  {Aharonian}, {An}, {Axikegu}, {Bai}, {Bai}, {Bao}, {Bastieri}, {Bi}, {Bi},
  {Cai}, {Cai}, {Cao}, {Chang}, {Chang}, {Chen}, {Chen}, {Chen}, {Chen},
  {Chen}, {Chen}, {Chen}, {Chen}, {Chen}, {Chen}, {Chen}, {Chen}, {Chen},
  {Chen}, {Cheng}, {Cheng}, {Cui}, {Cui}, {Cui}, {D'Ettorre Piazzoli}, {Dai},
  {Dai}, {Dai}, {Danzengluobu}, {Della Volpe}, {Dong}, {Duan}, {Fan}, {Fan},
  {Fan}, {Fang}, {Fang}, {Feng}, {Feng}, {Feng}, {Feng}, {Gao}, {Gao}, {Gao},
  {Gao}, {Gao}, {Ge}, {Geng}, {Gong}, {Gou}, {Gu}, {Guo}, {Guo}, {Guo}, {Guo},
  {Guo}, {Han}, {He}, {He}, {He}, {He}, {He}, {He}, {Heller}, {Hor}, {Hou},
  {Hou}, {Hu}, {Hu}, {Hu}, {Hu}, {Huang}, {Huang}, {Huang}, {Huang}, {Huang},
  {Huang}, {Ji}, {Ji}, {Jia}, {Jiang}, {Jiang}, {Jin}, {Ke}, {Kuleshov},
  {Levochkin}, {Li}, {Li}, {Li}, {Li}, {Li}, {Li}, {Li}, {Li}, {Li}, {Li},
  {Li}, {Li}, {Li}, {Li}, {Li}, {Li}, {Li}, {Li}, {Liang}, {Liang}, {Lin},
  {Liu}, {Liu}, {Liu}, {Liu}, {Liu}, {Liu}, {Liu}, {Liu}, {Liu}, {Liu}, {Liu},
  {Liu}, {Liu}, {Liu}, {Liu}, {Liu}, {Long}, {Lu}, {Lv}, {Ma}, {Ma}, {Ma},
  {Mao}, {Masood}, {Min}, {Mitthumsiri}, {Montaruli}, {Nan}, {Pang},
  {Pattarakijwanich}, {Pei}, {Qi}, {Qi}, {Qiao}, {Qin}, {Ruffolo}, {Rulev},
  {Saiz}, {Shao}, {Shchegolev}, {Sheng}, {Shi}, {Song}, {Stenkin}, {Stepanov},
  {Su}, {Sun}, {Sun}, {Sun}, {Tam}, {Tang}, {Tian}, {Wang}, {Wang}, {Wang},
  {Wang}, {Wang}, {Wang}, {Wang}, {Wang}, {Wang}, {Wang}, {Wang}, {Wang},
  {Wang}, {Wang}, {Wang}, {Wang}, {Wang}, {Wang}, {Wang}, {Wang}, {Wang},
  {Wang}, {Wei}, {Wei}, {Wei}, {Wen}, {Wu}, {Wu}, {Wu}, {Wu}, {Wu}, {Xi},
  {Xia}, {Xia}, {Xiang}, {Xiao}, {Xiao}, {Xiao}, {Xin}, {Xin}, {Xing}, {Xu},
  {Xu}, {Xue}, {Yan}, {Yan}, {Yang}, {Yang}, {Yang}, {Yang}, {Yang}, {Yang},
  {Yang}, {Yao}, {Yao}, {Ye}, {Yin}, {Yin}, {You}, {You}, {Yu}, {Yuan}, {Zeng},
  {Zeng}, {Zeng}, {Zeng}, {Zha}, {Zhai}, {Zhang}, {Zhang}, {Zhang}, {Zhang},
  {Zhang}, {Zhang}, {Zhang}, {Zhang}, {Zhang}, {Zhang}, {Zhang}, {Zhang},
  {Zhang}, {Zhang}, {Zhang}, {Zhang}, {Zhang}, {Zhang}, {Zhang}, {Zhao},
  {Zhao}, {Zhao}, {Zhao}, {Zhao}, {Zheng}, {Zheng}, {Zhou}, {Zhou}, {Zhou},
  {Zhou}, {Zhou}, {Zhou}, {Zhu}, {Zhu}, {Zhu}, {Zhu}, \&
  {Zuo}}]{2021Sci...373..425L}
{LHAASO Collaboration}, {Cao}, Z., {Aharonian}, F., {et~al.} 2021, Science,
  373, 425, \dodoi{10.1126/science.abg5137}

\bibitem[{{Lv} {et~al.}(2018){Lv}, {He}, {Sheng}, {Liu}, {Chen}, {Liu}, {Hou},
  {Zhao}, {Zhang}, {Wu}, {Wang}, \& {Lhaaso
  Collaboration}}]{2018APh...100...22L}
{Lv}, H., {He}, H., {Sheng}, X., {et~al.} 2018, Astroparticle Physics, 100, 22,
  \dodoi{10.1016/j.astropartphys.2018.02.011}

\bibitem[{{Ma} {et~al.}(2022){Ma}, {Bi}, {Cao}, {Chen}, {Chen}, {Cheng},
  {Gong}, {Gu}, {He}, {Hou}, {Huang}, {Huang}, {Liu}, {Shchegolev}, {Sheng},
  {Stenkin}, {Wu}, {Wu}, {Wu}, {Xiao}, {Yao}, {Zhang}, {Zhang}, \&
  {Zuo}}]{2022ChPhC..46c0001M}
{Ma}, X.-H., {Bi}, Y.-J., {Cao}, Z., {et~al.} 2022, Chinese Physics C, 46,
  030001, \dodoi{10.1088/1674-1137/ac3fa6}

\bibitem[{{Mattox} {et~al.}(1997){Mattox}, {Schachter}, {Molnar}, {Hartman}, \&
  {Patnaik}}]{1997ApJ...481...95M}
{Mattox}, J.~R., {Schachter}, J., {Molnar}, L., {Hartman}, R.~C., \& {Patnaik},
  A.~R. 1997, \apj, 481, 95, \dodoi{10.1086/304039}

\bibitem[{{Meagher} \& {VERITAS Collaboration}(2015)}]{2015ICRC...34..792M}
{Meagher}, K., \& {VERITAS Collaboration}. 2015, in International Cosmic Ray
  Conference, Vol.~34, 34th International Cosmic Ray Conference (ICRC2015),
  792, \dodoi{10.22323/1.236.0792}

\bibitem[{{Nowak} {et~al.}(2012){Nowak}, {Wilms}, {Pottschmidt}, {Schulz},
  {Maitra}, \& {Miller}}]{2012ApJ...744..107N}
{Nowak}, M.~A., {Wilms}, J., {Pottschmidt}, K., {et~al.} 2012, \apj, 744, 107,
  \dodoi{10.1088/0004-637X/744/2/107}

\bibitem[{{Planck Collaboration} {et~al.}(2014){Planck Collaboration},
  {Abergel}, {Ade}, {Aghanim}, {Alves}, {Aniano}, {Armitage-Caplan}, {Arnaud},
  {Ashdown}, {Atrio-Barandela}, {Aumont}, {Baccigalupi}, {Banday}, {Barreiro},
  {Bartlett}, {Battaner}, {Benabed}, {Beno{\^\i}t}, {Benoit-L{\'e}vy},
  {Bernard}, {Bersanelli}, {Bielewicz}, {Bobin}, {Bock}, {Bonaldi}, {Bond},
  {Borrill}, {Bouchet}, {Boulanger}, {Bridges}, {Bucher}, {Burigana}, {Butler},
  {Cardoso}, {Catalano}, {Chamballu}, {Chary}, {Chiang}, {Chiang},
  {Christensen}, {Church}, {Clemens}, {Clements}, {Colombi}, {Colombo},
  {Combet}, {Couchot}, {Coulais}, {Crill}, {Curto}, {Cuttaia}, {Danese},
  {Davies}, {Davis}, {de Bernardis}, {de Rosa}, {de Zotti}, {Delabrouille},
  {Delouis}, {D{\'e}sert}, {Dickinson}, {Diego}, {Dole}, {Donzelli},
  {Dor{\'e}}, {Douspis}, {Draine}, {Dupac}, {Efstathiou}, {En{\ss}lin},
  {Eriksen}, {Falgarone}, {Finelli}, {Forni}, {Frailis}, {Fraisse},
  {Franceschi}, {Galeotta}, {Ganga}, {Ghosh}, {Giard}, {Giardino},
  {Giraud-H{\'e}raud}, {Gonz{\'a}lez-Nuevo}, {G{\'o}rski}, {Gratton},
  {Gregorio}, {Grenier}, {Gruppuso}, {Guillet}, {Hansen}, {Hanson}, {Harrison},
  {Helou}, {Henrot-Versill{\'e}}, {Hern{\'a}ndez-Monteagudo}, {Herranz},
  {Hildebrandt}, {Hivon}, {Hobson}, {Holmes}, {Hornstrup}, {Hovest},
  {Huffenberger}, {Jaffe}, {Jaffe}, {Jewell}, {Joncas}, {Jones}, {Juvela},
  {Keih{\"a}nen}, {Keskitalo}, {Kisner}, {Knoche}, {Knox}, {Kunz},
  {Kurki-Suonio}, {Lagache}, {L{\"a}hteenm{\"a}ki}, {Lamarre}, {Lasenby},
  {Laureijs}, {Lawrence}, {Leonardi}, {Le{\'o}n-Tavares}, {Lesgourgues},
  {Levrier}, {Liguori}, {Lilje}, {Linden-V{\o}rnle}, {L{\'o}pez-Caniego},
  {Lubin}, {Mac{\'\i}as-P{\'e}rez}, {Maffei}, {Maino}, {Mandolesi}, {Maris},
  {Marshall}, {Martin}, {Mart{\'\i}nez-Gonz{\'a}lez}, {Masi}, {Massardi},
  {Matarrese}, {Matthai}, {Mazzotta}, {McGehee}, {Melchiorri}, {Mendes},
  {Mennella}, {Migliaccio}, {Mitra}, {Miville-Desch{\^e}nes}, {Moneti},
  {Montier}, {Morgante}, {Mortlock}, {Munshi}, {Murphy}, {Naselsky}, {Nati},
  {Natoli}, {Netterfield}, {N{\o}rgaard-Nielsen}, {Noviello}, {Novikov},
  {Novikov}, {Osborne}, {Oxborrow}, {Paci}, {Pagano}, {Pajot}, {Paladini},
  {Paoletti}, {Pasian}, {Patanchon}, {Perdereau}, {Perotto}, {Perrotta},
  {Piacentini}, {Piat}, {Pierpaoli}, {Pietrobon}, {Plaszczynski},
  {Pointecouteau}, {Polenta}, {Ponthieu}, {Popa}, {Poutanen}, {Pratt},
  {Pr{\'e}zeau}, {Prunet}, {Puget}, {Rachen}, {Reach}, {Rebolo}, {Reinecke},
  {Remazeilles}, {Renault}, {Ricciardi}, {Riller}, {Ristorcelli}, {Rocha},
  {Rosset}, {Roudier}, {Rowan-Robinson}, {Rubi{\~n}o-Mart{\'\i}n}, {Rusholme},
  {Sandri}, {Santos}, {Savini}, {Scott}, {Seiffert}, {Shellard}, {Spencer},
  {Starck}, {Stolyarov}, {Stompor}, {Sudiwala}, {Sunyaev}, {Sureau}, {Sutton},
  {Suur-Uski}, {Sygnet}, {Tauber}, {Tavagnacco}, {Terenzi}, {Toffolatti},
  {Tomasi}, {Tristram}, {Tucci}, {Tuovinen}, {T{\"u}rler}, {Umana},
  {Valenziano}, {Valiviita}, {Van Tent}, {Verstraete}, {Vielva}, {Villa},
  {Vittorio}, {Wade}, {Wandelt}, {Welikala}, {Ysard}, {Yvon}, {Zacchei}, \&
  {Zonca}}]{2014A&A...571A..11P}
{Planck Collaboration}, {Abergel}, A., {Ade}, P.~A.~R., {et~al.} 2014, \aap,
  571, A11, \dodoi{10.1051/0004-6361/201323195}

\bibitem[{{Planck Collaboration} {et~al.}(2016){Planck Collaboration},
  {Aghanim}, {Ashdown}, {Aumont}, {Baccigalupi}, {Ballardini}, {Banday},
  {Barreiro}, {Bartolo}, {Basak}, {Benabed}, {Bernard}, {Bersanelli},
  {Bielewicz}, {Bonavera}, {Bond}, {Borrill}, {Bouchet}, {Boulanger},
  {Burigana}, {Calabrese}, {Cardoso}, {Carron}, {Chiang}, {Colombo}, {Comis},
  {Couchot}, {Coulais}, {Crill}, {Curto}, {Cuttaia}, {de Bernardis}, {de
  Zotti}, {Delabrouille}, {Di Valentino}, {Dickinson}, {Diego}, {Dor{\'e}},
  {Douspis}, {Ducout}, {Dupac}, {Dusini}, {Elsner}, {En{\ss}lin}, {Eriksen},
  {Falgarone}, {Fantaye}, {Finelli}, {Forastieri}, {Frailis}, {Fraisse},
  {Franceschi}, {Frolov}, {Galeotta}, {Galli}, {Ganga}, {G{\'e}nova-Santos},
  {Gerbino}, {Ghosh}, {Giraud-H{\'e}raud}, {Gonz{\'a}lez-Nuevo}, {G{\'o}rski},
  {Gruppuso}, {Gudmundsson}, {Hansen}, {Helou}, {Henrot-Versill{\'e}},
  {Herranz}, {Hivon}, {Huang}, {Jaffe}, {Jones}, {Keih{\"a}nen}, {Keskitalo},
  {Kiiveri}, {Kisner}, {Krachmalnicoff}, {Kunz}, {Kurki-Suonio}, {Lamarre},
  {Langer}, {Lasenby}, {Lattanzi}, {Lawrence}, {Le Jeune}, {Levrier}, {Lilje},
  {Lilley}, {Lindholm}, {L{\'o}pez-Caniego}, {Ma}, {Mac{\'\i}as-P{\'e}rez},
  {Maggio}, {Maino}, {Mandolesi}, {Mangilli}, {Maris}, {Martin},
  {Mart{\'\i}nez-Gonz{\'a}lez}, {Matarrese}, {Mauri}, {McEwen}, {Melchiorri},
  {Mennella}, {Migliaccio}, {Miville-Desch{\^e}nes}, {Molinari}, {Moneti},
  {Montier}, {Morgante}, {Moss}, {Natoli}, {Oxborrow}, {Pagano}, {Paoletti},
  {Patanchon}, {Perdereau}, {Perotto}, {Pettorino}, {Piacentini},
  {Plaszczynski}, {Polastri}, {Polenta}, {Puget}, {Rachen}, {Racine},
  {Reinecke}, {Remazeilles}, {Renzi}, {Rocha}, {Rosset}, {Rossetti}, {Roudier},
  {Rubi{\~n}o-Mart{\'\i}n}, {Ruiz-Granados}, {Salvati}, {Sandri}, {Savelainen},
  {Scott}, {Sirignano}, {Sirri}, {Soler}, {Spencer}, {Suur-Uski}, {Tauber},
  {Tavagnacco}, {Tenti}, {Toffolatti}, {Tomasi}, {Tristram}, {Trombetti},
  {Valiviita}, {Van Tent}, {Vielva}, {Villa}, {Vittorio}, {Wandelt}, {Wehus},
  {Zacchei}, \& {Zonca}}]{2016A&A...596A.109P}
{Planck Collaboration}, {Aghanim}, N., {Ashdown}, M., {et~al.} 2016, \aap, 596,
  A109, \dodoi{10.1051/0004-6361/201629022}

\bibitem[{{Staszak} \& {VERITAS Collaboration}(2015)}]{2015ICRC...34..868S}
{Staszak}, D., \& {VERITAS Collaboration}. 2015, in International Cosmic Ray
  Conference, Vol.~34, 34th International Cosmic Ray Conference (ICRC2015),
  868, \dodoi{10.22323/1.236.0868}

\bibitem[Stewart(2009)]{2009A&A...495..989S} Stewart, I.~M.\ 2009, \aap, 495, 989. \dodoi{10.1051/0004-6361:200811311}

\bibitem[{{Wakely} \& {Horan}(2008)}]{2008ICRC....3.1341W}
{Wakely}, S.~P., \& {Horan}, D. 2008, in International Cosmic Ray Conference,
  Vol.~3, International Cosmic Ray Conference, 1341--1344

\bibitem[{{Zyuzin} {et~al.}(2018){Zyuzin}, {Karpova}, \&
  {Shibanov}}]{2018MNRAS.476.2177Z}
{Zyuzin}, D.~A., {Karpova}, A.~V., \& {Shibanov}, Y.~A. 2018, \mnras, 476,
  2177, \dodoi{10.1093/mnras/sty359}

\end{thebibliography}

\startlongtable
\begin{deluxetable}{lcccccccccccccccccc}
\setlength{\tabcolsep}{0.02in}
\tablewidth{660pt}
\tabletypesize{\scriptsize}
\rotate
\tablecaption{1LHAASO source catalog \label{tab:1}}
\tablehead{
\colhead{Source name} &
\colhead{Components}&
\colhead{$\alpha_{2000}$} &
\colhead{$\delta_{2000}$} &
\colhead{$\sigma_{p,95,stat}$} &
\colhead{$r_{39}$} &
\colhead{TS} &
\colhead{$N_0$} &
\colhead{$\Gamma$} &
\colhead{TS$_{100}$}&
\colhead{Assoc.(Sep.[$^{\circ}$])}
}
\startdata
 1LHAASO J0007$+$5659u & KM2A & 1.86 & 57.00 & 0.12 & $<$0.18 & 86.5 & 0.33$\pm$0.05 & 3.10$\pm$0.20 & 43.6 & \\
 & WCDA & & & & & & $<$0.27 & & & \\
\hline
 1LHAASO J0007$+$7303u & KM2A & 1.91 & 73.07 & 0.07 & 0.17$\pm$0.03 & 361.0 & 3.41$\pm$0.27 & 3.40$\pm$0.12 & 171.6 &   CTA 1                 (0.12) \\
 & WCDA & 1.48 & 73.15 & 0.10 & $<$0.22 & 141.6 & 5.01$\pm$1.11 & 2.74$\pm$0.11 & & \\
\hline
 1LHAASO J0056$+$6346u & KM2A & 14.10 & 63.77 & 0.08 & 0.24$\pm$0.03 & 380.2 & 1.47$\pm$0.10 & 3.33$\pm$0.10 & 94.1 &  \\
 & WCDA & 13.78 & 63.96 & 0.15 & 0.33$\pm$0.07 & 106.1 & 1.45$\pm$0.41 & 2.35$\pm$0.13 & &  \\
\hline
 1LHAASO J0206$+$4302u & KM2A & 31.70 & 43.05 & 0.13 & $<$0.27 & 96.0 & 0.24$\pm$0.03 & 2.62$\pm$0.16 & 82.8 & \\
 & WCDA & & & & & & $<$0.09 & & & \\
\hline
 1LHAASO J0212$+$4254u & KM2A & 33.01 & 42.91 & 0.20 & $<$0.31 & 38.4 & 0.12$\pm$0.03 & 2.45$\pm$0.23 & 30.2 & \\
 & WCDA & & & & & & $<$0.07 & & & \\
\hline
 1LHAASO J0216$+$4237u & KM2A & 34.10 & 42.63 & 0.10 & $<$0.13 & 102.0 & 0.18$\pm$0.03 & 2.58$\pm$0.17 & 65.6 & \\
 & WCDA & & & & & & $<$0.20 & & & \\
\hline
 1LHAASO J0249$+$6022 & KM2A & 42.39 & 60.37 & 0.16 & 0.38$\pm$0.08 & 148.8 & 0.93$\pm$0.09 & 3.82$\pm$0.18 &  &  \\
 & WCDA & 41.52 & 60.49 & 0.40 & 0.71$\pm$0.10 & 53.3 & 1.96$\pm$0.51 & 2.52$\pm$0.16 & &  \\
\hline
 1LHAASO J0339$+$5307 & KM2A & 54.79 & 53.13 & 0.11 & $<$0.22 & 144.0 & 0.58$\pm$0.06 & 3.64$\pm$0.16 & & LHAASO J0341+5258 (0.37)\\
 & WCDA & & & & & & $<$0.21 & & & \\
\hline
 1LHAASO J0343$+$5254u* & KM2A & 55.79 & 52.91 & 0.08 & 0.20$\pm$0.02 & 388.1 & 1.07$\pm$0.07 & 3.53$\pm$0.10 & 20.2 &   LHAASO J0341+5258     (0.28) \\
 & WCDA & 55.34 & 53.05 & 0.18 & 0.33$\pm$0.05 & 94.1 & 0.29$\pm$0.13 & 1.70$\pm$0.19 & &  \\
\hline
 1LHAASO J0359$+$5406 & KM2A & 59.78 & 54.10 & 0.10 & 0.30$\pm$0.04 & 259.2 & 0.85$\pm$0.06 & 3.84$\pm$0.15 &  &  \\
 & WCDA & 59.68 & 54.21 & 0.14 & 0.22$\pm$0.05 & 59.3 & 0.18$\pm$0.12 & 1.74$\pm$0.28 & &  \\
\hline
1LHAASO J0428$+$5531* & WCDA & 67.23 & 55.53 & 0.36 & 1.18$\pm$0.12 & 153.8 & 4.73$\pm$0.65 & 2.66$\pm$0.10 & & \\
 & KM2A & 66.63 & 54.63 & 0.18 & 0.32$\pm$0.06 & 98.0 & 0.54$\pm$0.06 & 3.45$\pm$0.19 &  & \\
\hline
1LHAASO J0500$+$4454 & WCDA & 75.01 & 44.92 & 0.28 & 0.41$\pm$0.07 & 43.6 & 0.69$\pm$0.16 & 2.53$\pm$0.20 & & \\
 & KM2A & & & & & & $<$0.09 & & & \\
\hline
 1LHAASO J0534$+$3533 & KM2A & 83.53 & 35.56 & 0.18 & $<$0.36 & 60.8 & 0.19$\pm$0.03 & 4.89$\pm$0.53 & & \\
 & WCDA & 83.38 & 35.48 & 0.18 & $<$0.36 & 50.4 & 0.43$\pm$0.11 & 2.37$\pm$0.21 & &\\
\hline
1LHAASO J0534$+$2200u & WCDA & 83.623 & 22.012 & 0.004 & $<$0.04 & 73603.7 & 21.10$\pm$0.11 & 2.69$\pm$0.01 & &   Crab                  (0.01)\\
 & KM2A & 83.614 & 22.036 & 0.011 & $<$0.06 & 14328.1 & 6.23$\pm$0.10 & 3.19$\pm$0.03 & 2381.4 & \\
\hline
 1LHAASO J0542$+$2311u & KM2A & 85.71 & 23.20 & 0.14 & 0.98$\pm$0.05 & 745.3 & 2.93$\pm$0.12 & 3.74$\pm$0.09 & 21.2 &   HAWC J0543+233        (0.21) \\
 & WCDA* & 86.07 & 23.19 & 0.50 & 1.45$\pm$0.18 & 136.9 & 2.08$\pm$0.54 & 1.95$\pm$0.13 & &  \\
\hline
1LHAASO J0617$+$2234 & WCDA & 94.35 & 22.57 & 0.18 & 0.59$\pm$0.08 & 243.4 & 1.95$\pm$0.27 & 2.92$\pm$0.14 & &   IC 443                (0.14)\\
 & KM2A & & & & & & $<$0.17 & & & \\
\hline
 1LHAASO J0622$+$3754 & KM2A & 95.50 & 37.90 & 0.08 & 0.46$\pm$0.03 & 615.0 & 1.42$\pm$0.07 & 3.68$\pm$0.10 & &   LHAASO J0621+3755     (0.03) \\
 & WCDA & 95.67 & 37.93 & 0.29 & 0.50$\pm$0.09 & 59.3 & 0.39$\pm$0.17 & 1.82$\pm$0.22 & &  \\
\hline
 1LHAASO J0631$+$1040 & KM2A & 97.77 & 10.67 & 0.11 & $<$0.30 & 141.6 & 0.54$\pm$0.06 & 3.33$\pm$0.16 &  &  3HWC J0631+107(0.06) \\
 & WCDA & & & & & & $<$0.36 & & & \\
\hline
 1LHAASO J0634$+$1741u & KM2A & 98.57 & 17.69 & 0.10 & 0.89$\pm$0.04 & 1043.3 & 4.42$\pm$0.15 & 3.69$\pm$0.06 & 23.0 &   Geminga               (0.54) \\
 & WCDA & 98.51 & 17.72 & 0.28 & 1.16$\pm$0.17 & 193.2 & 1.53$\pm$0.65 & 1.65$\pm$0.15 & &  \\
\hline
 1LHAASO J0635$+$0619 & KM2A & 98.76 & 6.33 & 0.23 & 0.60$\pm$0.07 & 106.1 & 0.94$\pm$0.10 & 3.67$\pm$0.18 &  &   HAWC J0635+070        (0.67)\\
 & WCDA & & & & & & $<$0.90 & & & \\
\hline
 1LHAASO J0703$+$1405 & KM2A & 105.83 & 14.10 & 0.26 & 1.88$\pm$0.09 & 841.0 & 6.30$\pm$0.23 & 3.98$\pm$0.08 &  &   2HWC J0700+143        (0.72) \\
 & WCDA & 105.32 & 14.55 & 0.48 & 1.30$\pm$0.21 & 90.2 & 2.27$\pm$0.74 & 1.98$\pm$0.12 & &  \\
\hline
1LHAASO J1104$+$3810 & WCDA & 166.07 & 38.18 & 0.01 & $<$0.04 & 5343.6 & 3.68$\pm$0.10 & 3.41$\pm$0.03 & &   Mrk 421         (0.02)\\
 & KM2A & & & & & & $<$0.06 & & & \\
\hline
1LHAASO J1219$+$2915 & WCDA & 184.98 & 29.25 & 0.09 & $<$0.08 & 50.4 & 0.34$\pm$0.06 & 2.67$\pm$0.17 & & \\
 & KM2A & & & & & & $<$0.05 & & & \\
\hline
1LHAASO J1653$+$3943 & WCDA & 253.43 & 39.73 & 0.01 & $<$0.04 & 4121.6 & 3.77$\pm$0.08 & 2.94$\pm$0.02 & &   Mrk 501         (0.04)\\
 & KM2A & & & & & & $<$0.07 & & & \\
\hline
1LHAASO J1727$+$5016 & WCDA & 261.89 & 50.28 & 0.10 & $<$0.08 & 43.6 & 0.49$\pm$0.08 & 3.09$\pm$0.21 & &   1ES 1727+502          (0.14)\\
 & KM2A & & & & & & $<$0.03 & & & \\
\hline
 1LHAASO J1740$+$0948u & KM2A & 265.03 & 9.81 & 0.07 & $<$0.11 & 156.2 & 0.41$\pm$0.04 & 3.13$\pm$0.15 & 37.2 &  3HWC J1739+099(0.13) \\
 & WCDA & & & & & & $<$0.29 & & & \\
\hline
 1LHAASO J1809$-$1918u & KM2A & 272.38 & -19.30 & 0.13 & $<$0.22 & 134.6 & 9.46$\pm$1.27 & 3.51$\pm$0.26 & 88.4 &   HESS J1809-193        (0.24) \\
 & WCDA & 272.66 & -19.32 & 0.20 & 0.35$\pm$0.06 & 54.8 & 3.46$\pm$0.61 & 2.24$\pm$0.06 & & \\
\hline
 1LHAASO J1813$-$1245 & KM2A & 273.36 & -12.75 & 0.22 & $<$0.31 & 42.2 & 1.42$\pm$0.27 & 3.66$\pm$0.34 &  &   HESS J1813-126        (0.07)\\
 & WCDA & 273.35 & -12.73 & 0.21 & $<$0.32 & 37.2 & 2.44$\pm$0.44 & 2.61$\pm$0.08 & &\\
\hline
1LHAASO J1814$-$1719u* & WCDA & 273.69 & -17.33 & 0.27 & 0.71$\pm$0.07 & 118.8 & 26.50$\pm$2.46 & 2.83$\pm$0.06 & &   2HWC J1814-173        (0.16)\\
 & KM2A & 273.27 & -17.89 & 0.17 & $<$0.27 & 59.3 & 4.20$\pm$0.75 & 3.49$\pm$0.31 & 41.0 & \\
\hline
 1LHAASO J1814$-$1636u & KM2A & 273.72 & -16.62 & 0.30 & 0.68$\pm$0.08 & 132.2 & 11.90$\pm$1.30 & 3.74$\pm$0.20 & 50.4 &   2HWC J1814-173        (0.72)\\
 & WCDA & & & & & & $<$19.71 & & & \\
\hline
1LHAASO J1825$-$1418 & WCDA & 276.29 & -14.32 & 0.19 & 0.81$\pm$0.05 & 210.2 & 39.20$\pm$2.30 & 2.98$\pm$0.05 & &   HESS J1825-137        (0.56)\\
 & KM2A & 276.25 & -14.00 & 0.43 & 0.81$\pm$0.10 & 68.9 & 7.26$\pm$0.98 & 3.53$\pm$0.18 &  & \\
\hline
 1LHAASO J1825$-$1256u & KM2A & 276.44 & -12.94 & 0.08 & $<$0.20 & 259.2 & 5.08$\pm$0.42 & 3.33$\pm$0.13 & 121.0 &   HESS J1826-130        (0.14) \\
 & WCDA & 276.55 & -13.04 & 0.08 & 0.24$\pm$0.03 & 231.0 & 8.27$\pm$0.54 & 2.61$\pm$0.03 & & \\
\hline
 1LHAASO J1825$-$1337u & KM2A & 276.45 & -13.63 & 0.05 & $<$0.18 & 453.7 & 10.10$\pm$0.61 & 3.28$\pm$0.09 & 249.6 &   HESS J1825-137        (0.15) \\
 & WCDA & 276.55 & -13.73 & 0.05 & 0.17$\pm$0.02 & 289.0 & 10.40$\pm$0.56 & 2.55$\pm$0.03 & & \\
\hline
1LHAASO J1831$-$1007u* & WCDA & 277.75 & -10.12 & 0.16 & 0.78$\pm$0.04 & 376.4 & 17.80$\pm$1.06 & 2.71$\pm$0.04 & &   HESS J1831-098        (0.25)\\
 & KM2A & 277.81 & -9.83 & 0.12 & 0.26$\pm$0.05 & 125.4 & 2.56$\pm$0.27 & 3.30$\pm$0.14 & 53.3 & \\
\hline
 1LHAASO J1831$-$1028 & KM2A & 277.84 & -10.48 & 0.39 & 0.94$\pm$0.09 & 100.0 & 5.39$\pm$0.60 & 3.53$\pm$0.15 &  &   HESS J1833-105        (0.55)\\
 & WCDA & & & & & & $<$7.77 & & & \\
\hline
1LHAASO J1834$-$0831 & WCDA & 278.62 & -8.53 & 0.22 & 0.40$\pm$0.07 & 96.0 & 5.99$\pm$1.09 & 3.08$\pm$0.13 & &   HESS J1834-087        (0.24)\\
 & KM2A & 278.44 & -8.38 & 0.28 & 0.40$\pm$0.07 & 68.9 & 1.55$\pm$0.21 & 3.63$\pm$0.21 &  & \\
\hline
1LHAASO J1837$-$0654u & WCDA* & 279.39 & -6.90 & 0.06 & 0.34$\pm$0.01 & 1049.8 & 15.50$\pm$0.54 & 2.92$\pm$0.03 & &   HESS J1837-069        (0.05)\\
 & KM2A & 279.31 & -6.86 & 0.09 & 0.33$\pm$0.04 & 331.2 & 3.06$\pm$0.21 & 3.70$\pm$0.12 & 25.0 & \\
\hline
 1LHAASO J1839$-$0548u & KM2A & 279.79 & -5.81 & 0.06 & 0.22$\pm$0.02 & 364.8 & 3.03$\pm$0.20 & 3.24$\pm$0.09 & 127.7 &   LHAASO J1839-0545     (0.17) \\
 & WCDA* & 279.85 & -5.90 & 0.07 & 0.22$\pm$0.02 & 198.8 & 4.62$\pm$0.29 & 2.65$\pm$0.04 & &  \\
\hline
1LHAASO J1841$-$0519 & WCDA & 280.33 & -5.33 & 0.09 & 0.60$\pm$0.03 & 519.8 & 15.20$\pm$0.84 & 2.88$\pm$0.04 & &   HESS J1841-055        (0.25)\\
 & KM2A* & 280.21 & -5.23 & 0.37 & 0.62$\pm$0.12 & 82.8 & 2.10$\pm$0.25 & 3.85$\pm$0.20 &  & \\
\hline
 1LHAASO J1843$-$0335u & KM2A & 280.91 & -3.60 & 0.04 & 0.36$\pm$0.01 & 1892.2 & 6.19$\pm$0.20 & 3.44$\pm$0.06 & 295.8 &   HESS J1843-033        (0.06) \\
 & WCDA & 281.01 & -3.50 & 0.06 & 0.40$\pm$0.02 & 1162.8 & 9.01$\pm$0.43 & 2.58$\pm$0.03 & &  \\
\hline
1LHAASO J1848$-$0153u & WCDA*& 282.06 & -1.89 & 0.10 & 0.51$\pm$0.03 & 571.2 & 7.05$\pm$0.40 & 2.65$\pm$0.04 & &   HESS J1848-018        (0.11)\\
 & KM2A* & 282.02 & -1.78 & 0.11 & 0.56$\pm$0.03 & 542.9 & 3.29$\pm$0.17 & 3.69$\pm$0.10 & 36.0 & \\
\hline
 1LHAASO J1848$-$0001u & KM2A & 282.19 & -0.02 & 0.04 & $<$0.09 & 655.4 & 1.64$\pm$0.10 & 2.75$\pm$0.07 & 316.8 &   IGR J18490-0000       (0.08) \\
 & WCDA & & & & & & $<$0.95 & & & \\
\hline
1LHAASO J1850$-$0004u* & WCDA & 282.74 & -0.07 & 0.08 & 0.46$\pm$0.02 & 800.9 & 5.30$\pm$0.32 & 2.49$\pm$0.04 & &   HESS J1852-000        (0.36)\\
 & KM2A & 282.89 & -0.07 & 0.07 & 0.21$\pm$0.03 & 349.7 & 1.86$\pm$0.12 & 3.15$\pm$0.09 & 102.0 & \\
\hline
 1LHAASO J1852$+$0050u* & KM2A* & 283.10 & 0.84 & 0.26 & 0.85$\pm$0.06 & 275.6 & 3.22$\pm$0.22 & 3.64$\pm$0.12 & 22.1 &   2HWC J1852+013*       (0.55) \\
 & WCDA & 283.73 & 1.40 & 0.20 & 0.64$\pm$0.07 & 231.0 & 4.67$\pm$0.70 & 2.74$\pm$0.06 & &  \\
\hline
1LHAASO J1857$+$0245 & WCDA & 284.37 & 2.75 & 0.11 & 0.24$\pm$0.04 & 361.0 & 4.04$\pm$0.57 & 2.93$\pm$0.07 & &   HESS J1857+026        (0.11)\\
 & KM2A & & & & & & $<$0.32 & & & \\
\hline
 1LHAASO J1857$+$0203u & KM2A & 284.38 & 2.06 & 0.07 & 0.28$\pm$0.03 & 475.2 & 1.78$\pm$0.10 & 3.31$\pm$0.10 & 112.4 &   HESS J1858+020        (0.21) \\
 & WCDA & 284.50 & 1.98 & 0.11 & 0.19$\pm$0.03 & 187.7 & 1.68$\pm$0.43 & 2.46$\pm$0.11 & &  \\
\hline
 1LHAASO J1858$+$0330 & KM2A* & 284.59 & 3.51 & 0.12 & 0.43$\pm$0.04 & 299.3 & 1.56$\pm$0.10 & 3.78$\pm$0.15 &  &  \\
 & WCDA & 284.79 & 3.70 & 0.34 & 0.52$\pm$0.08 & 114.5 & 2.84$\pm$0.63 & 2.63$\pm$0.10 & &  \\
\hline
1LHAASO J1902$+$0648 & WCDA & 285.58 & 6.80 & 0.10 & $<$0.15 & 46.2 & 0.45$\pm$0.13 & 2.39$\pm$0.18 & & \\
 & KM2A & & & & & & $<$0.06 & & & \\
\hline
1LHAASO J1906$+$0712 & WCDA & 286.56 & 7.20 & 0.21 & 0.21$\pm$0.05 & 57.8 & 1.01$\pm$0.25 & 2.72$\pm$0.15 & & \\
 & KM2A & & & & & & $<$0.19 & & & \\
\hline
1LHAASO J1907$+$0826 & WCDA* & 286.96 & 8.44 & 0.31 & 0.43$\pm$0.08 & 51.8 & 1.34$\pm$0.29 & 2.62$\pm$0.14 & &   2HWC J1907+084*       (0.18)\\
 & KM2A & & & & & & $<$0.29 & & & \\
\hline
 1LHAASO J1908$+$0615u & KM2A & 287.05 & 6.26 & 0.03 & 0.36$\pm$0.01 & 3410.6 & 6.86$\pm$0.16 & 2.82$\pm$0.03 & 912.0 &   MGRO J1908+06         (0.07) \\
 & WCDA & 287.05 & 6.26 & 0.05 & 0.43$\pm$0.02 & 2070.2 & 7.97$\pm$0.54 & 2.42$\pm$0.03 & &  \\
\hline
 1LHAASO J1910$+$0516* & KM2A & 287.55 & 5.28 & 0.15 & $<$0.30 & 74.0 & 0.57$\pm$0.08 & 3.15$\pm$0.18 & &   SS 433 w1             (0.26) \\
 & WCDA & 287.88 & 5.07 & 0.38 & 0.29$\pm$0.09 & 36.0 & 0.86$\pm$0.29 & 2.54$\pm$0.15 & & \\
\hline
1LHAASO J1912$+$1014u & WCDA & 288.22 & 10.25 & 0.08 & 0.36$\pm$0.03 & 585.6 & 3.07$\pm$0.24 & 2.68$\pm$0.06 & &   HESS J1912+101        (0.10)\\
 & KM2A* & 288.38 & 10.50 & 0.13 & 0.50$\pm$0.04 & 346.0 & 1.52$\pm$0.10 & 3.26$\pm$0.11 & 68.9 & \\
\hline
 1LHAASO J1913$+$0501 & KM2A & 288.28 & 5.03 & 0.11 & $<$0.10 & 96.0 & 0.45$\pm$0.06 & 3.30$\pm$0.18 & &   SS 433 e1             (0.15) \\
 & WCDA & & & & & & $<$0.37 & & & \\
\hline
 1LHAASO J1914$+$1150u & KM2A* & 288.73 & 11.84 & 0.09 & 0.21$\pm$0.04 & 259.2 & 0.79$\pm$0.06 & 3.41$\pm$0.13 & 26.0 &   2HWC J1914+117*       (0.13) \\
 & WCDA* & 288.81 & 11.74 & 0.14 & 0.33$\pm$0.04 & 151.3 & 1.09$\pm$0.12 & 2.34$\pm$0.07 & &  \\
\hline
 1LHAASO J1919$+$1556 & KM2A & 289.78 & 15.93 & 0.32 & $<$0.44 & 37.2 & 0.24$\pm$0.04 & 4.71$\pm$0.53 &  &  3HWC J1918+159(0.09) \\
 & WCDA & & & & & & $<$0.03 & & & \\
\hline
1LHAASO J1922$+$1403 & WCDA & 290.70 & 14.06 & 0.07 & 0.18$\pm$0.02 & 256.0 & 1.37$\pm$0.10 & 2.62$\pm$0.07 & &   W 51                  (0.13)\\
 & KM2A & 290.73 & 14.11 & 0.07 & $<$0.10 & 158.8 & 0.45$\pm$0.04 & 3.79$\pm$0.20 &  & \\
\hline
1LHAASO J1924$+$1609 & WCDA* & 291.09 & 16.15 & 0.43 & 1.45$\pm$0.11 & 169.0 & 4.44$\pm$0.41 & 2.54$\pm$0.08 & &  3HWC J1923+169(0.86)\\
 & KM2A* & 290.53 & 15.71 & 0.50 & 1.22$\pm$0.20 & 68.9 & 1.35$\pm$0.17 & 3.61$\pm$0.22 &  & \\
\hline
 1LHAASO J1928$+$1813u & KM2A & 292.07 & 18.23 & 0.14 & 0.63$\pm$0.03 & 306.2 & 2.48$\pm$0.16 & 3.24$\pm$0.08 & 44.9 &   2HWC J1928+177        (0.46)\\
 & WCDA & & & & & & $<$0.48 & & & \\
\hline
1LHAASO J1928$+$1746u & WCDA & 292.14 & 17.78 & 0.07 & 0.17$\pm$0.02 & 196.0 & 0.79$\pm$0.05 & 2.22$\pm$0.05 & &   2HWC J1928+177        (0.01)\\
 & KM2A & 292.17 & 17.89 & 0.07 & $<$0.16 & 127.7 & 0.72$\pm$0.07 & 3.10$\pm$0.12 & 44.9 & \\
\hline
1LHAASO J1929$+$1846u* & WCDA & 292.34 & 18.77 & 0.10 & 0.49$\pm$0.02 & 416.2 & 2.48$\pm$0.11 & 2.37$\pm$0.04 & &   SNR G054.1+00.3       (0.29)\\
 & KM2A & 292.04 & 18.97 & 0.08 & $<$0.21 & 130.0 & 0.64$\pm$0.06 & 3.11$\pm$0.12 & 26.0 & \\
\hline
 1LHAASO J1931$+$1653 & KM2A & 292.79 & 16.90 & 0.11 & $<$0.10 & 51.8 & 0.22$\pm$0.04 & 3.15$\pm$0.25 &  & \\
 & WCDA & & & & & & $<$0.31 & & & \\
\hline
 1LHAASO J1937$+$2128 & KM2A* & 294.32 & 21.48 & 0.51 & 1.43$\pm$0.15 & 134.6 & 1.87$\pm$0.17 & 3.40$\pm$0.15 &  &  3HWC J1935+213(0.36) \\
 & WCDA* & 294.30 & 21.00 & 0.81 & 1.25$\pm$0.23 & 65.6 & 2.08$\pm$0.59 & 2.43$\pm$0.16 & &  \\
\hline
1LHAASO J1945$+$2424* & WCDA & 296.36 & 24.40 & 0.37 & 1.29$\pm$0.11 & 262.4 & 4.27$\pm$0.51 & 2.56$\pm$0.08 & &   2HWC J1949+244        (0.97)\\
 & KM2A & 297.42 & 23.97 & 0.20 & 0.36$\pm$0.06 & 72.2 & 0.40$\pm$0.05 & 3.93$\pm$0.30 &  & \\
\hline
 1LHAASO J1951$+$2608 & KM2A* & 297.94 & 26.15 & 0.42 & 1.00$\pm$0.11 & 100.0 & 1.13$\pm$0.12 & 3.43$\pm$0.17 &  &  3HWC J1951+266(0.47)\\
 & WCDA & & & & & & $<$0.48 & & & \\
\hline
1LHAASO J1952$+$2922 & WCDA & 298.05 & 29.38 & 0.07 & $<$0.12 & 104.0 & 0.55$\pm$0.05 & 2.52$\pm$0.10 & &   2HWC J1953+294        (0.21)\\
 & KM2A & & & & & & $<$0.09 & & & \\
\hline
 1LHAASO J1954$+$2836u & KM2A & 298.55 & 28.60 & 0.07 & $<$0.12 & 123.2 & 0.42$\pm$0.05 & 2.92$\pm$0.14 & 44.9 &   2HWC J1955+285        (0.24)\\
 & WCDA & 298.50 & 28.57 & 0.07 & $<$0.11 & 75.7 & 0.34$\pm$0.04 & 2.22$\pm$0.09 & &\\
\hline
1LHAASO J1954$+$3253 & WCDA & 298.63 & 32.88 & 0.09 & 0.17$\pm$0.03 & 144.0 & 0.62$\pm$0.06 & 2.45$\pm$0.09 & & \\
 & KM2A & & & & & & $<$0.04 & & & \\
\hline
1LHAASO J1956$+$2921 & WCDA & 299.24 & 29.35 & 0.38 & 0.99$\pm$0.07 & 161.3 & 1.47$\pm$0.16 & 2.03$\pm$0.06 & &   LHAASO J1956+2845     (0.63)\\
 & KM2A & 298.84 & 28.92 & 0.23 & 0.78$\pm$0.05 & 151.3 & 1.62$\pm$0.14 & 3.42$\pm$0.12 &  & \\
\hline
 1LHAASO J1959$+$2846u & KM2A & 299.78 & 28.78 & 0.09 & 0.29$\pm$0.03 & 213.2 & 0.84$\pm$0.07 & 2.90$\pm$0.10 & 74.0 &\\
 & WCDA & & & & & & $<$0.38 & & & \\
\hline
 1LHAASO J1959$+$1129u & KM2A & 299.82 & 11.49 & 0.10 & $<$0.21 & 94.1 & 0.27$\pm$0.04 & 2.69$\pm$0.17 & 60.8 & \\
 & WCDA & & & & & & $<$0.11 & & & \\
\hline
1LHAASO J2002$+$3244u & WCDA & 300.64 & 32.74 & 0.09 & $<$0.16 & 74.0 & 0.34$\pm$0.06 & 2.21$\pm$0.11 & & \\
 & KM2A & 300.60 & 32.64 & 0.11 & $<$0.08 & 43.6 & 0.15$\pm$0.03 & 2.70$\pm$0.22 & 28.1 & \\
\hline
1LHAASO J2005$+$3415* & WCDA & 301.30 & 34.25 & 0.16 & 0.74$\pm$0.04 & 388.1 & 3.53$\pm$0.21 & 2.58$\pm$0.05 & &   2HWC J2006+341        (0.22)\\
 & KM2A & 301.81 & 33.87 & 0.15 & 0.33$\pm$0.05 & 127.7 & 0.56$\pm$0.05 & 3.79$\pm$0.21 &  & \\
\hline
 1LHAASO J2005$+$3050 & KM2A & 301.45 & 30.85 & 0.13 & 0.27$\pm$0.05 & 110.2 & 0.46$\pm$0.05 & 3.62$\pm$0.21 & &  3HWC J2005+311(0.32) \\
 & WCDA & 301.37 & 30.99 & 0.13 & 0.21$\pm$0.04 & 60.8 & 0.29$\pm$0.05 & 1.99$\pm$0.10 & &  \\
\hline
 1LHAASO J2018$+$3643u & KM2A & 304.65 & 36.72 & 0.04 & 0.24$\pm$0.01 & 2362.0 & 3.93$\pm$0.11 & 3.46$\pm$0.05 & 139.2 &   MGRO J2019+37         (0.11) \\
 & WCDA & 304.61 & 36.75 & 0.03 & 0.26$\pm$0.01 & 1608.0 & 2.19$\pm$0.06 & 1.94$\pm$0.02 & &  \\
\hline
1LHAASO J2020$+$4034 & WCDA & 305.03 & 40.57 & 0.08 & 0.38$\pm$0.03 & 458.0 & 2.36$\pm$0.16 & 2.91$\pm$0.06 & &   VER J2019+407         (0.19)\\
 & KM2A & 305.20 & 40.43 & 0.14 & $<$0.30 & 70.6 & 0.35$\pm$0.05 & 3.56$\pm$0.23 &  & \\
\hline
1LHAASO J2020$+$3638 & WCDA & 305.14 & 36.63 & 0.34 & 1.27$\pm$0.11 & 88.4 & 3.59$\pm$0.43 & 2.62$\pm$0.09 & &   VER J2019+368         (0.28)\\
 & KM2A & & & & & & $<$0.18 & & & \\
\hline
 1LHAASO J2020$+$3649u & KM2A & 305.23 & 36.82 & 0.04 & 0.12$\pm$0.02 & 912.0 & 2.29$\pm$0.09 & 3.31$\pm$0.06 & 86.5 &   VER J2019+368         (0.30) \\
 & WCDA & 305.25 & 36.91 & 0.18 & 0.15$\pm$0.02 & 357.2 & 0.55$\pm$0.16 & 1.78$\pm$0.18 & &  \\
\hline
 1LHAASO J2027$+$3657 & KM2A & 306.88 & 36.95 & 0.23 & 0.38$\pm$0.05 & 84.6 & 0.50$\pm$0.06 & 3.21$\pm$0.17 &  &\\
 & WCDA & & & & & & $<$0.29 & & & \\
\hline
 1LHAASO J2028$+$3352 & KM2A & 307.21 & 33.88 & 0.86 & 1.70$\pm$0.23 & 75.7 & 1.61$\pm$0.19 & 3.38$\pm$0.19 &  &\\
 & WCDA & & & & & & $<$0.77 & & & \\
\hline
1LHAASO J2031$+$4052u* & WCDA & 307.90 & 40.88 & 0.19 & 0.25$\pm$0.05 & 57.8 & 0.77$\pm$0.12 & 2.81$\pm$0.16 & &   LHAASO J2032+4102     (0.20)\\
 & KM2A & 308.14 & 40.88 & 0.13 & $<$0.08 & 33.6 & 0.08$\pm$0.02 & 2.13$\pm$0.27 & 38.4 & \\
\hline
 1LHAASO J2031$+$4127u & KM2A & 307.95 & 41.46 & 0.03 & 0.22$\pm$0.01 & 1953.6 & 2.56$\pm$0.08 & 3.45$\pm$0.06 & 136.9 &   TeV J2032+4130        (0.12) \\
 & WCDA & 307.85 & 41.60 & 0.04 & 0.27$\pm$0.01 & 1521.0 & 3.07$\pm$0.12 & 2.29$\pm$0.03 & &  \\
\hline
 1LHAASO J2047$+$4434 & KM2A* & 311.92 & 44.58 & 0.23 & 0.42$\pm$0.09 & 62.4 & 0.46$\pm$0.06 & 3.17$\pm$0.20 &  &\\
 & WCDA & & & & & & $<$0.33 & & & \\
\hline
 1LHAASO J2108$+$5153u & KM2A & 317.10 & 51.90 & 0.04 & 0.19$\pm$0.02 & 942.5 & 1.38$\pm$0.07 & 2.97$\pm$0.07 & 252.8 &   LHAASO J2108+5157     (0.06) \\
 & WCDA* & 316.83 & 51.95 & 0.09 & 0.14$\pm$0.03 & 81.0 & 0.11$\pm$0.09 & 1.56$\pm$0.34& &  \\
\hline
 1LHAASO J2200$+$5643u & KM2A & 330.08 & 56.73 & 0.13 & 0.54$\pm$0.05 & 368.6 & 1.70$\pm$0.10 & 3.44$\pm$0.10 & 38.4 &  \\
 & WCDA & 330.38 & 56.73 & 0.20 & 0.43$\pm$0.07 & 75.7 & 0.38$\pm$0.24 & 1.77$\pm$0.28 & &  \\
\hline
 1LHAASO J2228$+$6100u & KM2A & 337.01 & 61.00 & 0.04 & 0.35$\pm$0.01 & 2180.9 & 4.76$\pm$0.14 & 2.95$\pm$0.04 & 605.2 &   SNR G106.3+02.7       (0.13) \\
 & WCDA & 336.79 & 61.02 & 0.05 & 0.25$\pm$0.02 & 576.0 & 2.37$\pm$0.16 & 2.26$\pm$0.04 & &  \\
\hline
1LHAASO J2229$+$5927u & WCDA & 337.26 & 59.45 & 0.36 & 1.98$\pm$0.10 & 228.0 & 14.90$\pm$0.99 & 2.67$\pm$0.05 & & \\
 & KM2A* & 337.88 & 59.55 & 0.46 & 1.74$\pm$0.16 & 163.8 & 4.43$\pm$0.36 & 3.53$\pm$0.11 & 31.4 & \\
\hline
 1LHAASO J2238$+$5900 & KM2A & 339.54 & 59.00 & 0.09 & 0.43$\pm$0.03 & 353.4 & 2.03$\pm$0.12 & 3.55$\pm$0.09 & &  \\
 & WCDA* & 339.40 & 58.92 & 0.18 & 0.51$\pm$0.04 & 110.2 & 1.91$\pm$0.26 & 2.39$\pm$0.07 & &  \\
\hline
1LHAASO J2323$+$5854 & WCDA & 350.80 & 58.90 & 0.14 & $<$0.24 & 50.4 & 1.10$\pm$0.17 & 3.18$\pm$0.16 & &   Cassiopeia A          (0.10)\\
 & KM2A & & & & & & $<$0.15 & & & \\
\hline
1LHAASO J2346$+$5138 & WCDA & 356.70 & 51.64 & 0.12 & $<$0.16 & 47.6 & 0.57$\pm$0.09 & 3.26$\pm$0.17 & &   1ES 2344+514          (0.08)\\
 & KM2A & & & & & & $<$0.07 & & & \\
\hline
\enddata
\tablecomments{The first column lists the LHAASO catalog name. The name designation is 1LHAASO JHHMM+DDMM, where the 1 refers to this being the first LHAASO catalog,   JHHMM+DDMM is according to the source location. For a source observed by both WCDA and KM2A, the source name corresponds to the position of that with higher significance level.  If the source is an UHE source with TS$_{100} > 20$, we add  ``u" at the end of the name, i.e., 1LHAASO JHHMM+DDMMu. For dubiously merged sources, we label the names with ``*". }
\tablecomments{The 2nd column lists the component name, represented by the detector name. If the component suffers  a significant GDE impact in our test, we label the component with ``*".  }
\tablecomments{The 3rd $-$ 5th columns list the position parameters for the source components. $\alpha_{2000}$ and $\delta_{2000}$ are the right ascension and declination at J2000.0 epoch. $\sigma_{p,95,stat}$ is the 95\% statistical uncertainty of the source component position. The units are degrees.}
\tablecomments{The 6th column lists the extension of the source components.  $r_{39}$ is the 39\% containment radius of the 2D-Gaussian model. For the point-like source components, the 95\% statistical upper limits (i.e., $< r_{39,ul,stat}$ ) are shown. } 
\tablecomments{The 7th column lists the TS values of the source components. It is distributed as $\chi^2$ with 5 dof. } 
\tablecomments{The 8th $-$ 9th columns list the parameters of the power-law SED of the source components. The  power-law shape is defined by 
$dN/dE=N_{0}(E/E_{0})^{-\Gamma}$, where $N_{0}$ is the differential flux, $E_{0}$ is the reference energy which is 3 TeV for the WCDA component and 50 TeV for the KM2A component, and $\Gamma$ is the photon spectral index. $N_{0}$ is in a units of $10^{-13}$ cm$^{-2}$ s$^{-1}$ TeV$^{-1}$ and  $10^{-16}$ cm$^{-2}$ s$^{-1}$ TeV$^{-1}$  for the WCDA  the KM2A components, respectively. For a source with  single component detection, we give the 95\% statistical upper limits of $N_{0}$ of the non-detected component with the same position, extension and photon index.}
\tablecomments{The 10th column lists the TS values of  the UHE sources at energies $E >$ 100 TeV. } 
\tablecomments{The 11th column lists the closest known TeV source counterpart within the searching region (as described in Sec.~\ref{asso}).  The angular separation between the TeV counterpart and the LHAASO source is shown in parentheses.} 
\end{deluxetable}

\clearpage
\startlongtable
\begin{deluxetable}{lcccccc}
\setlength{\tabcolsep}{0.02in}
\tablewidth{660pt}
\tabletypesize{\scriptsize}
\rotate
\tablecaption{Associations of the New TeV Sources in 1LHAASO Catalog \label{tab:2}}
\tablehead{
\colhead{Name} &
\colhead{Sep.[$^{\circ}$]} &
\colhead{Assoc.} &
\colhead{Description} 
}
\startdata
1LHAASO J0007$+$5659u & &  &   \\
\hline
1LHAASO J0206$+$4302u & &  &   \\
\hline
1LHAASO J0212$+$4254u & &  &   \\
\hline
1LHAASO J1937$+$2128 & &  &   \\
\hline
1LHAASO J1959$+$1129u & &  &   \\
\hline
1LHAASO J2200$+$5643u & &  &   \\
\hline
1LHAASO J2229$+$5927u & &  &   \\
\hline
1LHAASO J0056$+$6346u & 0.38 & 4FGL J0057.9+6326 &  \\
\hline
1LHAASO J0500$+$4454 & 0.32 & 4FGL J0501.7+4459 &  \\
\hline
1LHAASO J1858$+$0330 & 0.31 & 4FGL J1857.9+0313c & LQAC 284+003;bcu; \\
& 0.41 & 4FGL J1858.0+0354 & \\
\hline
1LHAASO J1902$+$0648 & 0.12 & 4FGL J1902.5+0654 &  \\
\hline
1LHAASO J1924$+$1609 & 0.12 & 4FGL J1924.3+1601c &  \\
& 0.29 & 4FGL J1925.4+1616 & \\
& 0.33 & 4FGL J1924.3+1628 & \\
\hline
1LHAASO J1931$+$1653 & 0.05 & 4FGL J1931.1+1656 &  \\
\hline
1LHAASO J2027$+$3657 & 0.40 & 4FGL J2026.5+3718c & NVSS J202726+372249;unk; \\
\hline
1LHAASO J2047$+$4434 & 0.32 & 4FGL J2049.3+4440c &  \\
\hline
1LHAASO J0216$+$4237u & 0.33 & ATNF PSR J0218+4232 & $\dot{E}=2.44\times 10^{35} \rm\ erg\ s^{-1}$,$\tau_{c}=476000.0\rm\ kyr$,$d=3.15 \rm\ kpc$ \\
 & 0.33 & 4FGL J0218.1+4232 & PSR J0218+4232;MSP;\\
\hline
1LHAASO J0249$+$6022 & 0.16 & ATNF PSR J0248+6021 & $\dot{E}=2.13\times 10^{35} \rm\ erg\ s^{-1}$,$\tau_{c}=62.4\rm\ kyr$,$d=2.00 \rm\ kpc$ \\
 & 0.14 & 4FGL J0248.4+6021 & PSR J0248+6021;PSR;\\
\hline
1LHAASO J0359$+$5406 & 0.12 & SNRcat G148.1+00.8 & \textcolor{blue}{http://SNRcat.physics.umanitoba.ca/SNRrecord.php?id=G148.1p00.8} \\
& 0.12 & ATNF PSR B0355+54 & $\dot{E}=4.54\times 10^{34} \rm\ erg\ s^{-1}$,$\tau_{c}=564.0\rm\ kyr$,$d=1.00 \rm\ kpc$ \\
& 0.15 & ATNF PSR J0359+5414 & $\dot{E}=1.32\times 10^{36} \rm\ erg\ s^{-1}$,$\tau_{c}=75.2\rm\ kyr$ \\
 & 0.15 & 4FGL J0359.4+5414 & PSR J0359+5414;PSR;\\
\hline
1LHAASO J0428$+$5531 & 0.26 & SNRcat G150.3+04.5 & \textcolor{blue}{http://SNRcat.physics.umanitoba.ca/SNRrecord.php?id=G150.3p04.5} \\
 & 0.48 & 4FGL J0425.6+5522e & SNR G150.3+04.5;SNR;$r_{68}$=1.359 Gaussian;\\
\hline
1LHAASO J0534$+$3533 & 0.30 & SNRcat G172.8+01.5 & \textcolor{blue}{http://SNRcat.physics.umanitoba.ca/SNRrecord.php?id=G172.8p01.5} \\
\hline
1LHAASO J1814$-$1636u & 0.31 & SNRcat G014.3+00.1 & \textcolor{blue}{http://SNRcat.physics.umanitoba.ca/SNRrecord.php?id=G005.7m00.1} \\
 & 0.43 & SNRcat G014.1-00.1 & \textcolor{blue}{http://SNRcat.physics.umanitoba.ca/SNRrecord.php?id=G005.7m00.1} \\
 & 0.44 & 4FGL J1816.2-1654c & PMN J1816-1649;unk;\\
\hline
1LHAASO J1831$-$1028 & 0.30 & SNRcat G021.0-00.4 & \textcolor{blue}{http://SNRcat.physics.umanitoba.ca/SNRrecord.php?id=G053.6m02.2} \\
 & 0.36 & SNRcat G021.5-00.1 & \textcolor{blue}{http://SNRcat.physics.umanitoba.ca/SNRrecord.php?id=G053.6m02.2} \\
 & 0.49 & SNRcat G021.8-00.6 & \textcolor{blue}{http://SNRcat.physics.umanitoba.ca/SNRrecord.php?id=G053.6m02.2} \\
\hline
1LHAASO J1852$+$0050u & 0.17 & SNRcat G033.6+00.1 & \textcolor{blue}{http://SNRcat.physics.umanitoba.ca/SNRrecord.php?id=G033.6p00.1} \\
& 0.31 & ATNF PSR J1853+0056 & $\dot{E}=4.03\times 10^{34} \rm\ erg\ s^{-1}$,$\tau_{c}=204.0\rm\ kyr$,$d=3.84 \rm\ kpc$\\
 & 0.21 & 4FGL J1852.4+0037e & Kes 79;spp;0.63 Disk;\\
\hline
1LHAASO J1906$+$0712 & 0.34 & SNRcat G041.1-00.3 & \textcolor{blue}{http://SNRcat.physics.umanitoba.ca/SNRrecord.php?id=G041.1m00.3} \\
& 0.19 & ATNF PSR J1906+0722 & $\dot{E}=1.02\times 10^{36} \rm\ erg\ s^{-1}$,$\tau_{c}=49.2\rm\ kyr$\\
& 0.19 & 4FGL J1906.9+0712 & \\
& 0.20 & 4FGL J1906.4+0723 & PSR J1906+0722;PSR;\\
& 0.37 & 4FGL J1907.6+0703 & 3C 397;snr;\\
\hline
1LHAASO J1928$+$1813u & 0.39 & SNRcat G053.4+00.0 & \textcolor{blue}{http://SNRcat.physics.umanitoba.ca/SNRrecord.php?id=G053.4p00.0} \\
& 0.47 & ATNF PSR J1928+1746 & $\dot{E}=1.60\times 10^{36} \rm\ erg\ s^{-1}$,$\tau_{c}=82.6\rm\ kyr$,$d=4.34 \rm\ kpc$\\
& 0.22 & 4FGL J1928.4+1801c & MG2 J192830+1759;unk;\\
& 0.48 & 4FGL J1929.8+1832 & 1RXS J193012.1+183201;unk;\\
\hline
1LHAASO J1954$+$3253 & 0.25 & SNRcat G069.0+02.7 & \textcolor{blue}{http://SNRcat.physics.umanitoba.ca/SNRrecord.php?id=G069.0p02.7} \\
& 0.33 & ATNF PSR B1951+32 & $\dot{E}=3.74\times 10^{36} \rm\ erg\ s^{-1}$,$\tau_{c}=107.0\rm\ kyr$,$d=3.00 \rm\ kpc$\\
& 0.32 & 4FGL J1952.9+3252 & PSR J1952+3252;PSR;\\
& 0.49 & 4FGL J1955.1+3321 & SNR G069.0+02.7;spp;\\
\hline
1LHAASO J1956$+$2921 & 0.36 & SNRcat G066.0-00.0 & \textcolor{blue}{http://SNRcat.physics.umanitoba.ca/SNRrecord.php?id=G066.0m00.0} \\
\hline
1LHAASO J1959$+$2846u & 0.15 & SNRcat G065.8-00.5 & \textcolor{blue}{http://SNRcat.physics.umanitoba.ca/SNRrecord.php?id=G001.9p00.3} \\
 & 0.39 & SNRcat G066.0-00.0 & \textcolor{blue}{http://SNRcat.physics.umanitoba.ca/SNRrecord.php?id=G001.9p00.3} \\
& 0.10 & ATNF PSR J1958+2846 & $\dot{E}=3.42\times 10^{35} \rm\ erg\ s^{-1}$,$\tau_{c}=21.7\rm\ kyr$,$d=1.95 \rm\ kpc$\\
 & 0.09 & 4FGL J1958.7+2846 & PSR J1958+2846;PSR;\\
\hline
1LHAASO J2002$+$3244u & 0.04 & SNRcat G069.7+01.0 & \textcolor{blue}{http://SNRcat.physics.umanitoba.ca/SNRrecord.php?id=G069.7p01.0} \\
& 0.47 & ATNF PSR B2000+32 & $\dot{E}=1.23\times 10^{34} \rm\ erg\ s^{-1}$,$\tau_{c}=105.0\rm\ kyr$,$d=6.46 \rm\ kpc$\\
 & 0.05 & 4FGL J2002.3+3246 & SNR G069.7+01.0;spp;\\
\hline
1LHAASO J2028$+$3352 & 0.36 & ATNF PSR J2028+3332 & $\dot{E}=3.48\times 10^{34} \rm\ erg\ s^{-1}$,$\tau_{c}=576.0\rm\ kyr$ \\
& 0.36 & 4FGL J2028.3+3331 & PSR J2028+3332;PSR;\\
& 0.40 & 4FGL J2027.0+3343 & 1RXS J202658.5+334253;unk;\\
\hline
1LHAASO J2238$+$5900 & 0.07 & ATNF PSR J2238+5903 & $\dot{E}=8.89\times 10^{35} \rm\ erg\ s^{-1}$,$\tau_{c}=26.6\rm\ kyr$,$d=2.83 \rm\ kpc$ \\
 & 0.07 & 4FGL J2238.5+5903 & PSR J2238+5903;PSR;\\
 \hline
1LHAASO J1219$+$2915 & 0.05 &  NGC 4278 & AGN; z=0.002; \\
\hline
\hline
\enddata
\tablecomments{In the description of the 4FGL counterparts, `unk' represents $|b| < 10°$ sources solely associated with the likelihood-ratio method from large radio and X-ray surveys, `bcu' is a blazar candidate of uncertain type, `spp' is a supernova remnant or pulsar wind nebula, `PSR' is a gamma-ray pulsar identified by pulsations, and `snr' is the Supernova remnant~\citep{2022ApJS..260...53A}.}
\end{deluxetable}

\clearpage
\startlongtable
\begin{deluxetable}{lccccccc}
\setlength{\tabcolsep}{0.05in}
\rotate
\tablecaption{1LHAASO sources associated pulsars\label{tab:pulsar}}
\tablehead{
\colhead{Source name} &
\colhead{PSR name} &
\colhead{Sep.($^{\circ}$)}&
\colhead{d (kpc)} &
\colhead{$\tau_c$ (kyr)} &
\colhead{$\dot{E}$ ($\rm erg\ s^{-1}$)} &
\colhead{$P_c$} &
\colhead{Identified type in TeVCat} 
}
\startdata
1LHAASO J0007+7303u & PSR J0007+7303 & 0.05 & 1.40 & 14 & 4.5e+35  & 7.3e-05 & PWN\\
1LHAASO J0216+4237u & PSR J0218+4232 & 0.33 & 3.15 & 476000 & 2.4e+35  & 3.6e-03 \\
1LHAASO J0249+6022 & PSR J0248+6021 & 0.16 & 2.00 & 62 & 2.1e+35  & 1.5e-03 \\
1LHAASO J0359+5406 & PSR J0359+5414 & 0.15 & - & 75 & 1.3e+36  & 7.2e-04 \\
1LHAASO J0534+2200u & PSR J0534+2200 & 0.01 & 2.00 & 1 & 4.5e+38  & 3.2e-06 & PWN  \\
1LHAASO J0542+2311u & PSR J0543+2329 & 0.30 & 1.56 & 253 & 4.1e+34  & 8.3e-03 \\
1LHAASO J0622+3754 & PSR J0622+3749 & 0.09 & - & 208 & 2.7e+34  & 2.5e-04 & PWN/TeV Halo\\
1LHAASO J0631+1040 & PSR J0631+1037 & 0.11 & 2.10 & 44 & 1.7e+35  & 3.5e-04 & PWN \\
1LHAASO J0634+1741u & PSR J0633+1746 & 0.12 & 0.19 & 342 & 3.3e+34  & 1.3e-03 & PWN/TeV Halo\\
1LHAASO J0635+0619 & PSR J0633+0632 & 0.39 & 1.35 & 59 & 1.2e+35  & 9.4e-03 \\
1LHAASO J1740+0948u & PSR J1740+1000 & 0.21 & 1.23 & 114 & 2.3e+35  & 1.4e-03 \\
1LHAASO J1809-1918u & PSR J1809-1917 & 0.05 & 3.27 & 51 & 1.8e+36  & 6.2e-04 \\
1LHAASO J1813-1245 & PSR J1813-1245 & 0.01 & 2.63 & 43 & 6.2e+36  & 6.3e-06 \\
1LHAASO J1825-1256u & PSR J1826-1256 & 0.09 & 1.55 & 14 & 3.6e+36  & 1.6e-03 \\
1LHAASO J1825-1337u & PSR J1826-1334 & 0.11 & 3.61 & 21 & 2.8e+36  & 2.8e-03 & PWN/TeV Halo \\
1LHAASO J1837-0654u & PSR J1838-0655 & 0.12 & 6.60 & 23 & 5.6e+36  & 2.2e-03 & PWN \\
1LHAASO J1839-0548u & PSR J1838-0537 & 0.20 & - & 5 & 6.0e+36  & 6.1e-03 \\
1LHAASO J1848-0001u & PSR J1849-0001 & 0.06 & - & 43 & 9.8e+36  & 1.2e-04 & PWN \\
1LHAASO J1857+0245 & PSR J1856+0245 & 0.16 & 6.32 & 21 & 4.6e+36  & 3.1e-03 & PWN \\
1LHAASO J1906+0712 & PSR J1906+0722 & 0.19 & - & 49 & 1.0e+36  & 5.9e-03 \\
1LHAASO J1908+0615u & PSR J1907+0602 & 0.23 & 2.37 & 20 & 2.8e+36  & 6.8e-03 \\
1LHAASO J1912+1014u & PSR J1913+1011 & 0.13 & 4.61 & 169 & 2.9e+36  & 1.5e-03 \\
1LHAASO J1914+1150u & PSR J1915+1150 & 0.09 & 14.01 & 116 & 5.4e+35  & 1.8e-03 \\
1LHAASO J1928+1746u & PSR J1928+1746 & 0.04 & 4.34 & 83 & 1.6e+36  & 1.6e-04 \\
1LHAASO J1929+1846u  & PSR J1930+1852 & 0.29 & 7.00 & 3 & 1.2e+37  & 2.6e-03 & PWN \\
1LHAASO J1954+2836u & PSR J1954+2836 & 0.01 & 1.96 & 69 & 1.1e+36  & 1.6e-05 & PWN\\
1LHAASO J1954+3253 & PSR J1952+3252 & 0.33 & 3.00 & 107 & 3.7e+36  & 6.7e-03 \\
1LHAASO J1959+2846u & PSR J1958+2845 & 0.10 & 1.95 & 22 & 3.4e+35  & 2.8e-03 & PWN \\
1LHAASO J2005+3415 & PSR J2004+3429 & 0.25 & 10.78 & 18 & 5.8e+35  & 9.9e-03 \\
1LHAASO J2005+3050 & PSR J2006+3102 & 0.20 & 6.04 & 104 & 2.2e+35  & 9.2e-03 \\
1LHAASO J2020+3649u & PSR J2021+3651 & 0.05 & 1.80 & 17 & 3.4e+36  & 1.5e-04 & PWN \\
1LHAASO J2028+3352 & PSR J2028+3332 & 0.36 & - & 576 & 3.5e+34  & 8.0e-03 \\
1LHAASO J2031+4127u & PSR J2032+4127 & 0.08 & 1.33 & 201 & 1.5e+35  & 1.0e-03 &PWN \\
1LHAASO J2228+6100u & PSR J2229+6114 & 0.27 & 3.00 & 10 & 2.2e+37  & 2.2e-03 &PWN\\
1LHAASO J2238+5900 & PSR J2238+5903 & 0.07 & 2.83 & 27 & 8.9e+35  & 3.0e-04 \\
\enddata
\end{deluxetable}
\end{document}